\documentclass[twocolumn,english,aps,prx,twocolum,superscriptaddress,bibnotes,amsmath,amssymb,floatfix]{revtex4-1}
\usepackage[colorlinks=true,citecolor=blue,linkcolor=magenta]{hyperref}

\usepackage{changes}
\usepackage{soul}
\usepackage{comment}
\usepackage[utf8]{inputenc}
\usepackage[english]{babel}
\usepackage{amsmath,amsfonts,amssymb}
\usepackage[T1]{fontenc}
\usepackage{url}
\usepackage{subfiles}

\usepackage{amsmath}
\usepackage{amsfonts}
\usepackage{amssymb}
\usepackage[version=3]{mhchem}

\usepackage{epstopdf}
\usepackage{graphicx}
\graphicspath{{./Figures/}}

\usepackage[utf8]{inputenc}
\usepackage[english]{babel}
\usepackage{amsmath,amsfonts,amssymb}
\usepackage[T1]{fontenc}
\usepackage{url}

\usepackage{amsmath}
\usepackage{amsfonts}
\usepackage{amssymb}
\usepackage[version=3]{mhchem}

\usepackage{epstopdf}
\usepackage{graphicx}
\graphicspath{{./Figures/}}

\begin{document}

\title{A hybrid integrated dual-microcomb source}

\author{Nikita Yu. Dmitriev}
\thanks{These authors contributed equally to this work.}
\affiliation{Russian Quantum Center, Moscow, 143026, Russia}
\affiliation{Moscow Institute of Physics and Technology (MIPT), Dolgoprudny, Moscow Region, 141701, Russia}

\author{Sergey N. Koptyaev}
\thanks{These authors contributed equally to this work.}
\affiliation{Samsung R\&D Institute Russia, SAIT-Russia Laboratory, Moscow, 127018, Russia}

\author{Andrey S. Voloshin}
\thanks{These authors contributed equally to this work.}
\affiliation{Institute of Physics, Swiss Federal Institute of Technology Lausanne (EPFL), CH-1015 Lausanne, Switzerland}

\author{Nikita M. Kondratiev}
\thanks{These authors contributed equally to this work.}
\affiliation{Russian Quantum Center, Moscow, 143026, Russia}

\author{Valery E. Lobanov}
\thanks{These authors contributed equally to this work.}
\affiliation{Russian Quantum Center, Moscow, 143026, Russia}

\author{Kirill N. Min'kov}
\affiliation{Russian Quantum Center, Moscow, 143026, Russia}

\author{Maxim V. Ryabko}
\affiliation{Samsung R\&D Institute Russia, SAIT-Russia Laboratory, Moscow, 127018, Russia}

\author{Stanislav V. Polonsky}
\affiliation{Samsung R\&D Institute Russia, SAIT-Russia Laboratory, Moscow, 127018, Russia}

\author{Igor A. Bilenko}
\affiliation{Russian Quantum Center, Moscow, 143026, Russia}
\affiliation{Faculty of Physics, M.V. Lomonosov Moscow State University, 119991 Moscow, Russia}

\maketitle

\section*{Abstract}

\noindent{\textbf{\noindent
Dual-comb interferometry is based on self-heterodyning two optical frequency combs, with corresponding mapping of the optical spectrum into the radio-frequency domain. The dual-comb enables diverse applications, including metrology, fast high-precision spectroscopy with high signal-to-noise ratio, distance ranging, and coherent optical communications. However, current dual-frequency-comb systems are designed for research applications and typically rely on scientific equipment and bulky mode-locked lasers. Here we demonstrate for the first time a fully integrated power-efficient dual-microcomb source that is electrically driven and allows turnkey operation. Our implementation uses commercially available components, including distributed-feedback and Fabry--Perot laser diodes, and silicon nitride photonic circuits with microresonators fabricated in commercial multi-project wafer runs. Our devices are therefore unique in terms of size, weight, power consumption, and cost. Laser-diode self-injection locking relaxes the requirements on microresonator spectral purity and Q-factor, so that we can generate soliton microcombs resilient to thermal frequency drift and with pump-to-comb sideband efficiency of up to 40\% at mW power levels. We demonstrate down-conversion of the optical spectrum from 1400 nm to 1700 nm into the radio-frequency domain, which is valuable for fast wide-band Fourier spectroscopy, which was previously not available with chip-scale devices. Our findings pave the way for further
integration of miniature microcomb-based sensors and devices for high-volume applications, thus opening up the prospect of innovative products that redefine the market of industrial and consumer mobile and wearable devices and sensors.}}

\section{Introduction}

\begin{figure*}[t!]
\centering
\includegraphics[width=\textwidth]{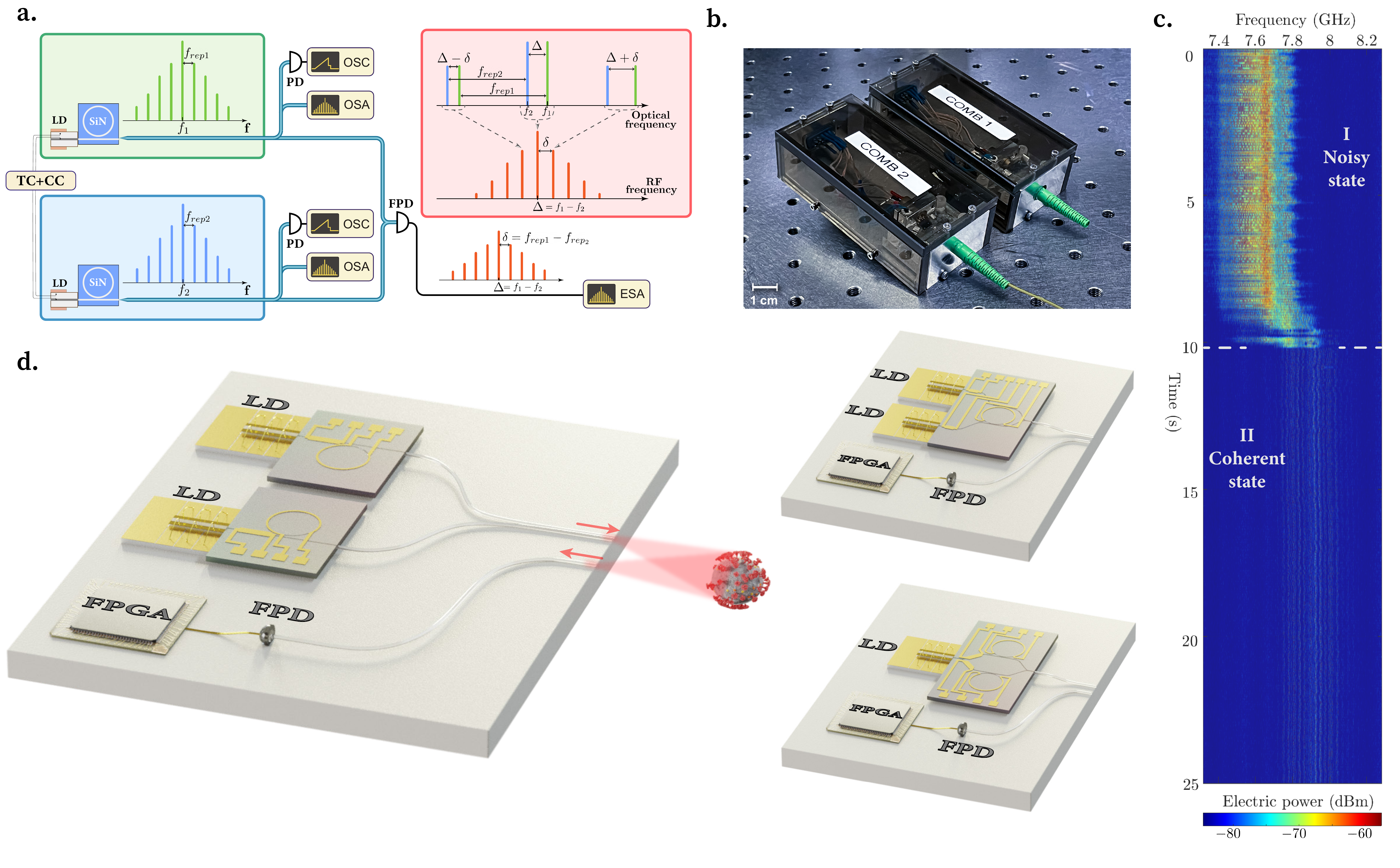}
\caption{Principle of the hybrid integrated dual-microcomb source \textbf{a.} Sketch of the experimental setup illustrating integrated dual-comb signal generation based on the two separate soliton microcomb sources (marked with green and blue rectangles): LD -- semiconductor laser diode; SiN -- photonic chip with high-Q silicon nitride microresonator; TC and CC -- temperature and current controllers; FPD -- fast photo detector; OSA -- optical spectrum analyzer; ESA -- electrical spectrum analyzer.\textbf{b.} Photograph of the portable turnkey dual-comb source comprising two standalone matched integrated soliton microcomb sources \textbf{c.} Evolution of the experimentally observed dual-comb signal based on two matched soliton microcombs during the search for the operating point.  
Demonstration of the dual-comb signal transition from a noisy (Area I) to a soliton state (Area II) by fine-tuning the LD current. \textbf{d.} Various concepts of dual-comb spectrometers: two separate  microcomb-generating photonic chips pumped with two LDs; single photonic chip generating two microcombs pumped with two LDs; and single photonic chip generating two microcombs pumped with a single LD. FPGA is a field-programmable gate array.}
\label{fig:fig1}
\end{figure*}

Over the past few decades, optical frequency combs have become a versatile tool for addressing scientific and technical challenges \cite{Udem2002,Fortier2019,Diddams-2021}. One of the most promising and widely employed applications is the use of a double (or, dual) optical combs for the efficient transfer of signals from the optical domain into the radio frequency (RF) range, thus greatly simplifying data acquisition and subsequent processing. For instance, dual-comb spectroscopy is a remarkable form of Fourier spectroscopy, enabling ultra-fast measurement of broadband optical absorption spectra that provide fingerprints of specific materials or their quantity in the sample using a single photodetector, without the need for moving components, and only a few basic optical components \cite{Suh2016,Coddington:16,Ideguchi:17,Picque2019,baumann2019dual,koptyaev2019optical}. The basic idea of the dual-comb technique is to combine two coherent optical frequency combs with shifted pump lines ($f_{1}$,$f_{2}$) and slightly different line spacing in the frequency domain ($f_{rep1}$,$f_{rep2}$) (Fig. \ref{fig:fig1}(a)). Thereby, the optical spectrum of the combs transmitted through the matter is down-converted at the photodetector into the RF band for measurement. The resulting signal is a RF frequency comb with $\delta = |f_{rep 1} - f_{rep 2}|$ line spacing, a central line located at $\Delta = |f_{1}-f_{2}|$, and line amplitudes uniquely defined by the corresponding lines of the optical combs.
These characertistics make dual-comb techniques highly attractive for myriad practical applications, such as ultra-broadband near-IR spectroscopy \cite{Okubo:14,Okubo-2015-2}, near-field microscopy for sub-wavelength spatial resolution \cite{Brehm:06,vonRibbeck:08}, precision metrology of molecular-line center frequencies \cite{ZOLOT2013}, greenhouse-gas monitoring \cite{Baumann:2011,Rieker:14,Zhu:2015}, combustion diagnosis \cite{SCHROEDER:2017}, and distance ranging (LIDAR) \cite{Coddington2009,Suh2018,Trocha:2018,Nurnberg:21,riemensberger2020massively}.

There exist a number of rapidly developing approaches for implementing dual-comb techniques. Various dual-comb systems are based on conventional fibre mode-locked lasers (MLLs) \cite{coddington2010coherent,hoghooghi202111,Kayes:18,Link:16}. The use of mode-locked integrated external-cavity surface emitting lasers (MIXSELs) potentially allows dual-comb signal generation with a single cavity, exploiting different polarisation states \cite{link2015dual,nurnberg2019single}. Hybrid THz dual-comb spectrometers based on quantum cascade lasers (QCLs) \cite{Consolino2020} and coherently averaged dual-comb spectrometers \cite{komagata2021coherently} demonstrated fast and high-accuracy spectroscopy measurements in the mid-wavelength infrared (MWIR) and long-wavelength infrared (LWIR) ranges. These results have been obtained in laboratory settings, which means that many of these dual-comb systems are only partially integrated and rather bulky and complicated, requiring a range of auxiliary equipment and technical expertise. As a consequence, they are not suitable for industrial and consumer applications, despite outstanding performance in laboratory environments. Therefore, integration and device miniaturisation are pressing issues that need to be addressed before any of the approaches can be utilized in industrial-grade device. Fully integrated dual-comb systems hold the promise to unlock major applications, including airborne and spaceborne sensors, distance ranging with unprecedented speed and resolution, and compact spectroscopy sensors. In turn, such devices could become a core technology for consumer and wearable applications, including non-invasive spectroscopic sensors.

The most promising platform for a fully integrated dual-comb source is silicon-based integrated photonics. In recent years, this platform has experienced major advances and reached considerable maturity \cite{pfeiffer2016photonic,liu2018ultralow,liu2021high}. Today, the performance of low-loss silicon-based photonic systems has become comparable with that of free-space optic systems. In addition, high-level compatibility with CMOS fabrication processes \cite{hochberg2010towards,hochberg2013silicon,weimann2017silicon} and also with the III--V semiconductor platform \cite{xiang2021laser,xiang2021high,cuyvers2021low} have been demonstrated. Recent progress in silicon photonics, combined with the self-injection locking (SIL) effect \cite{Dahmani:87, Laurent1989,Oraevsky2001,Liang:10,liang2015high,liang2015ultralow,Kondratiev:17,Galiev2020,Voloshin2021}, enabled optical microcomb generation using semiconductor laser diodes (LDs) instead of bulky narrow-linewidth lasers, thereby greatly simplifying the process of microcomb generation and paving the way for the development of fully integrated chip-scale single microcomb sources based on high-Q microresonators (MRs)  \cite{xiang2021laser,xiang2021high,shen2019integrated,raja2019electrically,stern2018battery}.
Proof-of-concept experiments based on passive high-Q MRs achieved broadband optical spectrum down-conversion to the RF range, generating ultra-wide frequency combs with widely variable (from GHz to THz) line spacing in the near-infrared (NIR), SWIR, and visible-wavelength ranges in bulk and on-chip structures  \cite{Kippenberg:2011,Lin:17,PASQUAZI20181,Gaeta2019,Kovach:20,lin2020broadband,Bao:19,Dutt:2018,Wang:18,yu2018silicon,pavlov2017soliton}. Also, the recently demonstrated scanning dual-comb spectroscopy (SDCS) technique allows additionally to increase the resolution of spectroscopic systems based on high-Q MRs \cite{lin2020broadband}.

Here, we report a feasibility study of dual-comb integration and introduce the first hybrid integrated dual-microcomb source for the SWIR range based on commercially available low-cost components (Fig. \ref{fig:fig1}(b)). With the assembled prototype we have successfully down-converted a 300-nm wide optical spectrum to a 600-MHz wide RF signal. Our findings establish that electrically driven soliton microcombs comprising integrated SiN high-Q MRs combined with SIL semiconductor LDs (Fig. \ref{fig:fig1}(a)) are a promising technology platform for highly integrated energy-efficient dual-comb sources covering wide spectral ranges. Specifically, we demonstrate that SIL provides up to 40{\%} pump-to-comb sideband conversion efficiency ($\eta_{p2c}$) for bright solitons. 
In the light of such a high conversion value we consider in detail the $\eta_{p2c}$ efficiency of SIL-enabled generation of bright dissipative Kerr solitons in photonic chip-based microresonators, and its dependence on key parameters. In this way we found that SIL relaxes the requirements on MR properties such as Q-factor, spectral purity, the number of mode crossings, the width of a so-called `soliton step', making it possible to generate microcombs with the majority of commercial photonic chip-based MRs featuring moderate Q-factors. SIL greatly facilitates tuning to the soliton regime, due to the compensation of the thermal effects inevitable in systems where a free-running laser is used as a pump source. Consequently, SIL enables soliton microcomb generation in cases where it is otherwise not possible with optically isolated external-cavity diode lasers (ECDLs).

The here-presented prototype of the integrated dual-microcomb source, based on laser diode self-injection locked to a microresonator, allows to combine all the benefits of Fourier-transform infrared broadband spectroscopy in a chip-scale spectroscopic sensor and looks promising as a platform for future mobile and wearable devices.
The demonstrated versatile approach to dual-microcomb source integration provides various design options (Fig\ref{fig:fig1}(d)), offering interesting perspectives in terms of device miniaturization and performance, in particular with a view to broadband infrared dual-comb sensors for high-volume applications. With further integration, there is a clear route to satisfying the so-called SWaP-C (Size, Weight, Power and Cost) requirements that are of central importance for industrial, airborne, space, and consumer applications.

\section{Results}

\subsection{Hybrid integrated platform for optical dual-microcomb source}

Our hybrid integrated dual-microcomb source comprises specially matched (see Methods and Supplementary Note 1) microcomb sources. Each consists of a thermally stabilized SiN photonic chip with a high-Q MR, a butt-coupled semiconductor LD, and an output lensed fiber (Fig. \ref{fig:fig1} (a,b)). This versatile approach enables fast prototyping by testing different photonic-chip designs and 
various types of LDs.

The MRs based on CMOS-compatible SiN photonic chips that we used in our experiments were fabricated in commercial multi-project wafer (MPW) runs. 
We use two sets of chips with MRs of two diameters, corresponding to $\sim$150 GHz and to $\sim$1 THz free spectral range (FSR) with integrated microheaters enabling grid matching of the eigenfrequencies of different MRs by tuning their FSR (spacing between fundamental modes in the frequency domain). Also, each chip has edge waveguide couplers, ensuring insertion losses as low as -1.1 dB on both sides of a chip for coupling light in and out.

Fabry--Perot (FP) and distributed feedback (DFB) laser diodes have been used for experiments and were compared in terms of performance for microcomb generation (see Methods and Supplementary Note 2). FP diodes  have a single spatial mode, 35-GHz longitudinal mode spacing, 1535-nm central wavelength, and $\sim$ 200 mW optical power at 500 mA of injection current.  DFB diodes have 1545 nm wavelength and optical power of $\sim$ 100 mW at 400 mA.

During the experiment we simultaneously monitored the optical spectra of the microcombs and the resulting RF dual-comb signal using an optical spectrum analyzer (OSA) and an electrical spectrum analyzer (ESA). Owing to the SIL effect, the assembled microcomb sources offer turnkey operation. This kind of turnkey operation was described in \cite{shen2019integrated} and also demonstrated in \cite{Jin2021}.
The spectrogram presented in Fig. \ref{fig:fig1}(c) shows dual-comb signal evolution while selecting the turnkey operating point by slow manual tuning of the LD injection current.
Away from the operating point, the optical spectrum consists of one coherent soliton microcomb and one chaotic non-coherent microcomb (MI state). The heterodyne beatnote signal of this state is noisy (regime I in Fig. \ref{fig:fig1}(c)). 
When the LD-current value reaches the operating point, the system locks to the state with both soliton microcombs, featuring high mutual coherence and low-noise RF beatnotes of the optical components (regime II in Fig. \ref{fig:fig1}(c)). The SIL mechanism compensates thermal effects and the microcomb sources lock to the comb states without additional manipulations, which are inevitable when pumping with a free-running laser. At the operating point, our dual-microcomb source quickly transits to a coherent state. However, it should be noted that due to the independent thermal stabilization of the two photonic chips, a relative thermal drift of the RF beat frequencies arises for observation periods exceeding several minutes.


\begin{figure*}[t!]
\centering
\includegraphics[width=\textwidth]{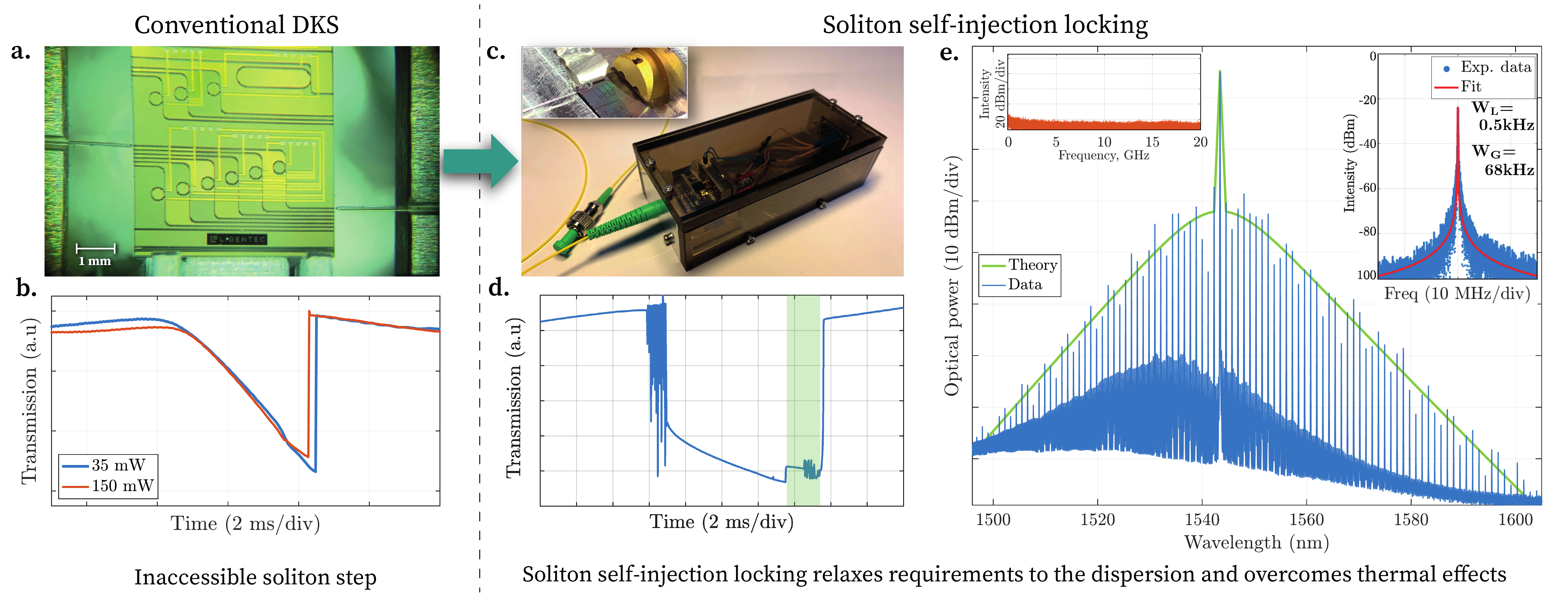}
\caption{Soliton microcomb generation using self-injection-locked laser-diode pumping in the case of an inaccessible conventional soliton step \textbf{a.} Photograph of the silicon nitride photonic chip used in the experiment 
\textbf{b.} Nonlinear-resonance shape for the different pump powers of the external optically isolated laser, demonstrating inaccessibility of the conventional soliton step. The measured nonlinear threshold of this microresonator is approximately 13 mW. \textbf{c.} Photograph of the assembled prototype of the microcomb source, based on the same photonic chip pumped with a butt-coupled Fabry--Perot laser diode. Inset: Inside view the prototype. \textbf{d.} Nonlinear-resonance shape for 35-mW pump power for the same resonance shown in (b) for the case of the self-injection locking effect (laser diode pumping). The area where the soliton exists is highlighted in green colour. \textbf{e.} Output of the microcomb source (blue line) and  theoretically predicted envelope of the single soliton state microcomb spectrum (green line) based on the measured microresonator parameters (see Supplementary Note 4). Left inset: RF spectrum of the output signal. Right inset: Beatnote signal of the generated microcomb line and a Toptica CTL-1550 tunable laser recorded with a resolution bandwidth (RBW) of 10kHz (blue points), together with a Voigt-profile fit (red line) with 0.5 kHz and 68 kHz Lorentzian and Gaussian widths, respectively.}
\label{fig:fig2}
\end{figure*}

\subsection{Highly efficient soliton microcombs for dual-microcomb source}


Current technology used in the fabrication of high-Q silicon nitride MRs, especially for commercially available runs, does not guarantee Q-factors higher than one million and a spectral purity of MRs sufficient high for sustainable soliton microcomb generation using external pumping with single-frequency narrow-linewidth lasers. The procedure of soliton comb excitation by means of an external pump using ECDL requires accessible soliton steps and is complicated by the need for additional equipment 
to achieve a high tuning rate and to overcome thermal instabilities \cite{herr2014temporal}. We note that in our experiments only high-FSR MRs (FSR=1 THz) provide easily accessible soliton steps and support soliton generation using ECDL. 

\begin{figure*}[t!]
\centering
\includegraphics[width=\textwidth]{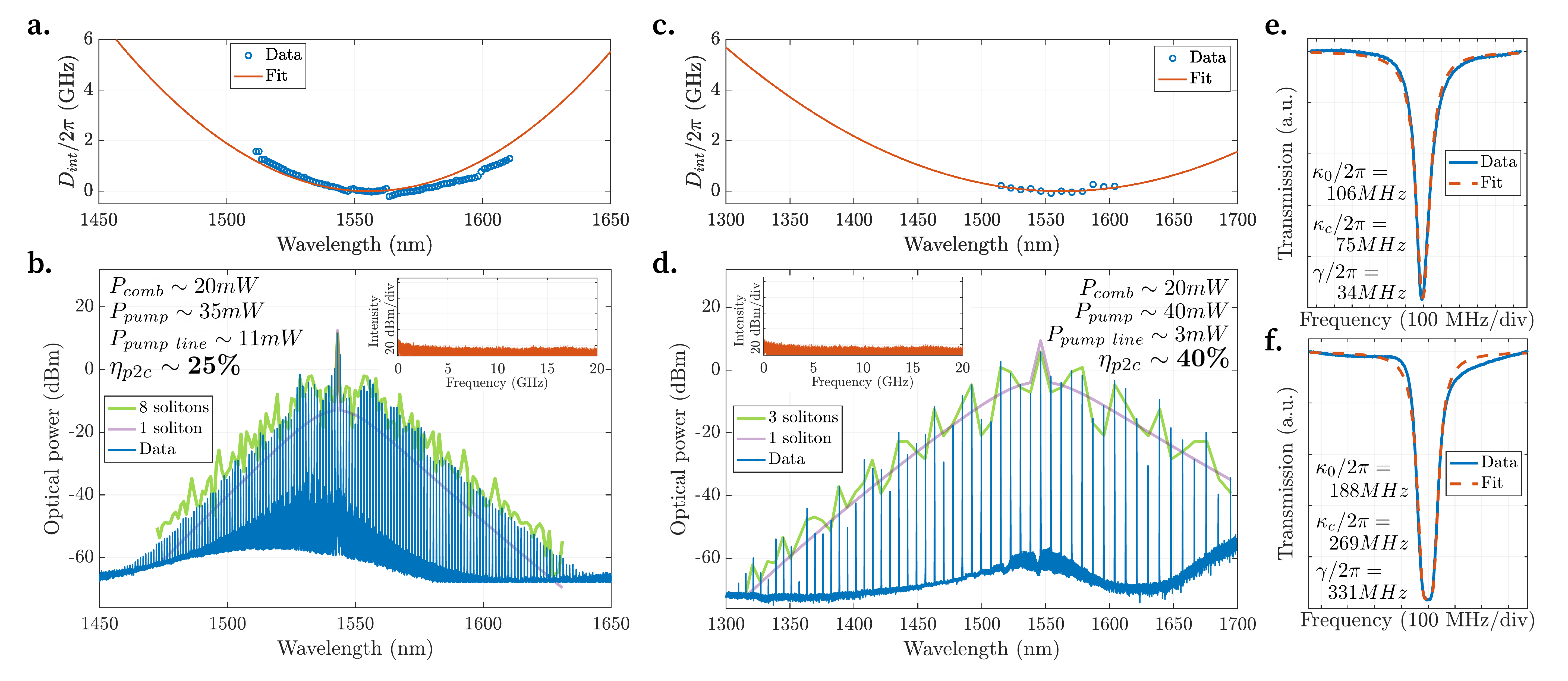}
\caption{High-power and broadband microcombs in 150 GHz and 1 THz microchips. \textbf{a,c.} Measured and fitted anomalous dispersion landscape in a $Si_3N_4$ microresonator with FSR = 143.6 and 999.8 GHz, and estimated second-order dispersion coefficient $D_2/2\pi$ $\approx$ 1.38 and $\approx$ 14.3 MHz, respectively. \textbf{b,d.} Optical spectra of generated high-power microcomb with $\sim$150 GHz \textbf{(b)} and $\sim$1 THz \textbf{(d)} repetition rate (blue lines) and values of key parameters: total comb power ($P_{comb}$), laser diode power ($P_{pump}$), the power of the microcomb central line (pump line) ($P_{pump\ line}$), and pump-to-comb sideband conversion efficiency ($\eta_{p2c}$). Theoretically predicted envelope for corresponding microcombs based on measured microresonator parameters (see Supplementary Note 4) for single-soliton state (green line) and multi-soliton state (purple line). \textbf{e,f.} Experimentally measured resonances in the linear regime (blue line) and loaded Lorentzian profile fits (red dashed line) for $\sim$150 GHz \textbf{(e)} and $\sim$1 THz \textbf{(f)} microresonators. The estimated intrinsic loss ($\kappa_0$), coupling rate ($\kappa_c$)  and backward-wave coupling rate ($\gamma$) correspond to a coupling coefficient $\eta$ = 0.42 and 0.59, and Q-factor $\sim 1.8 \times 10^6$ and $\sim 1 \times 10^6$ for 150 GHz and 1 THz microresonators, respectively. }

\label{fig:fig3}
\end{figure*}

We have not been able to observe soliton generation with 150 GHz MRs using the external amplified ECDL providing more than 150 mW of in-chip pump power. This power level is more than ten times higher than the parametric instability threshold, corresponding to the normalized pump amplitude $f=\sqrt{P_{pump}/P_{th}}>3$, where $P_{pump}$ is the optical power of the amplified ECDL reduced by losses for the butt-coupling of the lensed fiber with the chip (power in the bus waveguide) and $P_{th}$ is the nonlinearity threshold power \cite{herr2014temporal}. Apparently, the influence of thermal processes, high-order dispersion (Fig. \ref{fig:fig3}(a)) and avoided mode crossing points shortens a soliton step  and makes it inaccessible (see Fig. \ref{fig:fig2}(b)),  \cite{PhysRevLett.113.123901,Kim:21, briles2017,Joshi:16,Kuse:20}. 

However, the fully integrated SIL scheme (Fig. \ref{fig:fig2}(c)) 
provides outstanding turnkey operation without any additional equipment.  The same MR pumped by the self-injection locked LD allows to clearly observe a soliton step in the locked state, even for $f \sim 1.6$ (Fig. \ref{fig:fig2}(d,e)). Indeed, most of the thermal effects are suppressed as the laser frequency is locked to the MR and the laser--microresonator detuning $\zeta_{\rm eff}$ is fixed. If the MR frequency fluctuates due to the thermal effects, the generation frequency also changes, keeping the comb-generation regime stable \cite{KondratievSPIE21}.
In the SIL regime the laser--microresonator detuning becomes fixed to the value $\zeta_{\rm eff}^0\approx -3(f^2/2)^{1/3}+(f^2/2)^{-1/3}$ (for small backscattering and $f>1$), which lies inside the soliton-generation region $\zeta_{\rm eff} \in[-3(f^2/4)^{1/3}+(f^2/4)^{-1/3}/4; -\pi^2f^2/8]$ (see Supplementary Note 3).

In addition, the detuning control is much more robust in the SIL regime as the effective detuning does not change while the LD is tuned within the locking range. More precisely, the speed of the laser-frequency tuning is effectively reduced by the factor of the stabilization coefficient $K_0$ (see Supplementary Note 3).


\begin{figure*}[t!]
\centering
\includegraphics[width=\textwidth]{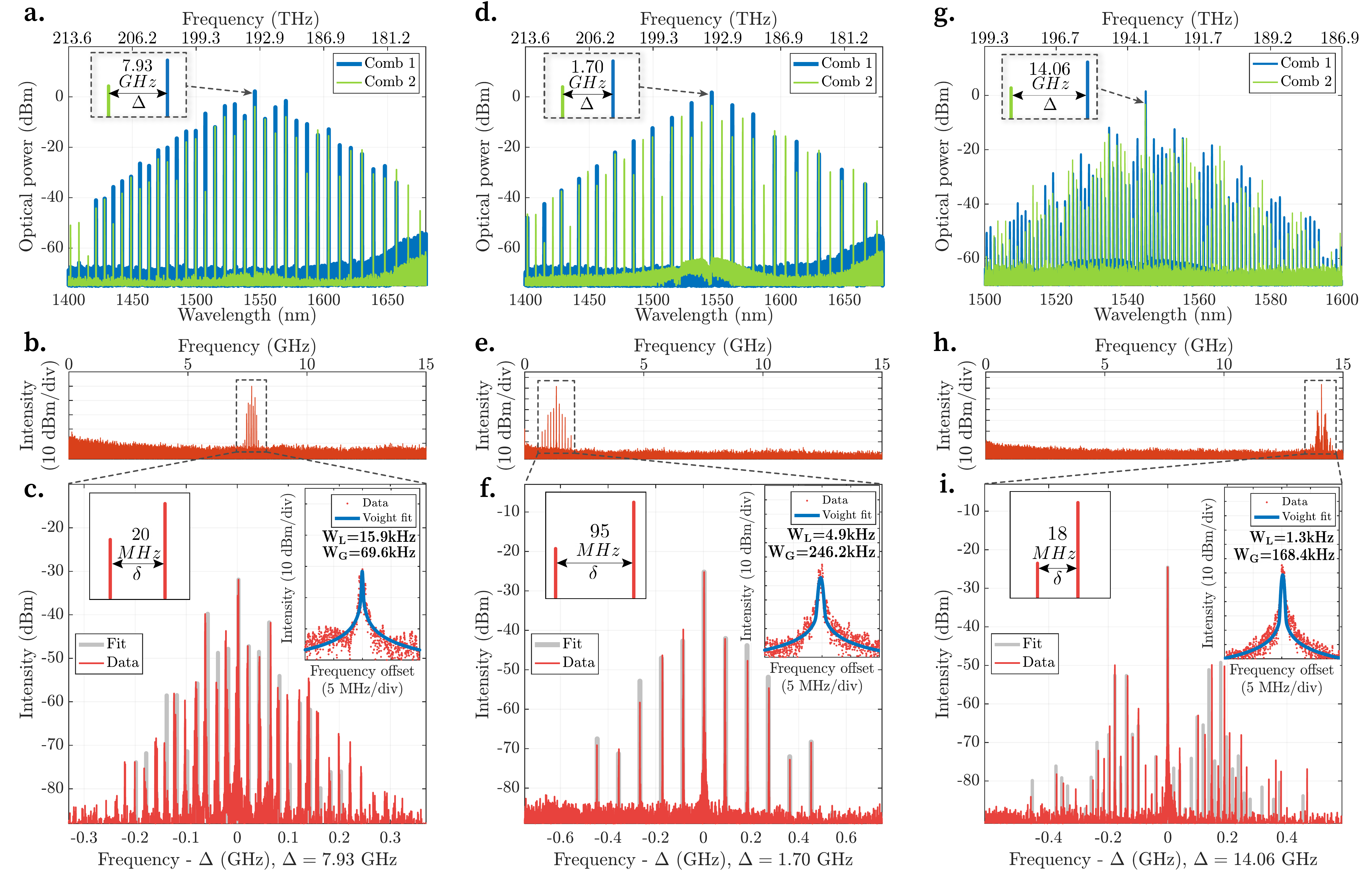}
\caption{ Dual-microcomb signals \textbf{Top row:} Optical spectra of the generated microcombs that were combined for dual-microcomb signal generation. Inset: Zoom-in on the central line area, with the distance between them ($\Delta$) indicated. \textbf{Bottom row:} The dual-microcomb signal is an RF comb with central line at $\Delta$ and repetition rate $\delta = |f_{rep1}-f_{rep2}|$, where $FSR_i$ are the repetition rates of the combined optical combs. The top full-span spectrum (from 0 to 15 GHz) demonstrates the absence of low-frequency noise, showing a strong dual-microcomb signal. The bottom spectrum is a zoom-in on the central (Frequency - $\Delta$) dual-microcomb signal area. The red line is the experimentally observed signal and the grey line the predicted signal based on the optical spectra above. Inset: estimation of the RF-comb linewidth from a Voigt-profile fit. \textbf{a,b,c.} Dual-comb signal obtained by combining two optical frequency combs with $f_{rep1}\sim f_{rep2}\sim$1 THz and $\delta = |f_{rep1}-f_{rep2}|\approx$20 MHz. The distance between the pump lines of the optical microcombs is $\Delta=$7.93 GHz.  \textbf{d,e,f.} Dual-microcomb signal obtained by combining two optical frequency combs with repetition rates $f_{rep1}\sim$1 THz, $f_{rep2}\sim$2 THz and $\delta =|2f_{rep1}-f_{rep2}|\approx$95 MHz. The distance between the pump lines of the optical microcombs is $\Delta=$1.70 GHz. \textbf{g,h,i.} Dual-microcomb signal obtained by combining two optical frequency combs with $f_{rep1}\sim f_{rep2}\sim$150 GHz  and $\delta =|f_{rep1}-f_{rep2}|\approx$18 MHz. The distance between the pump lines of the optical microcombs is $\Delta=$14.06 GHz.}
\label{fig:fig4}
\end{figure*}

The assembled prototype works as a turnkey device, and thereby greatly simplifies the process of comb generation and significantly improves its  stability. Once calibrated to define the operating point, this device can generate microcombs immediately after being turned on. 

Microcomb spectra and the microresonator parameters are shown in Fig. \ref{fig:fig3}. Notably, there were no optical elements filtering the pump line. Despite the moderate Q-factor ($\sim 10^6$),  
the generated combs are broadband. The spectrum width of the microcombs with 1-THz line spacing reaches 500 nm, with 20 lines of power > -20 dBm. The width of the 150-GHz microcomb spectrum exceeds 200 nm, with 30 lines of  power > -20 dBm. The generated microcombs also feature high optical power per line and therefore a high signal-to-noise ratio, and demonstrates high pump-to-comb sideband power conversion efficiency ($\eta_{\rm p2c}$). The latter can be expressed as $\eta_{\rm p2c} = (P_{comb}-P_{pump\ line})/P_{pump}$, where $P_{comb}$ is the total generated microcomb power in the output fiber, $P_{pump\ line}$ is the power of the microcomb central line (pump line) in the output fibre, and $P_{pump}$ is the optical power of the free-running laser diode for the same injection current value reduced by the coupling losses (power in the bus waveguide). For the 150-GHz microcomb, $P_{comb}\approx$ 20 mW, $P_{pump\ line}\approx$ 11 mW, and $P_{pump}\approx$ 35 mW; for the 1-THz microcomb, $P_{comb} \approx$ 20 mW, $P_{pump\ line}\approx$ 3 mW, and $P_{pump}\approx$ 40 mW. Evaluated $\eta_{\rm p2c}$ efficiency values for the 150-GHz and 1-THz FSR microcombs are 25$\%$ and 40$\%$, respectively. Before this work, comparable efficiency has been demonstrated for the generation of dark microcombs only \cite{Xue:17,Kim:19}. The high efficiencies obtained are in a good agreement with our estimates (see Supplementary Note 4) and are compatible with \cite{jang2021conversion}.

\subsection{Spectral characteristics of dual-microcomb source}

Combined microcomb optical spectra and the resulting RF dual-microcomb signals are shown in Figs \ref{fig:fig4} (a,d,g) and in \ref{fig:fig4}(b,c,e,f,h,i), respectively. Insets in the bottom row give information about the linewidth of the lines of the generated dual-microcomb signal, estimated with Voigt-profile fits ($\mathrm{W_L}$ is the Lorentzian linewidth and $\mathrm{W_G}$ the Gaussian linewidth). With 1-THz FSR microresonators we successfully down-converted $\sim$300-nm-wide optical spectra to 600-MHz-wide spectra in the RF range (Fig. \ref{fig:fig4} (a-f)); with 150-GHz FSR microresonators we achieved down-conversion from $\sim$100 nm to 800 MHz (Fig. \ref{fig:fig4}(g-i)).

By applying voltage to the microheaters we can adjust the microresonator temperature and thereby control the microcomb line spacing. One optical microcomb can be shifted relatively to the other to control the dual-microcomb signal, changing its central line position $\Delta$ and repetition rate $\delta$. This capability is illustrated in Fig. \ref{fig:fig4}(a--f). These two dual-microcomb signals were observed for the same pair of MRs, and using microheaters we changed the central-frequency  difference $\Delta$ from 7.93 GHz to 1.70 GHz.

Verification of the measurement data is conducted by comparing the experimental data with the theoretically predicted beatnote signal of the two combined optical microcombs. Knowing the MR parameters, including dispersion profile, and with the measured optical frequency combs, we can calculate the expected dual-comb spectrum profile in the RF domain. These simulated dual-microcomb spectra are shown as grey lines in Fig. \ref{fig:fig4}(c,f,i).



\section{Discussion}


We found that SIL relaxes the requirements on the microresonator Q-factor and its spectral purity, and makes widely available microresonators commercially fabricated in MPW runs suitable for being used in on-chip dual-microcomb sources. Note that soliton-comb excitation with such  microresonators pumped by a tunable optically isolated ECDL (not SIL) was unsuccessful, because of the required soliton steps could not be accessed (Fig. \ref{fig:fig2}(b)). We connect this failure with the influence of the thermal processes, high-order dispersion effects, and avoided mode crossing points, which makes the soliton step shorter and therefore inaccessible. When exploing the SIL effect, the frequencies of the laser diode and the microresonator are connected and fluctuate in a correlated manner, resulting in microcomb generation with higher tolerance to thermal drift (Fig. \ref{fig:fig2}(d)). 

The demonstrated pump-to-comb sideband conversion efficiency of up to 40\% can be explained by the matching between Q-factor and pump power, taking into account the following considerations. First, the comb power does not depend directly on the pump power (see Supplementary Note 4). Although both the maximal and the locked detunings do depend on the pump power, these dependencies are weaker, and the pump-to-soliton (and pump-to-comb sideband) power conversion efficiency decreases with it. This creates the illusion that using a low-power laser and reducing the threshold we can make a highly energy-efficient device.
However, the second point is that the comb power does depend on the threshold power of the parametric instability. Therefore, using the strategy described above, we end up with a negligible output signal. This brings us to the rather counter-intuitive conclusion that for best performance, the threshold power should be increased (meaning lower Q-factors or less nonlinearity, for example) and matched with the used laser pump power. This immediately brings up a trade-off problem, as the lower Q-factor means less stabilization and wider beatnote or even the the self-injection locking regime cannor be reached at all.
Another way to increase the pump-to-comb sideband power conversion efficiency is increasing the second-order dispersion coefficient. This presents, however, a trade-off problem for the comb width, which should be solved separately for the desired application.

We also can see an additional mechanism to increase the comb power (and subsequently the power efficiency). In our system, multi-soliton states can emerge; however, the number of solitons has a maximum for a given dispersion value \cite{Karpov2019} and the pump-to-comb sideband conversion efficiency ($\eta_{p2c}$) also saturates with it (see Supplementary Note 4). The multi-soliton option was realized for the 1-THz microresonator  (see Fig. \ref{fig:fig3}d), where we found good correspondence of the measured spectrum with the theoretical prediction for 3-soliton states (the soliton positions were optimized to fit the data) for the experimentally estimated parameters ($f$, $D_2$,$\kappa$) and SIL detuning value ($\zeta_{\rm eff}^0$). We note that while the form of the spectrum is highly dependent on the intersoliton distances, the total comb power (and the conversion efficiency) does not. For the 150-GHz comb we saw a slightly different picture. By increasing the number of solitons we were able to match the total comb output power, but the comb envelope at the sides show much smoother behaviour than the multi-soliton state can provide (see green curve in Fig. \ref{fig:fig3}b). At the same time, the comb power is too high for a single-soliton state (see purple curve in Fig. \ref{fig:fig3}d). Such comb enhancement can be attributed to the comb-line amplification inside the active medium of the laser or so-far unexplored effects of the multi-frequency locking, while the non-smooth envelope near the pump -- to the dispersion distortions from the mode-crossings \cite{PhysRevLett.113.123901}.

A comparison of the semiconductor FP and DFB laser diodes highlights the benefits of DFB in terms of predictable wavelength of locking and more convenient matching of two combs, while the FP is much more powerful and cheaper, and hence more promising with a view to practical applications.

The early integrated dual-microcomb source prototype presented here is still affected by the relative thermal drift of the  microresonators on the two separate photonic chips. Various design options (Fig. \ref{fig:fig1}(d)) could help to overcome the drift, paving the way to more compact devices. A first improvement could be the combination of the two microresonators on the same photonic chip positioned on a common temperature-stabilized substrate with two pumping laser diodes. This option would simplify comb matching owing to small fabrication errors for the two microresonators at close distance, thus enhancing the stability of the dual-comb beatnotes. The next improvement for higher stability and smaller size is using the same microresonator for the generation of two soliton combs propagating in the same or opposite directions along the ring. Two laser diodes self-injection locked into the same microresonator provide the highest mutual coherence and generated microcombs would lead to dual-comb beatnotes with the lowest phase noise. To reduce the number of components that need to be aligned during the integration, keeping small size and high stability, the design option with just one laser diode and one photonic chip with two microresonator could be considered. A laser diode locked to one of the microresonators would provide the first microcomb and the second resonator tuned relative to the frequency of the locked laser would provide the second one. The discussed design options based on the microcomb generation using self-injection locking should improve the performance of the integrated dual-microcomb source, and will be explored in future research.

Taking into account recent advances of III--V heterogeneous integration with silicon nitride photonic waveguides \cite{xiang2021laser}, deeper integration of the proposed dual-microcomb source can lead to reaching tiny chip-scale sizes. Also, the demonstrated possibility to use FP laser diodes might allow using just a gain section without an additional laser cavity in future generations of the chip-based dual-comb sources.

In conclusion, obtained results demonstrate that SIL Kerr microcombs based on silicon photonics can successfully compete with other on-chip optical-comb sources and outperform them owing to the unique combination of power efficiency with mWs comb power, a wider spectrum, and low phase noise.

\vspace{0.5cm}
\noindent\textbf{Methods}
\medskip
\begin{footnotesize}

\noindent \textbf{Silicon nitride chips characterisation}:
The SiN photonic chip-based microresonators used in our experiments were fabricated by Ligentec SA, Switzerland.
The pumped microresonator resonance is measured using an external tunable laser Toptica CTL1550, and fitted taking backscattering into account \cite{Gorodetsky:00} to obtain the intrinsic loss $\kappa_{\mathrm{0}}$, the coupling rate $\kappa_c$, and the backward-wave coupling rate $\gamma$ (mode splitting). Based on these data, we evaluated the full resonance linewidth $\kappa =\kappa_{\mathrm{0}}+\kappa_c$, the pump coupling efficiency $\eta={\kappa_c}/{\kappa}$, and the normalized backscattering coefficient $\beta=\gamma/\kappa$.

For nonlinearity-threshold estimation the experimental setup with the external laser Toptica CTL1550 and a booster NKT Photonics Koheras Boostik HP E15 was used. Gradually increasing the pump power, we simultaneously monitor the optical spectrum and the resonance shape using an OSA (Yokogawa AQ6370D) and an oscilloscope (Keysight DSO-X 3024A). First, we reach the thermal nonlinearity threshold ($ P_{\mathrm{th}}^{\mathrm{Thermal}} $), where the resonance shape becomes triangular. By further increasing the pump power, we reach the parametric-instability threshold ($P_ {\mathrm{th}}$), accompanied by sideband generation.

Dispersion characteristics of microresonators were measured using the original experimental setup based on the tunable laser Toptica CTL1550 and a calibrated fibre Mach–-Zehnder interferometer (MZI) with a FSR of 102 MHz.

The results obtained for the 150-GHz and 1-THz FSR microresonators, respectively, were as follows: $\kappa_{\mathrm{0}}/2\pi = $ 106 and 188 MHz, $\kappa_c/2\pi = $ 75 and 269 MHz, $\gamma/2\pi = $ 34 and 331 MHz, $\eta = $ 0.42 and 0.59, Q-factor = $1.8 \times 10^6$ and $1 \times 10^6$, $P_ {\mathrm{th}} = $ 14 and 11 mW, $ P_{\mathrm{th}}^{\mathrm{Thermal}} = $3.7 and 2.5 mW, $ D_1/2\pi = $ 143.6 and 999.8 GHz, and $ D_2/2\pi =$ 1.38 and 14.3 MHz.

\noindent \textbf{Microcomb source}:
After chip characterisation we have successfully generated microcombs with all available chips and compared the reproducibility of their parameters (see Supplementary Note 1). As a result, we have matched photonic chips for RF dual-microcomb signal generation and defined the operating points for all laser diodes and photonic chip pairs, which were used for the assembly process. The operating point is defined by the following parameters: laser-diode temperature, injection current, and microheater voltage.

In addition, we have compared two types of laser diodes, DFB and FP (Seminex Corp., USA), in terms of their suitability for being used for microcomb generation (see Supplementary Note 2).

We have also estimated the linewidth of the microcomb components using the heterodyne technique. The beatnote signal between the external laser and the microcomb line is presented in the inset of Fig. \ref{fig:fig3}(e, f). The Lorentzian width of the beatnote signal (estimated using a Voigt-profile fit) is 0.5 kHz. The Gaussian width is about 68 kHz, which can be improved by further reduction of the technical noises.


\noindent \textbf{Funding Information}: 
This work was supported by the Russian Science Foundation (grant 20-12-00344).

\noindent \textbf{Acknowledgments}: 
V.E.L. acknowledges personal support from the Foundation for the Advancement of Theoretical Physics and Mathematics “BASIS.”
A.S.V. is supported by the EU H2020 research and innovation programme under the Marie Sklodowska-Curie grant agreement No 101033663 (RaMSoM).
N.Yu.D. was partially supported by the Samsung Research Center in Moscow. The authors thank H.-S. Lee and Y.-G. Roh from the Samsung Advanced Institute of Technologies for help in establishing the project and its further support. Authors express special thanks to Alexander Gorodnitskiy, graduate student of MIPT and Sofya Agafonova, student of MIPT for assistance at an early stage of experiments.

\noindent \textbf{Author contribution}: A.S.V., S.N.K., N.Yu.D. conducted the experiment. N.M.K., V.E.L. developed a theoretical model and performed numerical simulations. All authors analyzed the data and prepared the manuscript. I.A.B., M.V.R. and S.V.P. initiated the collaboration and supervised the project. 

\noindent \textbf{Data Availability Statement}: 
Data underlying the results presented in this paper are not publicly available at this time but may be obtained from the authors upon reasonable request.

\noindent \textbf{Competing interests}: 
S.N.K., A.S.V., V.E.L., M.V.R., S.V.P. and I.A.B. are listed as co-authors in joint Samsung Electronics and Russian Quantum Center US  patent  no. US10224688B2, which is related to the technology reported in this article.


The authors declare no competing interests.

\end{footnotesize}

\bibliography{bibliography}

\begin{thebibliography}{76}%
\makeatletter
\providecommand \@ifxundefined [1]{%
 \@ifx{#1\undefined}
}%
\providecommand \@ifnum [1]{%
 \ifnum #1\expandafter \@firstoftwo
 \else \expandafter \@secondoftwo
 \fi
}%
\providecommand \@ifx [1]{%
 \ifx #1\expandafter \@firstoftwo
 \else \expandafter \@secondoftwo
 \fi
}%
\providecommand \natexlab [1]{#1}%
\providecommand \enquote  [1]{``#1''}%
\providecommand \bibnamefont  [1]{#1}%
\providecommand \bibfnamefont [1]{#1}%
\providecommand \citenamefont [1]{#1}%
\providecommand \href@noop [0]{\@secondoftwo}%
\providecommand \href [0]{\begingroup \@sanitize@url \@href}%
\providecommand \@href[1]{\@@startlink{#1}\@@href}%
\providecommand \@@href[1]{\endgroup#1\@@endlink}%
\providecommand \@sanitize@url [0]{\catcode `\\12\catcode `\$12\catcode
  `\&12\catcode `\#12\catcode `\^12\catcode `\_12\catcode `\%12\relax}%
\providecommand \@@startlink[1]{}%
\providecommand \@@endlink[0]{}%
\providecommand \url  [0]{\begingroup\@sanitize@url \@url }%
\providecommand \@url [1]{\endgroup\@href {#1}{\urlprefix }}%
\providecommand \urlprefix  [0]{URL }%
\providecommand \Eprint [0]{\href }%
\providecommand \doibase [0]{http://dx.doi.org/}%
\providecommand \selectlanguage [0]{\@gobble}%
\providecommand \bibinfo  [0]{\@secondoftwo}%
\providecommand \bibfield  [0]{\@secondoftwo}%
\providecommand \translation [1]{[#1]}%
\providecommand \BibitemOpen [0]{}%
\providecommand \bibitemStop [0]{}%
\providecommand \bibitemNoStop [0]{.\EOS\space}%
\providecommand \EOS [0]{\spacefactor3000\relax}%
\providecommand \BibitemShut  [1]{\csname bibitem#1\endcsname}%
\let\auto@bib@innerbib\@empty
\bibitem [{\citenamefont {Udem}\ \emph {et~al.}(2002)\citenamefont {Udem},
  \citenamefont {Holzwarth},\ and\ \citenamefont {H{\"a}nsch}}]{Udem2002}%
  \BibitemOpen
  \bibfield  {author} {\bibinfo {author} {\bibfnamefont {T.}~\bibnamefont
  {Udem}}, \bibinfo {author} {\bibfnamefont {R.}~\bibnamefont {Holzwarth}}, \
  and\ \bibinfo {author} {\bibfnamefont {T.~W.}\ \bibnamefont {H{\"a}nsch}},\
  }\href {\doibase 10.1038/416233a} {\bibfield  {journal} {\bibinfo  {journal}
  {Nature}\ }\textbf {\bibinfo {volume} {416}},\ \bibinfo {pages} {233}
  (\bibinfo {year} {2002})}\BibitemShut {NoStop}%
\bibitem [{\citenamefont {Fortier}\ and\ \citenamefont
  {Baumann}(2019)}]{Fortier2019}%
  \BibitemOpen
  \bibfield  {author} {\bibinfo {author} {\bibfnamefont {T.}~\bibnamefont
  {Fortier}}\ and\ \bibinfo {author} {\bibfnamefont {E.}~\bibnamefont
  {Baumann}},\ }\href {\doibase 10.1038/s42005-019-0249-y} {\bibfield
  {journal} {\bibinfo  {journal} {Communications Physics}\ }\textbf {\bibinfo
  {volume} {2}},\ \bibinfo {pages} {153} (\bibinfo {year} {2019})}\BibitemShut
  {NoStop}%
\bibitem [{\citenamefont {Diddams}\ \emph {et~al.}(2020)\citenamefont
  {Diddams}, \citenamefont {Vahala},\ and\ \citenamefont
  {Udem}}]{Diddams-2021}%
  \BibitemOpen
  \bibfield  {author} {\bibinfo {author} {\bibfnamefont {S.~A.}\ \bibnamefont
  {Diddams}}, \bibinfo {author} {\bibfnamefont {K.}~\bibnamefont {Vahala}}, \
  and\ \bibinfo {author} {\bibfnamefont {T.}~\bibnamefont {Udem}},\ }\href
  {\doibase 10.1126/science.aay3676} {\bibfield  {journal} {\bibinfo  {journal}
  {Science}\ }\textbf {\bibinfo {volume} {369}},\ \bibinfo {pages} {eaay3676}
  (\bibinfo {year} {2020})}\BibitemShut {NoStop}%
\bibitem [{\citenamefont {Suh}\ \emph {et~al.}(2016)\citenamefont {Suh},
  \citenamefont {Yang}, \citenamefont {Yang}, \citenamefont {Yi},\ and\
  \citenamefont {Vahala}}]{Suh2016}%
  \BibitemOpen
  \bibfield  {author} {\bibinfo {author} {\bibfnamefont {M.-G.}\ \bibnamefont
  {Suh}}, \bibinfo {author} {\bibfnamefont {Q.-F.}\ \bibnamefont {Yang}},
  \bibinfo {author} {\bibfnamefont {K.~Y.}\ \bibnamefont {Yang}}, \bibinfo
  {author} {\bibfnamefont {X.}~\bibnamefont {Yi}}, \ and\ \bibinfo {author}
  {\bibfnamefont {K.~J.}\ \bibnamefont {Vahala}},\ }\href {\doibase
  10.1126/science.aah6516} {\bibfield  {journal} {\bibinfo  {journal}
  {Science}\ }\textbf {\bibinfo {volume} {354}},\ \bibinfo {pages} {600}
  (\bibinfo {year} {2016})}\BibitemShut {NoStop}%
\bibitem [{\citenamefont {Coddington}\ \emph {et~al.}(2016)\citenamefont
  {Coddington}, \citenamefont {Newbury},\ and\ \citenamefont
  {Swann}}]{Coddington:16}%
  \BibitemOpen
  \bibfield  {author} {\bibinfo {author} {\bibfnamefont {I.}~\bibnamefont
  {Coddington}}, \bibinfo {author} {\bibfnamefont {N.}~\bibnamefont {Newbury}},
  \ and\ \bibinfo {author} {\bibfnamefont {W.}~\bibnamefont {Swann}},\ }\href
  {\doibase 10.1364/OPTICA.3.000414} {\bibfield  {journal} {\bibinfo  {journal}
  {Optica}\ }\textbf {\bibinfo {volume} {3}},\ \bibinfo {pages} {414} (\bibinfo
  {year} {2016})}\BibitemShut {NoStop}%
\bibitem [{\citenamefont {Ideguchi}(2017)}]{Ideguchi:17}%
  \BibitemOpen
  \bibfield  {author} {\bibinfo {author} {\bibfnamefont {T.}~\bibnamefont
  {Ideguchi}},\ }\href {\doibase 10.1364/OPN.28.1.000032} {\bibfield  {journal}
  {\bibinfo  {journal} {Opt. Photon. News}\ }\textbf {\bibinfo {volume} {28}},\
  \bibinfo {pages} {32} (\bibinfo {year} {2017})}\BibitemShut {NoStop}%
\bibitem [{\citenamefont {Picqu{\'e}}\ and\ \citenamefont
  {H{\"a}nsch}(2019)}]{Picque2019}%
  \BibitemOpen
  \bibfield  {author} {\bibinfo {author} {\bibfnamefont {N.}~\bibnamefont
  {Picqu{\'e}}}\ and\ \bibinfo {author} {\bibfnamefont {T.~W.}\ \bibnamefont
  {H{\"a}nsch}},\ }\href {\doibase 10.1038/s41566-018-0347-5} {\bibfield
  {journal} {\bibinfo  {journal} {Nature Photonics}\ }\textbf {\bibinfo
  {volume} {13}},\ \bibinfo {pages} {146} (\bibinfo {year} {2019})}\BibitemShut
  {NoStop}%
\bibitem [{\citenamefont {Baumann}\ \emph {et~al.}(2019)\citenamefont
  {Baumann}, \citenamefont {Hoenig}, \citenamefont {Perez}, \citenamefont
  {Colacion}, \citenamefont {Giorgetta}, \citenamefont {Cossel}, \citenamefont
  {Ycas}, \citenamefont {Carlson}, \citenamefont {Hickstein}, \citenamefont
  {Srinivasan} \emph {et~al.}}]{baumann2019dual}%
  \BibitemOpen
  \bibfield  {author} {\bibinfo {author} {\bibfnamefont {E.}~\bibnamefont
  {Baumann}}, \bibinfo {author} {\bibfnamefont {E.~V.}\ \bibnamefont {Hoenig}},
  \bibinfo {author} {\bibfnamefont {E.~F.}\ \bibnamefont {Perez}}, \bibinfo
  {author} {\bibfnamefont {G.~M.}\ \bibnamefont {Colacion}}, \bibinfo {author}
  {\bibfnamefont {F.~R.}\ \bibnamefont {Giorgetta}}, \bibinfo {author}
  {\bibfnamefont {K.~C.}\ \bibnamefont {Cossel}}, \bibinfo {author}
  {\bibfnamefont {G.}~\bibnamefont {Ycas}}, \bibinfo {author} {\bibfnamefont
  {D.~R.}\ \bibnamefont {Carlson}}, \bibinfo {author} {\bibfnamefont {D.~D.}\
  \bibnamefont {Hickstein}}, \bibinfo {author} {\bibfnamefont {K.}~\bibnamefont
  {Srinivasan}},  \emph {et~al.},\ }\href
  {https://www.osapublishing.org/oe/fulltext.cfm?uri=oe-27-8-11869&id=408989}
  {\bibfield  {journal} {\bibinfo  {journal} {Optics express}\ }\textbf
  {\bibinfo {volume} {27}},\ \bibinfo {pages} {11869} (\bibinfo {year}
  {2019})}\BibitemShut {NoStop}%
\bibitem [{\citenamefont {Koptyaev}\ \emph {et~al.}(2019)\citenamefont
  {Koptyaev}, \citenamefont {Lihachev}, \citenamefont {Pavlov}, \citenamefont
  {Shchekin}, \citenamefont {Bilenko}, \citenamefont {Riabko}, \citenamefont
  {Gorodetsky}, \citenamefont {Polonsky}, \citenamefont {Voloshin},
  \citenamefont {Lantsov} \emph {et~al.}}]{koptyaev2019optical}%
  \BibitemOpen
  \bibfield  {author} {\bibinfo {author} {\bibfnamefont {S.~N.}\ \bibnamefont
  {Koptyaev}}, \bibinfo {author} {\bibfnamefont {G.~V.}\ \bibnamefont
  {Lihachev}}, \bibinfo {author} {\bibfnamefont {N.~G.}\ \bibnamefont
  {Pavlov}}, \bibinfo {author} {\bibfnamefont {A.~A.}\ \bibnamefont
  {Shchekin}}, \bibinfo {author} {\bibfnamefont {I.~A.}\ \bibnamefont
  {Bilenko}}, \bibinfo {author} {\bibfnamefont {M.~V.}\ \bibnamefont {Riabko}},
  \bibinfo {author} {\bibfnamefont {M.~L.}\ \bibnamefont {Gorodetsky}},
  \bibinfo {author} {\bibfnamefont {S.~V.}\ \bibnamefont {Polonsky}}, \bibinfo
  {author} {\bibfnamefont {A.~S.}\ \bibnamefont {Voloshin}}, \bibinfo {author}
  {\bibfnamefont {A.~D.}\ \bibnamefont {Lantsov}},  \emph {et~al.},\ }\href
  {https://patents.google.com/patent/US10224688B2/en?q=voloshin&inventor=koptyaev&oq=voloshin+koptyaev}
  {\enquote {\bibinfo {title} {Optical dual-comb source apparatuses including
  optical microresonator},}\ } (\bibinfo {year} {2019}),\ \bibinfo {note} {uS
  Patent 10,224,688}\BibitemShut {NoStop}%
\bibitem [{\citenamefont {Okubo}\ \emph {et~al.}(2014)\citenamefont {Okubo},
  \citenamefont {Iwakuni}, \citenamefont {Inaba}, \citenamefont {Hosaka},
  \citenamefont {Onae}, \citenamefont {Sasada},\ and\ \citenamefont
  {Hong}}]{Okubo:14}%
  \BibitemOpen
  \bibfield  {author} {\bibinfo {author} {\bibfnamefont {S.}~\bibnamefont
  {Okubo}}, \bibinfo {author} {\bibfnamefont {K.}~\bibnamefont {Iwakuni}},
  \bibinfo {author} {\bibfnamefont {H.}~\bibnamefont {Inaba}}, \bibinfo
  {author} {\bibfnamefont {K.}~\bibnamefont {Hosaka}}, \bibinfo {author}
  {\bibfnamefont {A.}~\bibnamefont {Onae}}, \bibinfo {author} {\bibfnamefont
  {H.}~\bibnamefont {Sasada}}, \ and\ \bibinfo {author} {\bibfnamefont {F.-L.}\
  \bibnamefont {Hong}},\ }in\ \href {\doibase 10.1364/CLEO_SI.2014.STh1N.6}
  {\emph {\bibinfo {booktitle} {CLEO: 2014}}}\ (\bibinfo  {publisher} {Optical
  Society of America},\ \bibinfo {year} {2014})\ p.\ \bibinfo {pages}
  {STh1N.6}\BibitemShut {NoStop}%
\bibitem [{\citenamefont {Okubo}\ \emph {et~al.}(2015)\citenamefont {Okubo},
  \citenamefont {Iwakuni}, \citenamefont {Inaba}, \citenamefont {Hosaka},
  \citenamefont {Onae}, \citenamefont {Sasada},\ and\ \citenamefont
  {Hong}}]{Okubo-2015-2}%
  \BibitemOpen
  \bibfield  {author} {\bibinfo {author} {\bibfnamefont {S.}~\bibnamefont
  {Okubo}}, \bibinfo {author} {\bibfnamefont {K.}~\bibnamefont {Iwakuni}},
  \bibinfo {author} {\bibfnamefont {H.}~\bibnamefont {Inaba}}, \bibinfo
  {author} {\bibfnamefont {K.}~\bibnamefont {Hosaka}}, \bibinfo {author}
  {\bibfnamefont {A.}~\bibnamefont {Onae}}, \bibinfo {author} {\bibfnamefont
  {H.}~\bibnamefont {Sasada}}, \ and\ \bibinfo {author} {\bibfnamefont {F.-L.}\
  \bibnamefont {Hong}},\ }\href {\doibase 10.7567/apex.8.082402} {\bibfield
  {journal} {\bibinfo  {journal} {Applied Physics Express}\ }\textbf {\bibinfo
  {volume} {8}},\ \bibinfo {pages} {082402} (\bibinfo {year}
  {2015})}\BibitemShut {NoStop}%
\bibitem [{\citenamefont {Brehm}\ \emph {et~al.}(2006)\citenamefont {Brehm},
  \citenamefont {Schliesser},\ and\ \citenamefont {Keilmann}}]{Brehm:06}%
  \BibitemOpen
  \bibfield  {author} {\bibinfo {author} {\bibfnamefont {M.}~\bibnamefont
  {Brehm}}, \bibinfo {author} {\bibfnamefont {A.}~\bibnamefont {Schliesser}}, \
  and\ \bibinfo {author} {\bibfnamefont {F.}~\bibnamefont {Keilmann}},\ }\href
  {\doibase 10.1364/OE.14.011222} {\bibfield  {journal} {\bibinfo  {journal}
  {Opt. Express}\ }\textbf {\bibinfo {volume} {14}},\ \bibinfo {pages} {11222}
  (\bibinfo {year} {2006})}\BibitemShut {NoStop}%
\bibitem [{\citenamefont {von Ribbeck}\ \emph {et~al.}(2008)\citenamefont {von
  Ribbeck}, \citenamefont {Brehm}, \citenamefont {van~der Weide}, \citenamefont
  {Winnerl}, \citenamefont {Drachenko}, \citenamefont {Helm},\ and\
  \citenamefont {Keilmann}}]{vonRibbeck:08}%
  \BibitemOpen
  \bibfield  {author} {\bibinfo {author} {\bibfnamefont {H.-G.}\ \bibnamefont
  {von Ribbeck}}, \bibinfo {author} {\bibfnamefont {M.}~\bibnamefont {Brehm}},
  \bibinfo {author} {\bibfnamefont {D.}~\bibnamefont {van~der Weide}}, \bibinfo
  {author} {\bibfnamefont {S.}~\bibnamefont {Winnerl}}, \bibinfo {author}
  {\bibfnamefont {O.}~\bibnamefont {Drachenko}}, \bibinfo {author}
  {\bibfnamefont {M.}~\bibnamefont {Helm}}, \ and\ \bibinfo {author}
  {\bibfnamefont {F.}~\bibnamefont {Keilmann}},\ }\href {\doibase
  10.1364/OE.16.003430} {\bibfield  {journal} {\bibinfo  {journal} {Opt.
  Express}\ }\textbf {\bibinfo {volume} {16}},\ \bibinfo {pages} {3430}
  (\bibinfo {year} {2008})}\BibitemShut {NoStop}%
\bibitem [{\citenamefont {Zolot}\ \emph {et~al.}(2013)\citenamefont {Zolot},
  \citenamefont {Giorgetta}, \citenamefont {Baumann}, \citenamefont {Swann},
  \citenamefont {Coddington},\ and\ \citenamefont {Newbury}}]{ZOLOT2013}%
  \BibitemOpen
  \bibfield  {author} {\bibinfo {author} {\bibfnamefont {A.}~\bibnamefont
  {Zolot}}, \bibinfo {author} {\bibfnamefont {F.}~\bibnamefont {Giorgetta}},
  \bibinfo {author} {\bibfnamefont {E.}~\bibnamefont {Baumann}}, \bibinfo
  {author} {\bibfnamefont {W.}~\bibnamefont {Swann}}, \bibinfo {author}
  {\bibfnamefont {I.}~\bibnamefont {Coddington}}, \ and\ \bibinfo {author}
  {\bibfnamefont {N.}~\bibnamefont {Newbury}},\ }\href {\doibase
  https://doi.org/10.1016/j.jqsrt.2012.11.024} {\bibfield  {journal} {\bibinfo
  {journal} {Journal of Quantitative Spectroscopy and Radiative Transfer}\
  }\textbf {\bibinfo {volume} {118}},\ \bibinfo {pages} {26} (\bibinfo {year}
  {2013})}\BibitemShut {NoStop}%
\bibitem [{\citenamefont {Baumann}\ \emph {et~al.}(2011)\citenamefont
  {Baumann}, \citenamefont {Giorgetta}, \citenamefont {Swann}, \citenamefont
  {Zolot}, \citenamefont {Coddington},\ and\ \citenamefont
  {Newbury}}]{Baumann:2011}%
  \BibitemOpen
  \bibfield  {author} {\bibinfo {author} {\bibfnamefont {E.}~\bibnamefont
  {Baumann}}, \bibinfo {author} {\bibfnamefont {F.~R.}\ \bibnamefont
  {Giorgetta}}, \bibinfo {author} {\bibfnamefont {W.~C.}\ \bibnamefont
  {Swann}}, \bibinfo {author} {\bibfnamefont {A.~M.}\ \bibnamefont {Zolot}},
  \bibinfo {author} {\bibfnamefont {I.}~\bibnamefont {Coddington}}, \ and\
  \bibinfo {author} {\bibfnamefont {N.~R.}\ \bibnamefont {Newbury}},\ }\href
  {\doibase 10.1103/PhysRevA.84.062513} {\bibfield  {journal} {\bibinfo
  {journal} {Phys. Rev. A}\ }\textbf {\bibinfo {volume} {84}},\ \bibinfo
  {pages} {062513} (\bibinfo {year} {2011})}\BibitemShut {NoStop}%
\bibitem [{\citenamefont {Rieker}\ \emph {et~al.}(2014)\citenamefont {Rieker},
  \citenamefont {Giorgetta}, \citenamefont {Swann}, \citenamefont {Kofler},
  \citenamefont {Zolot}, \citenamefont {Sinclair}, \citenamefont {Baumann},
  \citenamefont {Cromer}, \citenamefont {Petron}, \citenamefont {Sweeney},
  \citenamefont {Tans}, \citenamefont {Coddington},\ and\ \citenamefont
  {Newbury}}]{Rieker:14}%
  \BibitemOpen
  \bibfield  {author} {\bibinfo {author} {\bibfnamefont {G.~B.}\ \bibnamefont
  {Rieker}}, \bibinfo {author} {\bibfnamefont {F.~R.}\ \bibnamefont
  {Giorgetta}}, \bibinfo {author} {\bibfnamefont {W.~C.}\ \bibnamefont
  {Swann}}, \bibinfo {author} {\bibfnamefont {J.}~\bibnamefont {Kofler}},
  \bibinfo {author} {\bibfnamefont {A.~M.}\ \bibnamefont {Zolot}}, \bibinfo
  {author} {\bibfnamefont {L.~C.}\ \bibnamefont {Sinclair}}, \bibinfo {author}
  {\bibfnamefont {E.}~\bibnamefont {Baumann}}, \bibinfo {author} {\bibfnamefont
  {C.}~\bibnamefont {Cromer}}, \bibinfo {author} {\bibfnamefont
  {G.}~\bibnamefont {Petron}}, \bibinfo {author} {\bibfnamefont
  {C.}~\bibnamefont {Sweeney}}, \bibinfo {author} {\bibfnamefont {P.~P.}\
  \bibnamefont {Tans}}, \bibinfo {author} {\bibfnamefont {I.}~\bibnamefont
  {Coddington}}, \ and\ \bibinfo {author} {\bibfnamefont {N.~R.}\ \bibnamefont
  {Newbury}},\ }\href {\doibase 10.1364/OPTICA.1.000290} {\bibfield  {journal}
  {\bibinfo  {journal} {Optica}\ }\textbf {\bibinfo {volume} {1}},\ \bibinfo
  {pages} {290} (\bibinfo {year} {2014})}\BibitemShut {NoStop}%
\bibitem [{\citenamefont {Zhu}\ \emph {et~al.}(2015)\citenamefont {Zhu},
  \citenamefont {Bicer}, \citenamefont {Askar}, \citenamefont {Bounds},
  \citenamefont {Kolomenskii}, \citenamefont {Kelessides}, \citenamefont
  {Amani},\ and\ \citenamefont {Schuessler}}]{Zhu:2015}%
  \BibitemOpen
  \bibfield  {author} {\bibinfo {author} {\bibfnamefont {F.}~\bibnamefont
  {Zhu}}, \bibinfo {author} {\bibfnamefont {A.}~\bibnamefont {Bicer}}, \bibinfo
  {author} {\bibfnamefont {R.}~\bibnamefont {Askar}}, \bibinfo {author}
  {\bibfnamefont {J.}~\bibnamefont {Bounds}}, \bibinfo {author} {\bibfnamefont
  {A.~A.}\ \bibnamefont {Kolomenskii}}, \bibinfo {author} {\bibfnamefont
  {V.}~\bibnamefont {Kelessides}}, \bibinfo {author} {\bibfnamefont
  {M.}~\bibnamefont {Amani}}, \ and\ \bibinfo {author} {\bibfnamefont {H.~A.}\
  \bibnamefont {Schuessler}},\ }\href {\doibase 10.1088/1612-2011/12/9/095701}
  {\ \textbf {\bibinfo {volume} {12}},\ \bibinfo {pages} {095701} (\bibinfo
  {year} {2015})}\BibitemShut {NoStop}%
\bibitem [{\citenamefont {Schroeder}\ \emph {et~al.}(2017)\citenamefont
  {Schroeder}, \citenamefont {Wright}, \citenamefont {Coburn}, \citenamefont
  {Sodergren}, \citenamefont {Cossel}, \citenamefont {Droste}, \citenamefont
  {Truong}, \citenamefont {Baumann}, \citenamefont {Giorgetta}, \citenamefont
  {Coddington}, \citenamefont {Newbury},\ and\ \citenamefont
  {Rieker}}]{SCHROEDER:2017}%
  \BibitemOpen
  \bibfield  {author} {\bibinfo {author} {\bibfnamefont {P.}~\bibnamefont
  {Schroeder}}, \bibinfo {author} {\bibfnamefont {R.}~\bibnamefont {Wright}},
  \bibinfo {author} {\bibfnamefont {S.}~\bibnamefont {Coburn}}, \bibinfo
  {author} {\bibfnamefont {B.}~\bibnamefont {Sodergren}}, \bibinfo {author}
  {\bibfnamefont {K.}~\bibnamefont {Cossel}}, \bibinfo {author} {\bibfnamefont
  {S.}~\bibnamefont {Droste}}, \bibinfo {author} {\bibfnamefont
  {G.}~\bibnamefont {Truong}}, \bibinfo {author} {\bibfnamefont
  {E.}~\bibnamefont {Baumann}}, \bibinfo {author} {\bibfnamefont
  {F.}~\bibnamefont {Giorgetta}}, \bibinfo {author} {\bibfnamefont
  {I.}~\bibnamefont {Coddington}}, \bibinfo {author} {\bibfnamefont
  {N.}~\bibnamefont {Newbury}}, \ and\ \bibinfo {author} {\bibfnamefont
  {G.}~\bibnamefont {Rieker}},\ }\href {\doibase
  https://doi.org/10.1016/j.proci.2016.06.032} {\bibfield  {journal} {\bibinfo
  {journal} {Proceedings of the Combustion Institute}\ }\textbf {\bibinfo
  {volume} {36}},\ \bibinfo {pages} {4565} (\bibinfo {year}
  {2017})}\BibitemShut {NoStop}%
\bibitem [{\citenamefont {Coddington}\ \emph {et~al.}(2009)\citenamefont
  {Coddington}, \citenamefont {Swann}, \citenamefont {Nenadovic},\ and\
  \citenamefont {Newbury}}]{Coddington2009}%
  \BibitemOpen
  \bibfield  {author} {\bibinfo {author} {\bibfnamefont {I.}~\bibnamefont
  {Coddington}}, \bibinfo {author} {\bibfnamefont {W.~C.}\ \bibnamefont
  {Swann}}, \bibinfo {author} {\bibfnamefont {L.}~\bibnamefont {Nenadovic}}, \
  and\ \bibinfo {author} {\bibfnamefont {N.~R.}\ \bibnamefont {Newbury}},\
  }\href {\doibase 10.1038/nphoton.2009.94} {\bibfield  {journal} {\bibinfo
  {journal} {Nature Photonics}\ }\textbf {\bibinfo {volume} {3}},\ \bibinfo
  {pages} {351} (\bibinfo {year} {2009})}\BibitemShut {NoStop}%
\bibitem [{\citenamefont {Suh}\ and\ \citenamefont {Vahala}(2018)}]{Suh2018}%
  \BibitemOpen
  \bibfield  {author} {\bibinfo {author} {\bibfnamefont {M.-G.}\ \bibnamefont
  {Suh}}\ and\ \bibinfo {author} {\bibfnamefont {K.~J.}\ \bibnamefont
  {Vahala}},\ }\href {\doibase 10.1126/science.aao1968} {\bibfield  {journal}
  {\bibinfo  {journal} {Science}\ }\textbf {\bibinfo {volume} {359}},\ \bibinfo
  {pages} {884} (\bibinfo {year} {2018})}\BibitemShut {NoStop}%
\bibitem [{\citenamefont {Trocha}\ \emph {et~al.}(2018)\citenamefont {Trocha},
  \citenamefont {Karpov}, \citenamefont {Ganin}, \citenamefont {Pfeiffer},
  \citenamefont {Kordts}, \citenamefont {Wolf}, \citenamefont {Krockenberger},
  \citenamefont {Marin-Palomo}, \citenamefont {Weimann}, \citenamefont
  {Randel}, \citenamefont {Freude}, \citenamefont {Kippenberg},\ and\
  \citenamefont {Koos}}]{Trocha:2018}%
  \BibitemOpen
  \bibfield  {author} {\bibinfo {author} {\bibfnamefont {P.}~\bibnamefont
  {Trocha}}, \bibinfo {author} {\bibfnamefont {M.}~\bibnamefont {Karpov}},
  \bibinfo {author} {\bibfnamefont {D.}~\bibnamefont {Ganin}}, \bibinfo
  {author} {\bibfnamefont {M.~H.~P.}\ \bibnamefont {Pfeiffer}}, \bibinfo
  {author} {\bibfnamefont {A.}~\bibnamefont {Kordts}}, \bibinfo {author}
  {\bibfnamefont {S.}~\bibnamefont {Wolf}}, \bibinfo {author} {\bibfnamefont
  {J.}~\bibnamefont {Krockenberger}}, \bibinfo {author} {\bibfnamefont
  {P.}~\bibnamefont {Marin-Palomo}}, \bibinfo {author} {\bibfnamefont
  {C.}~\bibnamefont {Weimann}}, \bibinfo {author} {\bibfnamefont
  {S.}~\bibnamefont {Randel}}, \bibinfo {author} {\bibfnamefont
  {W.}~\bibnamefont {Freude}}, \bibinfo {author} {\bibfnamefont {T.~J.}\
  \bibnamefont {Kippenberg}}, \ and\ \bibinfo {author} {\bibfnamefont
  {C.}~\bibnamefont {Koos}},\ }\href {\doibase 10.1126/science.aao3924}
  {\bibfield  {journal} {\bibinfo  {journal} {Science}\ }\textbf {\bibinfo
  {volume} {359}},\ \bibinfo {pages} {887} (\bibinfo {year}
  {2018})}\BibitemShut {NoStop}%
\bibitem [{\citenamefont {N\"{u}rnberg}\ \emph {et~al.}(2021)\citenamefont
  {N\"{u}rnberg}, \citenamefont {Willenberg}, \citenamefont {Phillips},\ and\
  \citenamefont {Keller}}]{Nurnberg:21}%
  \BibitemOpen
  \bibfield  {author} {\bibinfo {author} {\bibfnamefont {J.}~\bibnamefont
  {N\"{u}rnberg}}, \bibinfo {author} {\bibfnamefont {B.}~\bibnamefont
  {Willenberg}}, \bibinfo {author} {\bibfnamefont {C.~R.}\ \bibnamefont
  {Phillips}}, \ and\ \bibinfo {author} {\bibfnamefont {U.}~\bibnamefont
  {Keller}},\ }\href {\doibase 10.1364/OE.428051} {\bibfield  {journal}
  {\bibinfo  {journal} {Opt. Express}\ }\textbf {\bibinfo {volume} {29}},\
  \bibinfo {pages} {24910} (\bibinfo {year} {2021})}\BibitemShut {NoStop}%
\bibitem [{\citenamefont {Riemensberger}\ \emph {et~al.}(2020)\citenamefont
  {Riemensberger}, \citenamefont {Lukashchuk}, \citenamefont {Karpov},
  \citenamefont {Weng}, \citenamefont {Lucas}, \citenamefont {Liu},\ and\
  \citenamefont {Kippenberg}}]{riemensberger2020massively}%
  \BibitemOpen
  \bibfield  {author} {\bibinfo {author} {\bibfnamefont {J.}~\bibnamefont
  {Riemensberger}}, \bibinfo {author} {\bibfnamefont {A.}~\bibnamefont
  {Lukashchuk}}, \bibinfo {author} {\bibfnamefont {M.}~\bibnamefont {Karpov}},
  \bibinfo {author} {\bibfnamefont {W.}~\bibnamefont {Weng}}, \bibinfo {author}
  {\bibfnamefont {E.}~\bibnamefont {Lucas}}, \bibinfo {author} {\bibfnamefont
  {J.}~\bibnamefont {Liu}}, \ and\ \bibinfo {author} {\bibfnamefont {T.~J.}\
  \bibnamefont {Kippenberg}},\ }\href
  {https://www.nature.com/articles/s41586-020-2239-3} {\bibfield  {journal}
  {\bibinfo  {journal} {Nature}\ }\textbf {\bibinfo {volume} {581}},\ \bibinfo
  {pages} {164} (\bibinfo {year} {2020})}\BibitemShut {NoStop}%
\bibitem [{\citenamefont {Coddington}\ \emph {et~al.}(2010)\citenamefont
  {Coddington}, \citenamefont {Swann},\ and\ \citenamefont
  {Newbury}}]{coddington2010coherent}%
  \BibitemOpen
  \bibfield  {author} {\bibinfo {author} {\bibfnamefont {I.}~\bibnamefont
  {Coddington}}, \bibinfo {author} {\bibfnamefont {W.}~\bibnamefont {Swann}}, \
  and\ \bibinfo {author} {\bibfnamefont {N.}~\bibnamefont {Newbury}},\ }\href
  {https://journals.aps.org/pra/abstract/10.1103/PhysRevA.82.043817} {\bibfield
   {journal} {\bibinfo  {journal} {Physical Review A}\ }\textbf {\bibinfo
  {volume} {82}},\ \bibinfo {pages} {043817} (\bibinfo {year}
  {2010})}\BibitemShut {NoStop}%
\bibitem [{\citenamefont {Hoghooghi}\ \emph {et~al.}(2021)\citenamefont
  {Hoghooghi}, \citenamefont {Cole},\ and\ \citenamefont
  {Rieker}}]{hoghooghi202111}%
  \BibitemOpen
  \bibfield  {author} {\bibinfo {author} {\bibfnamefont {N.}~\bibnamefont
  {Hoghooghi}}, \bibinfo {author} {\bibfnamefont {R.~K.}\ \bibnamefont {Cole}},
  \ and\ \bibinfo {author} {\bibfnamefont {G.~B.}\ \bibnamefont {Rieker}},\
  }\href {https://link.springer.com/article/10.1007/s00340-020-07552-y}
  {\bibfield  {journal} {\bibinfo  {journal} {Applied Physics B}\ }\textbf
  {\bibinfo {volume} {127}},\ \bibinfo {pages} {1} (\bibinfo {year}
  {2021})}\BibitemShut {NoStop}%
\bibitem [{\citenamefont {Kayes}\ \emph {et~al.}(2018)\citenamefont {Kayes},
  \citenamefont {Abdukerim}, \citenamefont {Rekik},\ and\ \citenamefont
  {Rochette}}]{Kayes:18}%
  \BibitemOpen
  \bibfield  {author} {\bibinfo {author} {\bibfnamefont {M.~I.}\ \bibnamefont
  {Kayes}}, \bibinfo {author} {\bibfnamefont {N.}~\bibnamefont {Abdukerim}},
  \bibinfo {author} {\bibfnamefont {A.}~\bibnamefont {Rekik}}, \ and\ \bibinfo
  {author} {\bibfnamefont {M.}~\bibnamefont {Rochette}},\ }\href {\doibase
  10.1364/OL.43.005809} {\bibfield  {journal} {\bibinfo  {journal} {Opt.
  Lett.}\ }\textbf {\bibinfo {volume} {43}},\ \bibinfo {pages} {5809} (\bibinfo
  {year} {2018})}\BibitemShut {NoStop}%
\bibitem [{\citenamefont {Link}\ \emph {et~al.}(2016)\citenamefont {Link},
  \citenamefont {Klenner},\ and\ \citenamefont {Keller}}]{Link:16}%
  \BibitemOpen
  \bibfield  {author} {\bibinfo {author} {\bibfnamefont {S.~M.}\ \bibnamefont
  {Link}}, \bibinfo {author} {\bibfnamefont {A.}~\bibnamefont {Klenner}}, \
  and\ \bibinfo {author} {\bibfnamefont {U.}~\bibnamefont {Keller}},\ }\href
  {\doibase 10.1364/OE.24.001889} {\bibfield  {journal} {\bibinfo  {journal}
  {Opt. Express}\ }\textbf {\bibinfo {volume} {24}},\ \bibinfo {pages} {1889}
  (\bibinfo {year} {2016})}\BibitemShut {NoStop}%
\bibitem [{\citenamefont {Link}\ \emph {et~al.}(2015)\citenamefont {Link},
  \citenamefont {Zaugg}, \citenamefont {Klenner}, \citenamefont {Mangold},
  \citenamefont {Golling}, \citenamefont {Tilma},\ and\ \citenamefont
  {Keller}}]{link2015dual}%
  \BibitemOpen
  \bibfield  {author} {\bibinfo {author} {\bibfnamefont {S.}~\bibnamefont
  {Link}}, \bibinfo {author} {\bibfnamefont {C.}~\bibnamefont {Zaugg}},
  \bibinfo {author} {\bibfnamefont {A.}~\bibnamefont {Klenner}}, \bibinfo
  {author} {\bibfnamefont {M.}~\bibnamefont {Mangold}}, \bibinfo {author}
  {\bibfnamefont {M.}~\bibnamefont {Golling}}, \bibinfo {author} {\bibfnamefont
  {B.}~\bibnamefont {Tilma}}, \ and\ \bibinfo {author} {\bibfnamefont
  {U.}~\bibnamefont {Keller}},\ }in\ \href
  {https://www.spiedigitallibrary.org/conference-proceedings-of-spie/9349/93490Q/Dual-comb-MIXSEL/10.1117/12.2078010.short?SSO=1}
  {\emph {\bibinfo {booktitle} {Vertical External Cavity Surface Emitting
  Lasers (VECSELs) V}}},\ Vol.\ \bibinfo {volume} {9349}\ (\bibinfo
  {organization} {International Society for Optics and Photonics},\ \bibinfo
  {year} {2015})\ p.\ \bibinfo {pages} {93490Q}\BibitemShut {NoStop}%
\bibitem [{\citenamefont {N{\"u}rnberg}\ \emph {et~al.}(2019)\citenamefont
  {N{\"u}rnberg}, \citenamefont {Alfieri}, \citenamefont {Waldburger},
  \citenamefont {Kr{\"u}ger}, \citenamefont {Golling},\ and\ \citenamefont
  {Keller}}]{nurnberg2019single}%
  \BibitemOpen
  \bibfield  {author} {\bibinfo {author} {\bibfnamefont {J.}~\bibnamefont
  {N{\"u}rnberg}}, \bibinfo {author} {\bibfnamefont {C.~G.}\ \bibnamefont
  {Alfieri}}, \bibinfo {author} {\bibfnamefont {D.}~\bibnamefont {Waldburger}},
  \bibinfo {author} {\bibfnamefont {L.}~\bibnamefont {Kr{\"u}ger}}, \bibinfo
  {author} {\bibfnamefont {M.}~\bibnamefont {Golling}}, \ and\ \bibinfo
  {author} {\bibfnamefont {U.}~\bibnamefont {Keller}},\ }in\ \href
  {https://www.osapublishing.org/abstract.cfm?uri=CLEO_Europe-2019-ch_2_6}
  {\emph {\bibinfo {booktitle} {2019 Conference on Lasers and Electro-Optics
  Europe \& European Quantum Electronics Conference (CLEO/Europe-EQEC)}}}\
  (\bibinfo {organization} {IEEE},\ \bibinfo {year} {2019})\ pp.\ \bibinfo
  {pages} {1--1}\BibitemShut {NoStop}%
\bibitem [{\citenamefont {Consolino}\ \emph {et~al.}(2020)\citenamefont
  {Consolino}, \citenamefont {Nafa}, \citenamefont {De~Regis}, \citenamefont
  {Cappelli}, \citenamefont {Garrasi}, \citenamefont {Mezzapesa}, \citenamefont
  {Li}, \citenamefont {Davies}, \citenamefont {Linfield}, \citenamefont
  {Vitiello}, \citenamefont {Bartalini},\ and\ \citenamefont
  {De~Natale}}]{Consolino2020}%
  \BibitemOpen
  \bibfield  {author} {\bibinfo {author} {\bibfnamefont {L.}~\bibnamefont
  {Consolino}}, \bibinfo {author} {\bibfnamefont {M.}~\bibnamefont {Nafa}},
  \bibinfo {author} {\bibfnamefont {M.}~\bibnamefont {De~Regis}}, \bibinfo
  {author} {\bibfnamefont {F.}~\bibnamefont {Cappelli}}, \bibinfo {author}
  {\bibfnamefont {K.}~\bibnamefont {Garrasi}}, \bibinfo {author} {\bibfnamefont
  {F.~P.}\ \bibnamefont {Mezzapesa}}, \bibinfo {author} {\bibfnamefont
  {L.}~\bibnamefont {Li}}, \bibinfo {author} {\bibfnamefont {A.~G.}\
  \bibnamefont {Davies}}, \bibinfo {author} {\bibfnamefont {E.~H.}\
  \bibnamefont {Linfield}}, \bibinfo {author} {\bibfnamefont {M.~S.}\
  \bibnamefont {Vitiello}}, \bibinfo {author} {\bibfnamefont {S.}~\bibnamefont
  {Bartalini}}, \ and\ \bibinfo {author} {\bibfnamefont {P.}~\bibnamefont
  {De~Natale}},\ }\href {\doibase 10.1038/s42005-020-0344-0} {\bibfield
  {journal} {\bibinfo  {journal} {Communications Physics}\ }\textbf {\bibinfo
  {volume} {3}},\ \bibinfo {pages} {69} (\bibinfo {year} {2020})}\BibitemShut
  {NoStop}%
\bibitem [{\citenamefont {Komagata}\ \emph {et~al.}(2021)\citenamefont
  {Komagata}, \citenamefont {Shehzad}, \citenamefont {Terrasanta},
  \citenamefont {Brochard}, \citenamefont {Matthey}, \citenamefont {Gianella},
  \citenamefont {Jouy}, \citenamefont {Kapsalidis}, \citenamefont
  {Shahmohammadi}, \citenamefont {Beck} \emph
  {et~al.}}]{komagata2021coherently}%
  \BibitemOpen
  \bibfield  {author} {\bibinfo {author} {\bibfnamefont {K.}~\bibnamefont
  {Komagata}}, \bibinfo {author} {\bibfnamefont {A.}~\bibnamefont {Shehzad}},
  \bibinfo {author} {\bibfnamefont {G.}~\bibnamefont {Terrasanta}}, \bibinfo
  {author} {\bibfnamefont {P.}~\bibnamefont {Brochard}}, \bibinfo {author}
  {\bibfnamefont {R.}~\bibnamefont {Matthey}}, \bibinfo {author} {\bibfnamefont
  {M.}~\bibnamefont {Gianella}}, \bibinfo {author} {\bibfnamefont
  {P.}~\bibnamefont {Jouy}}, \bibinfo {author} {\bibfnamefont {F.}~\bibnamefont
  {Kapsalidis}}, \bibinfo {author} {\bibfnamefont {M.}~\bibnamefont
  {Shahmohammadi}}, \bibinfo {author} {\bibfnamefont {M.}~\bibnamefont {Beck}},
   \emph {et~al.},\ }\href
  {https://www.osapublishing.org/oe/fulltext.cfm?uri=oe-29-12-19126&id=451668}
  {\bibfield  {journal} {\bibinfo  {journal} {Optics Express}\ }\textbf
  {\bibinfo {volume} {29}},\ \bibinfo {pages} {19126} (\bibinfo {year}
  {2021})}\BibitemShut {NoStop}%
\bibitem [{\citenamefont {Pfeiffer}\ \emph {et~al.}(2016)\citenamefont
  {Pfeiffer}, \citenamefont {Kordts}, \citenamefont {Brasch}, \citenamefont
  {Zervas}, \citenamefont {Geiselmann}, \citenamefont {Jost},\ and\
  \citenamefont {Kippenberg}}]{pfeiffer2016photonic}%
  \BibitemOpen
  \bibfield  {author} {\bibinfo {author} {\bibfnamefont {M.~H.}\ \bibnamefont
  {Pfeiffer}}, \bibinfo {author} {\bibfnamefont {A.}~\bibnamefont {Kordts}},
  \bibinfo {author} {\bibfnamefont {V.}~\bibnamefont {Brasch}}, \bibinfo
  {author} {\bibfnamefont {M.}~\bibnamefont {Zervas}}, \bibinfo {author}
  {\bibfnamefont {M.}~\bibnamefont {Geiselmann}}, \bibinfo {author}
  {\bibfnamefont {J.~D.}\ \bibnamefont {Jost}}, \ and\ \bibinfo {author}
  {\bibfnamefont {T.~J.}\ \bibnamefont {Kippenberg}},\ }\href
  {https://www.osapublishing.org/optica/fulltext.cfm?uri=optica-3-1-20&id=335237}
  {\bibfield  {journal} {\bibinfo  {journal} {Optica}\ }\textbf {\bibinfo
  {volume} {3}},\ \bibinfo {pages} {20} (\bibinfo {year} {2016})}\BibitemShut
  {NoStop}%
\bibitem [{\citenamefont {Liu}\ \emph {et~al.}(2018)\citenamefont {Liu},
  \citenamefont {Raja}, \citenamefont {Karpov}, \citenamefont {Ghadiani},
  \citenamefont {Pfeiffer}, \citenamefont {Du}, \citenamefont {Engelsen},
  \citenamefont {Guo}, \citenamefont {Zervas},\ and\ \citenamefont
  {Kippenberg}}]{liu2018ultralow}%
  \BibitemOpen
  \bibfield  {author} {\bibinfo {author} {\bibfnamefont {J.}~\bibnamefont
  {Liu}}, \bibinfo {author} {\bibfnamefont {A.~S.}\ \bibnamefont {Raja}},
  \bibinfo {author} {\bibfnamefont {M.}~\bibnamefont {Karpov}}, \bibinfo
  {author} {\bibfnamefont {B.}~\bibnamefont {Ghadiani}}, \bibinfo {author}
  {\bibfnamefont {M.~H.}\ \bibnamefont {Pfeiffer}}, \bibinfo {author}
  {\bibfnamefont {B.}~\bibnamefont {Du}}, \bibinfo {author} {\bibfnamefont
  {N.~J.}\ \bibnamefont {Engelsen}}, \bibinfo {author} {\bibfnamefont
  {H.}~\bibnamefont {Guo}}, \bibinfo {author} {\bibfnamefont {M.}~\bibnamefont
  {Zervas}}, \ and\ \bibinfo {author} {\bibfnamefont {T.~J.}\ \bibnamefont
  {Kippenberg}},\ }\href
  {https://www.osapublishing.org/optica/fulltext.cfm?uri=optica-5-10-1347&id=399272}
  {\bibfield  {journal} {\bibinfo  {journal} {Optica}\ }\textbf {\bibinfo
  {volume} {5}},\ \bibinfo {pages} {1347} (\bibinfo {year} {2018})}\BibitemShut
  {NoStop}%
\bibitem [{\citenamefont {Liu}\ \emph {et~al.}(2021)\citenamefont {Liu},
  \citenamefont {Huang}, \citenamefont {Wang}, \citenamefont {He},
  \citenamefont {Raja}, \citenamefont {Liu}, \citenamefont {Engelsen},\ and\
  \citenamefont {Kippenberg}}]{liu2021high}%
  \BibitemOpen
  \bibfield  {author} {\bibinfo {author} {\bibfnamefont {J.}~\bibnamefont
  {Liu}}, \bibinfo {author} {\bibfnamefont {G.}~\bibnamefont {Huang}}, \bibinfo
  {author} {\bibfnamefont {R.~N.}\ \bibnamefont {Wang}}, \bibinfo {author}
  {\bibfnamefont {J.}~\bibnamefont {He}}, \bibinfo {author} {\bibfnamefont
  {A.~S.}\ \bibnamefont {Raja}}, \bibinfo {author} {\bibfnamefont
  {T.}~\bibnamefont {Liu}}, \bibinfo {author} {\bibfnamefont {N.~J.}\
  \bibnamefont {Engelsen}}, \ and\ \bibinfo {author} {\bibfnamefont {T.~J.}\
  \bibnamefont {Kippenberg}},\ }\href
  {https://www.nature.com/articles/s41467-021-21973-z} {\bibfield  {journal}
  {\bibinfo  {journal} {Nature communications}\ }\textbf {\bibinfo {volume}
  {12}},\ \bibinfo {pages} {1} (\bibinfo {year} {2021})}\BibitemShut {NoStop}%
\bibitem [{\citenamefont {Hochberg}\ and\ \citenamefont
  {Baehr-Jones}(2010)}]{hochberg2010towards}%
  \BibitemOpen
  \bibfield  {author} {\bibinfo {author} {\bibfnamefont {M.}~\bibnamefont
  {Hochberg}}\ and\ \bibinfo {author} {\bibfnamefont {T.}~\bibnamefont
  {Baehr-Jones}},\ }\href {https://www.nature.com/articles/nphoton.2010.172}
  {\bibfield  {journal} {\bibinfo  {journal} {Nature photonics}\ }\textbf
  {\bibinfo {volume} {4}},\ \bibinfo {pages} {492} (\bibinfo {year}
  {2010})}\BibitemShut {NoStop}%
\bibitem [{\citenamefont {Hochberg}\ \emph {et~al.}(2013)\citenamefont
  {Hochberg}, \citenamefont {Harris}, \citenamefont {Ding}, \citenamefont
  {Zhang}, \citenamefont {Novack}, \citenamefont {Xuan},\ and\ \citenamefont
  {Baehr-Jones}}]{hochberg2013silicon}%
  \BibitemOpen
  \bibfield  {author} {\bibinfo {author} {\bibfnamefont {M.}~\bibnamefont
  {Hochberg}}, \bibinfo {author} {\bibfnamefont {N.~C.}\ \bibnamefont
  {Harris}}, \bibinfo {author} {\bibfnamefont {R.}~\bibnamefont {Ding}},
  \bibinfo {author} {\bibfnamefont {Y.}~\bibnamefont {Zhang}}, \bibinfo
  {author} {\bibfnamefont {A.}~\bibnamefont {Novack}}, \bibinfo {author}
  {\bibfnamefont {Z.}~\bibnamefont {Xuan}}, \ and\ \bibinfo {author}
  {\bibfnamefont {T.}~\bibnamefont {Baehr-Jones}},\ }\href
  {https://ieeexplore.ieee.org/abstract/document/6449374} {\bibfield  {journal}
  {\bibinfo  {journal} {IEEE Solid-State Circuits Magazine}\ }\textbf {\bibinfo
  {volume} {5}},\ \bibinfo {pages} {48} (\bibinfo {year} {2013})}\BibitemShut
  {NoStop}%
\bibitem [{\citenamefont {Weimann}\ \emph {et~al.}(2017)\citenamefont
  {Weimann}, \citenamefont {Lauermann}, \citenamefont {Hoeller}, \citenamefont
  {Freude},\ and\ \citenamefont {Koos}}]{weimann2017silicon}%
  \BibitemOpen
  \bibfield  {author} {\bibinfo {author} {\bibfnamefont {C.}~\bibnamefont
  {Weimann}}, \bibinfo {author} {\bibfnamefont {M.}~\bibnamefont {Lauermann}},
  \bibinfo {author} {\bibfnamefont {F.}~\bibnamefont {Hoeller}}, \bibinfo
  {author} {\bibfnamefont {W.}~\bibnamefont {Freude}}, \ and\ \bibinfo {author}
  {\bibfnamefont {C.}~\bibnamefont {Koos}},\ }\href
  {https://www.osapublishing.org/oe/fulltext.cfm?uri=oe-25-24-30091&id=376976}
  {\bibfield  {journal} {\bibinfo  {journal} {Optics express}\ }\textbf
  {\bibinfo {volume} {25}},\ \bibinfo {pages} {30091} (\bibinfo {year}
  {2017})}\BibitemShut {NoStop}%
\bibitem [{\citenamefont {Xiang}\ \emph
  {et~al.}(2021{\natexlab{a}})\citenamefont {Xiang}, \citenamefont {Liu},
  \citenamefont {Guo}, \citenamefont {Chang}, \citenamefont {Wang},
  \citenamefont {Weng}, \citenamefont {Peters}, \citenamefont {Xie},
  \citenamefont {Zhang}, \citenamefont {Riemensberger} \emph
  {et~al.}}]{xiang2021laser}%
  \BibitemOpen
  \bibfield  {author} {\bibinfo {author} {\bibfnamefont {C.}~\bibnamefont
  {Xiang}}, \bibinfo {author} {\bibfnamefont {J.}~\bibnamefont {Liu}}, \bibinfo
  {author} {\bibfnamefont {J.}~\bibnamefont {Guo}}, \bibinfo {author}
  {\bibfnamefont {L.}~\bibnamefont {Chang}}, \bibinfo {author} {\bibfnamefont
  {R.~N.}\ \bibnamefont {Wang}}, \bibinfo {author} {\bibfnamefont
  {W.}~\bibnamefont {Weng}}, \bibinfo {author} {\bibfnamefont {J.}~\bibnamefont
  {Peters}}, \bibinfo {author} {\bibfnamefont {W.}~\bibnamefont {Xie}},
  \bibinfo {author} {\bibfnamefont {Z.}~\bibnamefont {Zhang}}, \bibinfo
  {author} {\bibfnamefont {J.}~\bibnamefont {Riemensberger}},  \emph {et~al.},\
  }\href {https://arxiv.org/abs/2103.02725} {\bibfield  {journal} {\bibinfo
  {journal} {arXiv preprint arXiv:2103.02725}\ } (\bibinfo {year}
  {2021}{\natexlab{a}})}\BibitemShut {NoStop}%
\bibitem [{\citenamefont {Xiang}\ \emph
  {et~al.}(2021{\natexlab{b}})\citenamefont {Xiang}, \citenamefont {Guo},
  \citenamefont {Jin}, \citenamefont {Wu}, \citenamefont {Peters},
  \citenamefont {Xie}, \citenamefont {Chang}, \citenamefont {Shen},
  \citenamefont {Wang}, \citenamefont {Yang} \emph {et~al.}}]{xiang2021high}%
  \BibitemOpen
  \bibfield  {author} {\bibinfo {author} {\bibfnamefont {C.}~\bibnamefont
  {Xiang}}, \bibinfo {author} {\bibfnamefont {J.}~\bibnamefont {Guo}}, \bibinfo
  {author} {\bibfnamefont {W.}~\bibnamefont {Jin}}, \bibinfo {author}
  {\bibfnamefont {L.}~\bibnamefont {Wu}}, \bibinfo {author} {\bibfnamefont
  {J.}~\bibnamefont {Peters}}, \bibinfo {author} {\bibfnamefont
  {W.}~\bibnamefont {Xie}}, \bibinfo {author} {\bibfnamefont {L.}~\bibnamefont
  {Chang}}, \bibinfo {author} {\bibfnamefont {B.}~\bibnamefont {Shen}},
  \bibinfo {author} {\bibfnamefont {H.}~\bibnamefont {Wang}}, \bibinfo {author}
  {\bibfnamefont {Q.-F.}\ \bibnamefont {Yang}},  \emph {et~al.},\ }\href
  {https://www.nature.com/articles/s41467-021-26804-9} {\bibfield  {journal}
  {\bibinfo  {journal} {Nature Communications}\ }\textbf {\bibinfo {volume}
  {12}},\ \bibinfo {pages} {1} (\bibinfo {year}
  {2021}{\natexlab{b}})}\BibitemShut {NoStop}%
\bibitem [{\citenamefont {Cuyvers}\ \emph {et~al.}(2021)\citenamefont
  {Cuyvers}, \citenamefont {Haq}, \citenamefont {Op~de Beeck}, \citenamefont
  {Poelman}, \citenamefont {Hermans}, \citenamefont {Wang}, \citenamefont
  {Gocalinska}, \citenamefont {Pelucchi}, \citenamefont {Corbett},
  \citenamefont {Roelkens} \emph {et~al.}}]{cuyvers2021low}%
  \BibitemOpen
  \bibfield  {author} {\bibinfo {author} {\bibfnamefont {S.}~\bibnamefont
  {Cuyvers}}, \bibinfo {author} {\bibfnamefont {B.}~\bibnamefont {Haq}},
  \bibinfo {author} {\bibfnamefont {C.}~\bibnamefont {Op~de Beeck}}, \bibinfo
  {author} {\bibfnamefont {S.}~\bibnamefont {Poelman}}, \bibinfo {author}
  {\bibfnamefont {A.}~\bibnamefont {Hermans}}, \bibinfo {author} {\bibfnamefont
  {Z.}~\bibnamefont {Wang}}, \bibinfo {author} {\bibfnamefont {A.}~\bibnamefont
  {Gocalinska}}, \bibinfo {author} {\bibfnamefont {E.}~\bibnamefont
  {Pelucchi}}, \bibinfo {author} {\bibfnamefont {B.}~\bibnamefont {Corbett}},
  \bibinfo {author} {\bibfnamefont {G.}~\bibnamefont {Roelkens}},  \emph
  {et~al.},\ }\href
  {https://onlinelibrary.wiley.com/doi/full/10.1002/lpor.202000485} {\bibfield
  {journal} {\bibinfo  {journal} {Laser \& Photonics Reviews}\ ,\ \bibinfo
  {pages} {2000485}} (\bibinfo {year} {2021})}\BibitemShut {NoStop}%
\bibitem [{\citenamefont {Dahmani}\ \emph {et~al.}(1987)\citenamefont
  {Dahmani}, \citenamefont {Hollberg},\ and\ \citenamefont
  {Drullinger}}]{Dahmani:87}%
  \BibitemOpen
  \bibfield  {author} {\bibinfo {author} {\bibfnamefont {B.}~\bibnamefont
  {Dahmani}}, \bibinfo {author} {\bibfnamefont {L.}~\bibnamefont {Hollberg}}, \
  and\ \bibinfo {author} {\bibfnamefont {R.}~\bibnamefont {Drullinger}},\
  }\href {\doibase 10.1364/OL.12.000876} {\bibfield  {journal} {\bibinfo
  {journal} {Opt. Lett.}\ }\textbf {\bibinfo {volume} {12}},\ \bibinfo {pages}
  {876} (\bibinfo {year} {1987})}\BibitemShut {NoStop}%
\bibitem [{\citenamefont {Laurent}\ \emph {et~al.}(1989)\citenamefont
  {Laurent}, \citenamefont {Clairon},\ and\ \citenamefont
  {Breant}}]{Laurent1989}%
  \BibitemOpen
  \bibfield  {author} {\bibinfo {author} {\bibfnamefont {P.}~\bibnamefont
  {Laurent}}, \bibinfo {author} {\bibfnamefont {A.}~\bibnamefont {Clairon}}, \
  and\ \bibinfo {author} {\bibfnamefont {C.}~\bibnamefont {Breant}},\ }\href
  {\doibase 10.1109/3.29238} {\bibfield  {journal} {\bibinfo  {journal} {IEEE
  Journ. Quant. El.}\ }\textbf {\bibinfo {volume} {25}},\ \bibinfo {pages}
  {1131} (\bibinfo {year} {1989})}\BibitemShut {NoStop}%
\bibitem [{\citenamefont {Oraevsky}\ \emph {et~al.}(2001)\citenamefont
  {Oraevsky}, \citenamefont {Yarovitsky},\ and\ \citenamefont
  {Velichansky}}]{Oraevsky2001}%
  \BibitemOpen
  \bibfield  {author} {\bibinfo {author} {\bibfnamefont {A.~N.}\ \bibnamefont
  {Oraevsky}}, \bibinfo {author} {\bibfnamefont {A.~V.}\ \bibnamefont
  {Yarovitsky}}, \ and\ \bibinfo {author} {\bibfnamefont {V.~L.}\ \bibnamefont
  {Velichansky}},\ }\href {\doibase 10.1070/qe2001v031n10abeh002073} {\ \textbf
  {\bibinfo {volume} {31}},\ \bibinfo {pages} {897} (\bibinfo {year}
  {2001})}\BibitemShut {NoStop}%
\bibitem [{\citenamefont {Liang}\ \emph {et~al.}(2010)\citenamefont {Liang},
  \citenamefont {Ilchenko}, \citenamefont {Savchenkov}, \citenamefont {Matsko},
  \citenamefont {Seidel},\ and\ \citenamefont {Maleki}}]{Liang:10}%
  \BibitemOpen
  \bibfield  {author} {\bibinfo {author} {\bibfnamefont {W.}~\bibnamefont
  {Liang}}, \bibinfo {author} {\bibfnamefont {V.~S.}\ \bibnamefont {Ilchenko}},
  \bibinfo {author} {\bibfnamefont {A.~A.}\ \bibnamefont {Savchenkov}},
  \bibinfo {author} {\bibfnamefont {A.~B.}\ \bibnamefont {Matsko}}, \bibinfo
  {author} {\bibfnamefont {D.}~\bibnamefont {Seidel}}, \ and\ \bibinfo {author}
  {\bibfnamefont {L.}~\bibnamefont {Maleki}},\ }\href {\doibase
  10.1364/OL.35.002822} {\bibfield  {journal} {\bibinfo  {journal} {Opt.
  Lett.}\ }\textbf {\bibinfo {volume} {35}},\ \bibinfo {pages} {2822} (\bibinfo
  {year} {2010})}\BibitemShut {NoStop}%
\bibitem [{\citenamefont {Liang}\ \emph
  {et~al.}(2015{\natexlab{a}})\citenamefont {Liang}, \citenamefont {Eliyahu},
  \citenamefont {Ilchenko}, \citenamefont {Savchenkov}, \citenamefont {Matsko},
  \citenamefont {Seidel},\ and\ \citenamefont {Maleki}}]{liang2015high}%
  \BibitemOpen
  \bibfield  {author} {\bibinfo {author} {\bibfnamefont {W.}~\bibnamefont
  {Liang}}, \bibinfo {author} {\bibfnamefont {D.}~\bibnamefont {Eliyahu}},
  \bibinfo {author} {\bibfnamefont {V.~S.}\ \bibnamefont {Ilchenko}}, \bibinfo
  {author} {\bibfnamefont {A.~A.}\ \bibnamefont {Savchenkov}}, \bibinfo
  {author} {\bibfnamefont {A.~B.}\ \bibnamefont {Matsko}}, \bibinfo {author}
  {\bibfnamefont {D.}~\bibnamefont {Seidel}}, \ and\ \bibinfo {author}
  {\bibfnamefont {L.}~\bibnamefont {Maleki}},\ }\href
  {https://doi.org/10.1038/ncomms8957} {\bibfield  {journal} {\bibinfo
  {journal} {Nature Communications}\ }\textbf {\bibinfo {volume} {6}},\
  \bibinfo {pages} {7957} (\bibinfo {year} {2015}{\natexlab{a}})}\BibitemShut
  {NoStop}%
\bibitem [{\citenamefont {Liang}\ \emph
  {et~al.}(2015{\natexlab{b}})\citenamefont {Liang}, \citenamefont {Ilchenko},
  \citenamefont {Eliyahu}, \citenamefont {Savchenkov}, \citenamefont {Matsko},
  \citenamefont {Seidel},\ and\ \citenamefont {Maleki}}]{liang2015ultralow}%
  \BibitemOpen
  \bibfield  {author} {\bibinfo {author} {\bibfnamefont {W.}~\bibnamefont
  {Liang}}, \bibinfo {author} {\bibfnamefont {V.~S.}\ \bibnamefont {Ilchenko}},
  \bibinfo {author} {\bibfnamefont {D.}~\bibnamefont {Eliyahu}}, \bibinfo
  {author} {\bibfnamefont {A.~A.}\ \bibnamefont {Savchenkov}}, \bibinfo
  {author} {\bibfnamefont {A.~B.}\ \bibnamefont {Matsko}}, \bibinfo {author}
  {\bibfnamefont {D.}~\bibnamefont {Seidel}}, \ and\ \bibinfo {author}
  {\bibfnamefont {L.}~\bibnamefont {Maleki}},\ }\href {\doibase
  10.1038/ncomms8371} {\bibfield  {journal} {\bibinfo  {journal} {Nature
  Communications}\ }\textbf {\bibinfo {volume} {6}},\ \bibinfo {pages} {7371}
  (\bibinfo {year} {2015}{\natexlab{b}})}\BibitemShut {NoStop}%
\bibitem [{\citenamefont {Kondratiev}\ \emph {et~al.}(2017)\citenamefont
  {Kondratiev}, \citenamefont {Lobanov}, \citenamefont {Cherenkov},
  \citenamefont {Voloshin}, \citenamefont {Pavlov}, \citenamefont {Koptyaev},\
  and\ \citenamefont {Gorodetsky}}]{Kondratiev:17}%
  \BibitemOpen
  \bibfield  {author} {\bibinfo {author} {\bibfnamefont {N.~M.}\ \bibnamefont
  {Kondratiev}}, \bibinfo {author} {\bibfnamefont {V.~E.}\ \bibnamefont
  {Lobanov}}, \bibinfo {author} {\bibfnamefont {A.~V.}\ \bibnamefont
  {Cherenkov}}, \bibinfo {author} {\bibfnamefont {A.~S.}\ \bibnamefont
  {Voloshin}}, \bibinfo {author} {\bibfnamefont {N.~G.}\ \bibnamefont
  {Pavlov}}, \bibinfo {author} {\bibfnamefont {S.}~\bibnamefont {Koptyaev}}, \
  and\ \bibinfo {author} {\bibfnamefont {M.~L.}\ \bibnamefont {Gorodetsky}},\
  }\href {\doibase 10.1364/OE.25.028167} {\bibfield  {journal} {\bibinfo
  {journal} {Opt. Express}\ }\textbf {\bibinfo {volume} {25}},\ \bibinfo
  {pages} {28167} (\bibinfo {year} {2017})}\BibitemShut {NoStop}%
\bibitem [{\citenamefont {Galiev}\ \emph {et~al.}(2020)\citenamefont {Galiev},
  \citenamefont {Kondratiev}, \citenamefont {Lobanov}, \citenamefont {Matsko},\
  and\ \citenamefont {Bilenko}}]{Galiev2020}%
  \BibitemOpen
  \bibfield  {author} {\bibinfo {author} {\bibfnamefont {R.~R.}\ \bibnamefont
  {Galiev}}, \bibinfo {author} {\bibfnamefont {N.~M.}\ \bibnamefont
  {Kondratiev}}, \bibinfo {author} {\bibfnamefont {V.~E.}\ \bibnamefont
  {Lobanov}}, \bibinfo {author} {\bibfnamefont {A.~B.}\ \bibnamefont {Matsko}},
  \ and\ \bibinfo {author} {\bibfnamefont {I.~A.}\ \bibnamefont {Bilenko}},\
  }\href {\doibase 10.1103/PhysRevApplied.14.014036} {\bibfield  {journal}
  {\bibinfo  {journal} {Phys. Rev. Applied}\ }\textbf {\bibinfo {volume}
  {14}},\ \bibinfo {pages} {014036} (\bibinfo {year} {2020})}\BibitemShut
  {NoStop}%
\bibitem [{\citenamefont {Voloshin}\ \emph {et~al.}(2021)\citenamefont
  {Voloshin}, \citenamefont {Kondratiev}, \citenamefont {Lihachev},
  \citenamefont {Liu}, \citenamefont {Lobanov}, \citenamefont {Dmitriev},
  \citenamefont {Weng}, \citenamefont {Kippenberg},\ and\ \citenamefont
  {Bilenko}}]{Voloshin2021}%
  \BibitemOpen
  \bibfield  {author} {\bibinfo {author} {\bibfnamefont {A.~S.}\ \bibnamefont
  {Voloshin}}, \bibinfo {author} {\bibfnamefont {N.~M.}\ \bibnamefont
  {Kondratiev}}, \bibinfo {author} {\bibfnamefont {G.~V.}\ \bibnamefont
  {Lihachev}}, \bibinfo {author} {\bibfnamefont {J.}~\bibnamefont {Liu}},
  \bibinfo {author} {\bibfnamefont {V.~E.}\ \bibnamefont {Lobanov}}, \bibinfo
  {author} {\bibfnamefont {N.~Y.}\ \bibnamefont {Dmitriev}}, \bibinfo {author}
  {\bibfnamefont {W.}~\bibnamefont {Weng}}, \bibinfo {author} {\bibfnamefont
  {T.~J.}\ \bibnamefont {Kippenberg}}, \ and\ \bibinfo {author} {\bibfnamefont
  {I.~A.}\ \bibnamefont {Bilenko}},\ }\href {\doibase
  10.1038/s41467-020-20196-y} {\bibfield  {journal} {\bibinfo  {journal}
  {Nature Communications}\ }\textbf {\bibinfo {volume} {12}},\ \bibinfo {pages}
  {235} (\bibinfo {year} {2021})}\BibitemShut {NoStop}%
\bibitem [{\citenamefont {Shen}\ \emph {et~al.}(2020)\citenamefont {Shen},
  \citenamefont {Chang}, \citenamefont {Liu}, \citenamefont {Wang},
  \citenamefont {Yang}, \citenamefont {Xiang}, \citenamefont {Wang},
  \citenamefont {He}, \citenamefont {Liu}, \citenamefont {Xie}, \citenamefont
  {Guo}, \citenamefont {Kinghorn}, \citenamefont {Wu}, \citenamefont {Ji},
  \citenamefont {Kippenberg}, \citenamefont {Vahala},\ and\ \citenamefont
  {Bowers}}]{shen2019integrated}%
  \BibitemOpen
  \bibfield  {author} {\bibinfo {author} {\bibfnamefont {B.}~\bibnamefont
  {Shen}}, \bibinfo {author} {\bibfnamefont {L.}~\bibnamefont {Chang}},
  \bibinfo {author} {\bibfnamefont {J.}~\bibnamefont {Liu}}, \bibinfo {author}
  {\bibfnamefont {H.}~\bibnamefont {Wang}}, \bibinfo {author} {\bibfnamefont
  {Q.-F.}\ \bibnamefont {Yang}}, \bibinfo {author} {\bibfnamefont
  {C.}~\bibnamefont {Xiang}}, \bibinfo {author} {\bibfnamefont {R.~N.}\
  \bibnamefont {Wang}}, \bibinfo {author} {\bibfnamefont {J.}~\bibnamefont
  {He}}, \bibinfo {author} {\bibfnamefont {T.}~\bibnamefont {Liu}}, \bibinfo
  {author} {\bibfnamefont {W.}~\bibnamefont {Xie}}, \bibinfo {author}
  {\bibfnamefont {J.}~\bibnamefont {Guo}}, \bibinfo {author} {\bibfnamefont
  {D.}~\bibnamefont {Kinghorn}}, \bibinfo {author} {\bibfnamefont
  {L.}~\bibnamefont {Wu}}, \bibinfo {author} {\bibfnamefont {Q.-X.}\
  \bibnamefont {Ji}}, \bibinfo {author} {\bibfnamefont {T.~J.}\ \bibnamefont
  {Kippenberg}}, \bibinfo {author} {\bibfnamefont {K.}~\bibnamefont {Vahala}},
  \ and\ \bibinfo {author} {\bibfnamefont {J.~E.}\ \bibnamefont {Bowers}},\
  }\href {\doibase 10.1038/s41586-020-2358-x} {\bibfield  {journal} {\bibinfo
  {journal} {Nature}\ }\textbf {\bibinfo {volume} {582}},\ \bibinfo {pages}
  {365} (\bibinfo {year} {2020})}\BibitemShut {NoStop}%
\bibitem [{\citenamefont {Raja}\ \emph {et~al.}(2019)\citenamefont {Raja},
  \citenamefont {Voloshin}, \citenamefont {Guo}, \citenamefont {Agafonova},
  \citenamefont {Liu}, \citenamefont {Gorodnitskiy}, \citenamefont {Karpov},
  \citenamefont {Pavlov}, \citenamefont {Lucas}, \citenamefont {Galiev},
  \citenamefont {Shitikov}, \citenamefont {Jost}, \citenamefont {Gorodetsky},\
  and\ \citenamefont {Kippenberg}}]{raja2019electrically}%
  \BibitemOpen
  \bibfield  {author} {\bibinfo {author} {\bibfnamefont {A.~S.}\ \bibnamefont
  {Raja}}, \bibinfo {author} {\bibfnamefont {A.~S.}\ \bibnamefont {Voloshin}},
  \bibinfo {author} {\bibfnamefont {H.}~\bibnamefont {Guo}}, \bibinfo {author}
  {\bibfnamefont {S.~E.}\ \bibnamefont {Agafonova}}, \bibinfo {author}
  {\bibfnamefont {J.}~\bibnamefont {Liu}}, \bibinfo {author} {\bibfnamefont
  {A.~S.}\ \bibnamefont {Gorodnitskiy}}, \bibinfo {author} {\bibfnamefont
  {M.}~\bibnamefont {Karpov}}, \bibinfo {author} {\bibfnamefont {N.~G.}\
  \bibnamefont {Pavlov}}, \bibinfo {author} {\bibfnamefont {E.}~\bibnamefont
  {Lucas}}, \bibinfo {author} {\bibfnamefont {R.~R.}\ \bibnamefont {Galiev}},
  \bibinfo {author} {\bibfnamefont {A.~E.}\ \bibnamefont {Shitikov}}, \bibinfo
  {author} {\bibfnamefont {J.~D.}\ \bibnamefont {Jost}}, \bibinfo {author}
  {\bibfnamefont {M.~L.}\ \bibnamefont {Gorodetsky}}, \ and\ \bibinfo {author}
  {\bibfnamefont {T.~J.}\ \bibnamefont {Kippenberg}},\ }\href {\doibase
  10.1038/s41467-019-08498-2} {\bibfield  {journal} {\bibinfo  {journal}
  {Nature Communications}\ }\textbf {\bibinfo {volume} {10}},\ \bibinfo {pages}
  {680} (\bibinfo {year} {2019})}\BibitemShut {NoStop}%
\bibitem [{\citenamefont {Stern}\ \emph {et~al.}(2018)\citenamefont {Stern},
  \citenamefont {Ji}, \citenamefont {Okawachi}, \citenamefont {Gaeta},\ and\
  \citenamefont {Lipson}}]{stern2018battery}%
  \BibitemOpen
  \bibfield  {author} {\bibinfo {author} {\bibfnamefont {B.}~\bibnamefont
  {Stern}}, \bibinfo {author} {\bibfnamefont {X.}~\bibnamefont {Ji}}, \bibinfo
  {author} {\bibfnamefont {Y.}~\bibnamefont {Okawachi}}, \bibinfo {author}
  {\bibfnamefont {A.~L.}\ \bibnamefont {Gaeta}}, \ and\ \bibinfo {author}
  {\bibfnamefont {M.}~\bibnamefont {Lipson}},\ }\href {\doibase
  10.1038/s41586-018-0598-9} {\bibfield  {journal} {\bibinfo  {journal}
  {Nature}\ }\textbf {\bibinfo {volume} {562}},\ \bibinfo {pages} {401}
  (\bibinfo {year} {2018})}\BibitemShut {NoStop}%
\bibitem [{\citenamefont {Kippenberg}\ \emph {et~al.}(2011)\citenamefont
  {Kippenberg}, \citenamefont {Holzwarth},\ and\ \citenamefont
  {Diddams}}]{Kippenberg:2011}%
  \BibitemOpen
  \bibfield  {author} {\bibinfo {author} {\bibfnamefont {T.~J.}\ \bibnamefont
  {Kippenberg}}, \bibinfo {author} {\bibfnamefont {R.}~\bibnamefont
  {Holzwarth}}, \ and\ \bibinfo {author} {\bibfnamefont {S.~A.}\ \bibnamefont
  {Diddams}},\ }\href {\doibase 10.1126/science.1193968} {\bibfield  {journal}
  {\bibinfo  {journal} {Science}\ }\textbf {\bibinfo {volume} {332}},\ \bibinfo
  {pages} {555} (\bibinfo {year} {2011})}\BibitemShut {NoStop}%
\bibitem [{\citenamefont {Lin}\ \emph {et~al.}(2017)\citenamefont {Lin},
  \citenamefont {Coillet},\ and\ \citenamefont {Chembo}}]{Lin:17}%
  \BibitemOpen
  \bibfield  {author} {\bibinfo {author} {\bibfnamefont {G.}~\bibnamefont
  {Lin}}, \bibinfo {author} {\bibfnamefont {A.}~\bibnamefont {Coillet}}, \ and\
  \bibinfo {author} {\bibfnamefont {Y.~K.}\ \bibnamefont {Chembo}},\ }\href
  {\doibase 10.1364/AOP.9.000828} {\bibfield  {journal} {\bibinfo  {journal}
  {Adv. Opt. Photon.}\ }\textbf {\bibinfo {volume} {9}},\ \bibinfo {pages}
  {828} (\bibinfo {year} {2017})}\BibitemShut {NoStop}%
\bibitem [{\citenamefont {Pasquazi}\ \emph {et~al.}(2018)\citenamefont
  {Pasquazi}, \citenamefont {Peccianti}, \citenamefont {Razzari}, \citenamefont
  {Moss}, \citenamefont {Coen}, \citenamefont {Erkintalo}, \citenamefont
  {Chembo}, \citenamefont {Hansson}, \citenamefont {Wabnitz}, \citenamefont
  {Del’Haye}, \citenamefont {Xue}, \citenamefont {Weiner},\ and\
  \citenamefont {Morandotti}}]{PASQUAZI20181}%
  \BibitemOpen
  \bibfield  {author} {\bibinfo {author} {\bibfnamefont {A.}~\bibnamefont
  {Pasquazi}}, \bibinfo {author} {\bibfnamefont {M.}~\bibnamefont {Peccianti}},
  \bibinfo {author} {\bibfnamefont {L.}~\bibnamefont {Razzari}}, \bibinfo
  {author} {\bibfnamefont {D.~J.}\ \bibnamefont {Moss}}, \bibinfo {author}
  {\bibfnamefont {S.}~\bibnamefont {Coen}}, \bibinfo {author} {\bibfnamefont
  {M.}~\bibnamefont {Erkintalo}}, \bibinfo {author} {\bibfnamefont {Y.~K.}\
  \bibnamefont {Chembo}}, \bibinfo {author} {\bibfnamefont {T.}~\bibnamefont
  {Hansson}}, \bibinfo {author} {\bibfnamefont {S.}~\bibnamefont {Wabnitz}},
  \bibinfo {author} {\bibfnamefont {P.}~\bibnamefont {Del’Haye}}, \bibinfo
  {author} {\bibfnamefont {X.}~\bibnamefont {Xue}}, \bibinfo {author}
  {\bibfnamefont {A.~M.}\ \bibnamefont {Weiner}}, \ and\ \bibinfo {author}
  {\bibfnamefont {R.}~\bibnamefont {Morandotti}},\ }\href {\doibase
  https://doi.org/10.1016/j.physrep.2017.08.004} {\bibfield  {journal}
  {\bibinfo  {journal} {Physics Reports}\ }\textbf {\bibinfo {volume} {729}},\
  \bibinfo {pages} {1 } (\bibinfo {year} {2018})}\BibitemShut {NoStop}%
\bibitem [{\citenamefont {Gaeta}\ \emph {et~al.}(2019)\citenamefont {Gaeta},
  \citenamefont {Lipson},\ and\ \citenamefont {Kippenberg}}]{Gaeta2019}%
  \BibitemOpen
  \bibfield  {author} {\bibinfo {author} {\bibfnamefont {A.~L.}\ \bibnamefont
  {Gaeta}}, \bibinfo {author} {\bibfnamefont {M.}~\bibnamefont {Lipson}}, \
  and\ \bibinfo {author} {\bibfnamefont {T.~J.}\ \bibnamefont {Kippenberg}},\
  }\href {\doibase 10.1038/s41566-019-0358-x} {\bibfield  {journal} {\bibinfo
  {journal} {Nature Photonics}\ }\textbf {\bibinfo {volume} {13}},\ \bibinfo
  {pages} {158} (\bibinfo {year} {2019})}\BibitemShut {NoStop}%
\bibitem [{\citenamefont {Kovach}\ \emph {et~al.}(2020)\citenamefont {Kovach},
  \citenamefont {Chen}, \citenamefont {He}, \citenamefont {Choi}, \citenamefont
  {Dogan}, \citenamefont {Ghasemkhani}, \citenamefont {Taheri},\ and\
  \citenamefont {Armani}}]{Kovach:20}%
  \BibitemOpen
  \bibfield  {author} {\bibinfo {author} {\bibfnamefont {A.}~\bibnamefont
  {Kovach}}, \bibinfo {author} {\bibfnamefont {D.}~\bibnamefont {Chen}},
  \bibinfo {author} {\bibfnamefont {J.}~\bibnamefont {He}}, \bibinfo {author}
  {\bibfnamefont {H.}~\bibnamefont {Choi}}, \bibinfo {author} {\bibfnamefont
  {A.~H.}\ \bibnamefont {Dogan}}, \bibinfo {author} {\bibfnamefont
  {M.}~\bibnamefont {Ghasemkhani}}, \bibinfo {author} {\bibfnamefont
  {H.}~\bibnamefont {Taheri}}, \ and\ \bibinfo {author} {\bibfnamefont {A.~M.}\
  \bibnamefont {Armani}},\ }\href {\doibase 10.1364/AOP.376924} {\bibfield
  {journal} {\bibinfo  {journal} {Adv. Opt. Photon.}\ }\textbf {\bibinfo
  {volume} {12}},\ \bibinfo {pages} {135} (\bibinfo {year} {2020})}\BibitemShut
  {NoStop}%
\bibitem [{\citenamefont {Lin}\ \emph {et~al.}(2020)\citenamefont {Lin},
  \citenamefont {Dutt}, \citenamefont {Joshi}, \citenamefont {Ji},
  \citenamefont {Phare}, \citenamefont {Okawachi}, \citenamefont {Gaeta},\ and\
  \citenamefont {Lipson}}]{lin2020broadband}%
  \BibitemOpen
  \bibfield  {author} {\bibinfo {author} {\bibfnamefont {T.}~\bibnamefont
  {Lin}}, \bibinfo {author} {\bibfnamefont {A.}~\bibnamefont {Dutt}}, \bibinfo
  {author} {\bibfnamefont {C.}~\bibnamefont {Joshi}}, \bibinfo {author}
  {\bibfnamefont {X.}~\bibnamefont {Ji}}, \bibinfo {author} {\bibfnamefont
  {C.~T.}\ \bibnamefont {Phare}}, \bibinfo {author} {\bibfnamefont
  {Y.}~\bibnamefont {Okawachi}}, \bibinfo {author} {\bibfnamefont {A.~L.}\
  \bibnamefont {Gaeta}}, \ and\ \bibinfo {author} {\bibfnamefont
  {M.}~\bibnamefont {Lipson}},\ }\href {https://arxiv.org/abs/2001.00869}
  {\bibfield  {journal} {\bibinfo  {journal} {arXiv preprint arXiv:2001.00869}\
  } (\bibinfo {year} {2020})}\BibitemShut {NoStop}%
\bibitem [{\citenamefont {Bao}\ \emph {et~al.}(2019)\citenamefont {Bao},
  \citenamefont {Suh},\ and\ \citenamefont {Vahala}}]{Bao:19}%
  \BibitemOpen
  \bibfield  {author} {\bibinfo {author} {\bibfnamefont {C.}~\bibnamefont
  {Bao}}, \bibinfo {author} {\bibfnamefont {M.-G.}\ \bibnamefont {Suh}}, \ and\
  \bibinfo {author} {\bibfnamefont {K.}~\bibnamefont {Vahala}},\ }\href
  {\doibase 10.1364/OPTICA.6.001110} {\bibfield  {journal} {\bibinfo  {journal}
  {Optica}\ }\textbf {\bibinfo {volume} {6}},\ \bibinfo {pages} {1110}
  (\bibinfo {year} {2019})}\BibitemShut {NoStop}%
\bibitem [{\citenamefont {Dutt}\ \emph {et~al.}(2018)\citenamefont {Dutt},
  \citenamefont {Joshi}, \citenamefont {Ji}, \citenamefont {Cardenas},
  \citenamefont {Okawachi}, \citenamefont {Luke}, \citenamefont {Gaeta},\ and\
  \citenamefont {Lipson}}]{Dutt:2018}%
  \BibitemOpen
  \bibfield  {author} {\bibinfo {author} {\bibfnamefont {A.}~\bibnamefont
  {Dutt}}, \bibinfo {author} {\bibfnamefont {C.}~\bibnamefont {Joshi}},
  \bibinfo {author} {\bibfnamefont {X.}~\bibnamefont {Ji}}, \bibinfo {author}
  {\bibfnamefont {J.}~\bibnamefont {Cardenas}}, \bibinfo {author}
  {\bibfnamefont {Y.}~\bibnamefont {Okawachi}}, \bibinfo {author}
  {\bibfnamefont {K.}~\bibnamefont {Luke}}, \bibinfo {author} {\bibfnamefont
  {A.~L.}\ \bibnamefont {Gaeta}}, \ and\ \bibinfo {author} {\bibfnamefont
  {M.}~\bibnamefont {Lipson}},\ }\href {\doibase 10.1126/sciadv.1701858}
  {\bibfield  {journal} {\bibinfo  {journal} {Science Advances}\ }\textbf
  {\bibinfo {volume} {4}},\ \bibinfo {pages} {e1701858} (\bibinfo {year}
  {2018})}\BibitemShut {NoStop}%
\bibitem [{\citenamefont {Wang}\ \emph {et~al.}(2018)\citenamefont {Wang},
  \citenamefont {Zhang}, \citenamefont {Lu}, \citenamefont {Chu}, \citenamefont
  {Little}, \citenamefont {Yang}, \citenamefont {Wang},\ and\ \citenamefont
  {Zhao}}]{Wang:18}%
  \BibitemOpen
  \bibfield  {author} {\bibinfo {author} {\bibfnamefont {W.}~\bibnamefont
  {Wang}}, \bibinfo {author} {\bibfnamefont {W.}~\bibnamefont {Zhang}},
  \bibinfo {author} {\bibfnamefont {Z.}~\bibnamefont {Lu}}, \bibinfo {author}
  {\bibfnamefont {S.~T.}\ \bibnamefont {Chu}}, \bibinfo {author} {\bibfnamefont
  {B.~E.}\ \bibnamefont {Little}}, \bibinfo {author} {\bibfnamefont
  {Q.}~\bibnamefont {Yang}}, \bibinfo {author} {\bibfnamefont {L.}~\bibnamefont
  {Wang}}, \ and\ \bibinfo {author} {\bibfnamefont {W.}~\bibnamefont {Zhao}},\
  }\href {\doibase 10.1364/PRJ.6.000363} {\bibfield  {journal} {\bibinfo
  {journal} {Photon. Res.}\ }\textbf {\bibinfo {volume} {6}},\ \bibinfo {pages}
  {363} (\bibinfo {year} {2018})}\BibitemShut {NoStop}%
\bibitem [{\citenamefont {Yu}\ \emph {et~al.}(2018)\citenamefont {Yu},
  \citenamefont {Okawachi}, \citenamefont {Griffith}, \citenamefont
  {Picqu{\'e}}, \citenamefont {Lipson},\ and\ \citenamefont
  {Gaeta}}]{yu2018silicon}%
  \BibitemOpen
  \bibfield  {author} {\bibinfo {author} {\bibfnamefont {M.}~\bibnamefont
  {Yu}}, \bibinfo {author} {\bibfnamefont {Y.}~\bibnamefont {Okawachi}},
  \bibinfo {author} {\bibfnamefont {A.~G.}\ \bibnamefont {Griffith}}, \bibinfo
  {author} {\bibfnamefont {N.}~\bibnamefont {Picqu{\'e}}}, \bibinfo {author}
  {\bibfnamefont {M.}~\bibnamefont {Lipson}}, \ and\ \bibinfo {author}
  {\bibfnamefont {A.~L.}\ \bibnamefont {Gaeta}},\ }\href
  {https://www.nature.com/articles/s41467-018-04350-1} {\bibfield  {journal}
  {\bibinfo  {journal} {Nature communications}\ }\textbf {\bibinfo {volume}
  {9}},\ \bibinfo {pages} {1} (\bibinfo {year} {2018})}\BibitemShut {NoStop}%
\bibitem [{\citenamefont {Pavlov}\ \emph {et~al.}(2017)\citenamefont {Pavlov},
  \citenamefont {Lihachev}, \citenamefont {Koptyaev}, \citenamefont {Lucas},
  \citenamefont {Karpov}, \citenamefont {Kondratiev}, \citenamefont {Bilenko},
  \citenamefont {Kippenberg},\ and\ \citenamefont
  {Gorodetsky}}]{pavlov2017soliton}%
  \BibitemOpen
  \bibfield  {author} {\bibinfo {author} {\bibfnamefont {N.}~\bibnamefont
  {Pavlov}}, \bibinfo {author} {\bibfnamefont {G.}~\bibnamefont {Lihachev}},
  \bibinfo {author} {\bibfnamefont {S.}~\bibnamefont {Koptyaev}}, \bibinfo
  {author} {\bibfnamefont {E.}~\bibnamefont {Lucas}}, \bibinfo {author}
  {\bibfnamefont {M.}~\bibnamefont {Karpov}}, \bibinfo {author} {\bibfnamefont
  {N.}~\bibnamefont {Kondratiev}}, \bibinfo {author} {\bibfnamefont
  {I.}~\bibnamefont {Bilenko}}, \bibinfo {author} {\bibfnamefont
  {T.}~\bibnamefont {Kippenberg}}, \ and\ \bibinfo {author} {\bibfnamefont
  {M.}~\bibnamefont {Gorodetsky}},\ }\href
  {https://www.osapublishing.org/ol/abstract.cfm?uri=ol-42-3-514} {\bibfield
  {journal} {\bibinfo  {journal} {Optics letters}\ }\textbf {\bibinfo {volume}
  {42}},\ \bibinfo {pages} {514} (\bibinfo {year} {2017})}\BibitemShut
  {NoStop}%
\bibitem [{\citenamefont {Jin}\ \emph {et~al.}(2021)\citenamefont {Jin},
  \citenamefont {Yang}, \citenamefont {Chang}, \citenamefont {Shen},
  \citenamefont {Wang}, \citenamefont {Leal}, \citenamefont {Wu}, \citenamefont
  {Gao}, \citenamefont {Feshali}, \citenamefont {Paniccia}, \citenamefont
  {Vahala},\ and\ \citenamefont {Bowers}}]{Jin2021}%
  \BibitemOpen
  \bibfield  {author} {\bibinfo {author} {\bibfnamefont {W.}~\bibnamefont
  {Jin}}, \bibinfo {author} {\bibfnamefont {Q.-F.}\ \bibnamefont {Yang}},
  \bibinfo {author} {\bibfnamefont {L.}~\bibnamefont {Chang}}, \bibinfo
  {author} {\bibfnamefont {B.}~\bibnamefont {Shen}}, \bibinfo {author}
  {\bibfnamefont {H.}~\bibnamefont {Wang}}, \bibinfo {author} {\bibfnamefont
  {M.~A.}\ \bibnamefont {Leal}}, \bibinfo {author} {\bibfnamefont
  {L.}~\bibnamefont {Wu}}, \bibinfo {author} {\bibfnamefont {M.}~\bibnamefont
  {Gao}}, \bibinfo {author} {\bibfnamefont {A.}~\bibnamefont {Feshali}},
  \bibinfo {author} {\bibfnamefont {M.}~\bibnamefont {Paniccia}}, \bibinfo
  {author} {\bibfnamefont {K.~J.}\ \bibnamefont {Vahala}}, \ and\ \bibinfo
  {author} {\bibfnamefont {J.~E.}\ \bibnamefont {Bowers}},\ }\href {\doibase
  10.1038/s41566-021-00761-7} {\bibfield  {journal} {\bibinfo  {journal}
  {Nature Photonics}\ }\textbf {\bibinfo {volume} {15}},\ \bibinfo {pages}
  {346} (\bibinfo {year} {2021})}\BibitemShut {NoStop}%
\bibitem [{\citenamefont {Herr}\ \emph
  {et~al.}(2014{\natexlab{a}})\citenamefont {Herr}, \citenamefont {Brasch},
  \citenamefont {Jost}, \citenamefont {Wang}, \citenamefont {Kondratiev},
  \citenamefont {Gorodetsky},\ and\ \citenamefont
  {Kippenberg}}]{herr2014temporal}%
  \BibitemOpen
  \bibfield  {author} {\bibinfo {author} {\bibfnamefont {T.}~\bibnamefont
  {Herr}}, \bibinfo {author} {\bibfnamefont {V.}~\bibnamefont {Brasch}},
  \bibinfo {author} {\bibfnamefont {J.~D.}\ \bibnamefont {Jost}}, \bibinfo
  {author} {\bibfnamefont {C.~Y.}\ \bibnamefont {Wang}}, \bibinfo {author}
  {\bibfnamefont {N.~M.}\ \bibnamefont {Kondratiev}}, \bibinfo {author}
  {\bibfnamefont {M.~L.}\ \bibnamefont {Gorodetsky}}, \ and\ \bibinfo {author}
  {\bibfnamefont {T.~J.}\ \bibnamefont {Kippenberg}},\ }\href {\doibase
  10.1038/nphoton.2013.343} {\bibfield  {journal} {\bibinfo  {journal} {Nat.
  Photon.}\ }\textbf {\bibinfo {volume} {8}},\ \bibinfo {pages} {145} (\bibinfo
  {year} {2014}{\natexlab{a}})}\BibitemShut {NoStop}%
\bibitem [{\citenamefont {Herr}\ \emph
  {et~al.}(2014{\natexlab{b}})\citenamefont {Herr}, \citenamefont {Brasch},
  \citenamefont {Jost}, \citenamefont {Mirgorodskiy}, \citenamefont {Lihachev},
  \citenamefont {Gorodetsky},\ and\ \citenamefont
  {Kippenberg}}]{PhysRevLett.113.123901}%
  \BibitemOpen
  \bibfield  {author} {\bibinfo {author} {\bibfnamefont {T.}~\bibnamefont
  {Herr}}, \bibinfo {author} {\bibfnamefont {V.}~\bibnamefont {Brasch}},
  \bibinfo {author} {\bibfnamefont {J.~D.}\ \bibnamefont {Jost}}, \bibinfo
  {author} {\bibfnamefont {I.}~\bibnamefont {Mirgorodskiy}}, \bibinfo {author}
  {\bibfnamefont {G.}~\bibnamefont {Lihachev}}, \bibinfo {author}
  {\bibfnamefont {M.~L.}\ \bibnamefont {Gorodetsky}}, \ and\ \bibinfo {author}
  {\bibfnamefont {T.~J.}\ \bibnamefont {Kippenberg}},\ }\href {\doibase
  10.1103/PhysRevLett.113.123901} {\bibfield  {journal} {\bibinfo  {journal}
  {Phys. Rev. Lett.}\ }\textbf {\bibinfo {volume} {113}},\ \bibinfo {pages}
  {123901} (\bibinfo {year} {2014}{\natexlab{b}})}\BibitemShut {NoStop}%
\bibitem [{\citenamefont {Kim}\ \emph {et~al.}(2021)\citenamefont {Kim},
  \citenamefont {Yvind},\ and\ \citenamefont {Pu}}]{Kim:21}%
  \BibitemOpen
  \bibfield  {author} {\bibinfo {author} {\bibfnamefont {C.}~\bibnamefont
  {Kim}}, \bibinfo {author} {\bibfnamefont {K.}~\bibnamefont {Yvind}}, \ and\
  \bibinfo {author} {\bibfnamefont {M.}~\bibnamefont {Pu}},\ }\href {\doibase
  10.1364/OL.431667} {\bibfield  {journal} {\bibinfo  {journal} {Opt. Lett.}\
  }\textbf {\bibinfo {volume} {46}},\ \bibinfo {pages} {3508} (\bibinfo {year}
  {2021})}\BibitemShut {NoStop}%
\bibitem [{\citenamefont {Li}\ \emph {et~al.}(2017)\citenamefont {Li},
  \citenamefont {Briles}, \citenamefont {Westly}, \citenamefont {Drake},
  \citenamefont {Stone}, \citenamefont {Ilic}, \citenamefont {Diddams},
  \citenamefont {Papp},\ and\ \citenamefont {Srinivasan}}]{briles2017}%
  \BibitemOpen
  \bibfield  {author} {\bibinfo {author} {\bibfnamefont {Q.}~\bibnamefont
  {Li}}, \bibinfo {author} {\bibfnamefont {T.~C.}\ \bibnamefont {Briles}},
  \bibinfo {author} {\bibfnamefont {D.~A.}\ \bibnamefont {Westly}}, \bibinfo
  {author} {\bibfnamefont {T.~E.}\ \bibnamefont {Drake}}, \bibinfo {author}
  {\bibfnamefont {J.~R.}\ \bibnamefont {Stone}}, \bibinfo {author}
  {\bibfnamefont {B.~R.}\ \bibnamefont {Ilic}}, \bibinfo {author}
  {\bibfnamefont {S.~A.}\ \bibnamefont {Diddams}}, \bibinfo {author}
  {\bibfnamefont {S.~B.}\ \bibnamefont {Papp}}, \ and\ \bibinfo {author}
  {\bibfnamefont {K.}~\bibnamefont {Srinivasan}},\ }\href {\doibase
  10.1364/optica.4.000193} {\ \textbf {\bibinfo {volume} {4}},\ \bibinfo
  {pages} {193} (\bibinfo {year} {2017})}\BibitemShut {NoStop}%
\bibitem [{\citenamefont {Joshi}\ \emph {et~al.}(2016)\citenamefont {Joshi},
  \citenamefont {Jang}, \citenamefont {Luke}, \citenamefont {Ji}, \citenamefont
  {Miller}, \citenamefont {Klenner}, \citenamefont {Okawachi}, \citenamefont
  {Lipson},\ and\ \citenamefont {Gaeta}}]{Joshi:16}%
  \BibitemOpen
  \bibfield  {author} {\bibinfo {author} {\bibfnamefont {C.}~\bibnamefont
  {Joshi}}, \bibinfo {author} {\bibfnamefont {J.~K.}\ \bibnamefont {Jang}},
  \bibinfo {author} {\bibfnamefont {K.}~\bibnamefont {Luke}}, \bibinfo {author}
  {\bibfnamefont {X.}~\bibnamefont {Ji}}, \bibinfo {author} {\bibfnamefont
  {S.~A.}\ \bibnamefont {Miller}}, \bibinfo {author} {\bibfnamefont
  {A.}~\bibnamefont {Klenner}}, \bibinfo {author} {\bibfnamefont
  {Y.}~\bibnamefont {Okawachi}}, \bibinfo {author} {\bibfnamefont
  {M.}~\bibnamefont {Lipson}}, \ and\ \bibinfo {author} {\bibfnamefont {A.~L.}\
  \bibnamefont {Gaeta}},\ }\href {\doibase 10.1364/OL.41.002565} {\bibfield
  {journal} {\bibinfo  {journal} {Opt. Lett.}\ }\textbf {\bibinfo {volume}
  {41}},\ \bibinfo {pages} {2565} (\bibinfo {year} {2016})}\BibitemShut
  {NoStop}%
\bibitem [{\citenamefont {Kuse}\ \emph {et~al.}(2020)\citenamefont {Kuse},
  \citenamefont {Tetsumoto}, \citenamefont {Navickaite}, \citenamefont
  {Geiselmann},\ and\ \citenamefont {Fermann}}]{Kuse:20}%
  \BibitemOpen
  \bibfield  {author} {\bibinfo {author} {\bibfnamefont {N.}~\bibnamefont
  {Kuse}}, \bibinfo {author} {\bibfnamefont {T.}~\bibnamefont {Tetsumoto}},
  \bibinfo {author} {\bibfnamefont {G.}~\bibnamefont {Navickaite}}, \bibinfo
  {author} {\bibfnamefont {M.}~\bibnamefont {Geiselmann}}, \ and\ \bibinfo
  {author} {\bibfnamefont {M.~E.}\ \bibnamefont {Fermann}},\ }\href {\doibase
  10.1364/OL.383036} {\bibfield  {journal} {\bibinfo  {journal} {Opt. Lett.}\
  }\textbf {\bibinfo {volume} {45}},\ \bibinfo {pages} {927} (\bibinfo {year}
  {2020})}\BibitemShut {NoStop}%
\bibitem [{\citenamefont {Kondratiev}\ \emph {et~al.}(2021)\citenamefont
  {Kondratiev}, \citenamefont {Galiev},\ and\ \citenamefont
  {Lobanov}}]{KondratievSPIE21}%
  \BibitemOpen
  \bibfield  {author} {\bibinfo {author} {\bibfnamefont {N.~M.}\ \bibnamefont
  {Kondratiev}}, \bibinfo {author} {\bibfnamefont {R.~R.}\ \bibnamefont
  {Galiev}}, \ and\ \bibinfo {author} {\bibfnamefont {V.~E.}\ \bibnamefont
  {Lobanov}},\ }in\ \href {\doibase 10.1117/12.2588931} {\emph {\bibinfo
  {booktitle} {Nonlinear Optics and Applications XII}}},\ Vol.\ \bibinfo
  {volume} {11770},\ \bibinfo {editor} {edited by\ \bibinfo {editor}
  {\bibfnamefont {M.}~\bibnamefont {Bertolotti}}, \bibinfo {editor}
  {\bibfnamefont {A.~V.}\ \bibnamefont {Zayats}}, \ and\ \bibinfo {editor}
  {\bibfnamefont {A.~M.}\ \bibnamefont {Zheltikov}}},\ \bibinfo {organization}
  {International Society for Optics and Photonics}\ (\bibinfo  {publisher}
  {SPIE},\ \bibinfo {year} {2021})\ pp.\ \bibinfo {pages} {53 --
  58}\BibitemShut {NoStop}%
\bibitem [{\citenamefont {Xue}\ \emph {et~al.}(2017)\citenamefont {Xue},
  \citenamefont {Wang}, \citenamefont {Xuan}, \citenamefont {Qi},\ and\
  \citenamefont {Weiner}}]{Xue:17}%
  \BibitemOpen
  \bibfield  {author} {\bibinfo {author} {\bibfnamefont {X.}~\bibnamefont
  {Xue}}, \bibinfo {author} {\bibfnamefont {P.-H.}\ \bibnamefont {Wang}},
  \bibinfo {author} {\bibfnamefont {Y.}~\bibnamefont {Xuan}}, \bibinfo {author}
  {\bibfnamefont {M.}~\bibnamefont {Qi}}, \ and\ \bibinfo {author}
  {\bibfnamefont {A.~M.}\ \bibnamefont {Weiner}},\ }\href {\doibase
  10.1002/lpor.201600276} {\bibfield  {journal} {\bibinfo  {journal} {Laser \&
  Photonics Reviews}\ }\textbf {\bibinfo {volume} {11}},\ \bibinfo {pages}
  {1600276} (\bibinfo {year} {2017})}\BibitemShut {NoStop}%
\bibitem [{\citenamefont {Kim}\ \emph {et~al.}(2019)\citenamefont {Kim},
  \citenamefont {Okawachi}, \citenamefont {Jang}, \citenamefont {Yu},
  \citenamefont {Ji}, \citenamefont {Zhao}, \citenamefont {Joshi},
  \citenamefont {Lipson},\ and\ \citenamefont {Gaeta}}]{Kim:19}%
  \BibitemOpen
  \bibfield  {author} {\bibinfo {author} {\bibfnamefont {B.~Y.}\ \bibnamefont
  {Kim}}, \bibinfo {author} {\bibfnamefont {Y.}~\bibnamefont {Okawachi}},
  \bibinfo {author} {\bibfnamefont {J.~K.}\ \bibnamefont {Jang}}, \bibinfo
  {author} {\bibfnamefont {M.}~\bibnamefont {Yu}}, \bibinfo {author}
  {\bibfnamefont {X.}~\bibnamefont {Ji}}, \bibinfo {author} {\bibfnamefont
  {Y.}~\bibnamefont {Zhao}}, \bibinfo {author} {\bibfnamefont {C.}~\bibnamefont
  {Joshi}}, \bibinfo {author} {\bibfnamefont {M.}~\bibnamefont {Lipson}}, \
  and\ \bibinfo {author} {\bibfnamefont {A.~L.}\ \bibnamefont {Gaeta}},\ }\href
  {\doibase 10.1364/OL.44.004475} {\bibfield  {journal} {\bibinfo  {journal}
  {Opt. Lett.}\ }\textbf {\bibinfo {volume} {44}},\ \bibinfo {pages} {4475}
  (\bibinfo {year} {2019})}\BibitemShut {NoStop}%
\bibitem [{\citenamefont {Jang}\ \emph {et~al.}(2021)\citenamefont {Jang},
  \citenamefont {Okawachi}, \citenamefont {Zhao}, \citenamefont {Ji},
  \citenamefont {Joshi}, \citenamefont {Lipson},\ and\ \citenamefont
  {Gaeta}}]{jang2021conversion}%
  \BibitemOpen
  \bibfield  {author} {\bibinfo {author} {\bibfnamefont {J.~K.}\ \bibnamefont
  {Jang}}, \bibinfo {author} {\bibfnamefont {Y.}~\bibnamefont {Okawachi}},
  \bibinfo {author} {\bibfnamefont {Y.}~\bibnamefont {Zhao}}, \bibinfo {author}
  {\bibfnamefont {X.}~\bibnamefont {Ji}}, \bibinfo {author} {\bibfnamefont
  {C.}~\bibnamefont {Joshi}}, \bibinfo {author} {\bibfnamefont
  {M.}~\bibnamefont {Lipson}}, \ and\ \bibinfo {author} {\bibfnamefont {A.~L.}\
  \bibnamefont {Gaeta}},\ }\href {\doibase 10.1364/OL.423654} {\bibfield
  {journal} {\bibinfo  {journal} {Opt. Lett.}\ }\textbf {\bibinfo {volume}
  {46}},\ \bibinfo {pages} {3657} (\bibinfo {year} {2021})}\BibitemShut
  {NoStop}%
\bibitem [{\citenamefont {Karpov}\ \emph {et~al.}(2019)\citenamefont {Karpov},
  \citenamefont {Pfeiffer}, \citenamefont {Guo}, \citenamefont {Weng},
  \citenamefont {Liu},\ and\ \citenamefont {Kippenberg}}]{Karpov2019}%
  \BibitemOpen
  \bibfield  {author} {\bibinfo {author} {\bibfnamefont {M.}~\bibnamefont
  {Karpov}}, \bibinfo {author} {\bibfnamefont {M.~H.~P.}\ \bibnamefont
  {Pfeiffer}}, \bibinfo {author} {\bibfnamefont {H.}~\bibnamefont {Guo}},
  \bibinfo {author} {\bibfnamefont {W.}~\bibnamefont {Weng}}, \bibinfo {author}
  {\bibfnamefont {J.}~\bibnamefont {Liu}}, \ and\ \bibinfo {author}
  {\bibfnamefont {T.~J.}\ \bibnamefont {Kippenberg}},\ }\href {\doibase
  10.1038/s41567-019-0635-0} {\bibfield  {journal} {\bibinfo  {journal} {Nature
  Physics}\ }\textbf {\bibinfo {volume} {15}},\ \bibinfo {pages} {1071}
  (\bibinfo {year} {2019})}\BibitemShut {NoStop}%
\bibitem [{\citenamefont {Gorodetsky}\ \emph {et~al.}(2000)\citenamefont
  {Gorodetsky}, \citenamefont {Pryamikov},\ and\ \citenamefont
  {Ilchenko}}]{Gorodetsky:00}%
  \BibitemOpen
  \bibfield  {author} {\bibinfo {author} {\bibfnamefont {M.~L.}\ \bibnamefont
  {Gorodetsky}}, \bibinfo {author} {\bibfnamefont {A.~D.}\ \bibnamefont
  {Pryamikov}}, \ and\ \bibinfo {author} {\bibfnamefont {V.~S.}\ \bibnamefont
  {Ilchenko}},\ }\href {\doibase 10.1364/JOSAB.17.001051} {\bibfield  {journal}
  {\bibinfo  {journal} {J. Opt. Soc. Am. B}\ }\textbf {\bibinfo {volume}
  {17}},\ \bibinfo {pages} {1051} (\bibinfo {year} {2000})}\BibitemShut
  {NoStop}%
\end{thebibliography}%


\begin{thebibliography}{14}%
\makeatletter
\providecommand \@ifxundefined [1]{%
 \@ifx{#1\undefined}
}%
\providecommand \@ifnum [1]{%
 \ifnum #1\expandafter \@firstoftwo
 \else \expandafter \@secondoftwo
 \fi
}%
\providecommand \@ifx [1]{%
 \ifx #1\expandafter \@firstoftwo
 \else \expandafter \@secondoftwo
 \fi
}%
\providecommand \natexlab [1]{#1}%
\providecommand \enquote  [1]{``#1''}%
\providecommand \bibnamefont  [1]{#1}%
\providecommand \bibfnamefont [1]{#1}%
\providecommand \citenamefont [1]{#1}%
\providecommand \href@noop [0]{\@secondoftwo}%
\providecommand \href [0]{\begingroup \@sanitize@url \@href}%
\providecommand \@href[1]{\@@startlink{#1}\@@href}%
\providecommand \@@href[1]{\endgroup#1\@@endlink}%
\providecommand \@sanitize@url [0]{\catcode `\\12\catcode `\$12\catcode
  `\&12\catcode `\#12\catcode `\^12\catcode `\_12\catcode `\%12\relax}%
\providecommand \@@startlink[1]{}%
\providecommand \@@endlink[0]{}%
\providecommand \url  [0]{\begingroup\@sanitize@url \@url }%
\providecommand \@url [1]{\endgroup\@href {#1}{\urlprefix }}%
\providecommand \urlprefix  [0]{URL }%
\providecommand \Eprint [0]{\href }%
\providecommand \doibase [0]{http://dx.doi.org/}%
\providecommand \selectlanguage [0]{\@gobble}%
\providecommand \bibinfo  [0]{\@secondoftwo}%
\providecommand \bibfield  [0]{\@secondoftwo}%
\providecommand \translation [1]{[#1]}%
\providecommand \BibitemOpen [0]{}%
\providecommand \bibitemStop [0]{}%
\providecommand \bibitemNoStop [0]{.\EOS\space}%
\providecommand \EOS [0]{\spacefactor3000\relax}%
\providecommand \BibitemShut  [1]{\csname bibitem#1\endcsname}%
\let\auto@bib@innerbib\@empty
\bibitem [{\citenamefont {Xue}\ \emph {et~al.}(2016)\citenamefont {Xue},
  \citenamefont {Xuan}, \citenamefont {Wang}, \citenamefont {Wang},
  \citenamefont {Liu}, \citenamefont {Niu}, \citenamefont {Leaird},
  \citenamefont {Qi},\ and\ \citenamefont {Weiner}}]{Xue:16}%
  \BibitemOpen
  \bibfield  {author} {\bibinfo {author} {\bibfnamefont {X.}~\bibnamefont
  {Xue}}, \bibinfo {author} {\bibfnamefont {Y.}~\bibnamefont {Xuan}}, \bibinfo
  {author} {\bibfnamefont {C.}~\bibnamefont {Wang}}, \bibinfo {author}
  {\bibfnamefont {P.-H.}\ \bibnamefont {Wang}}, \bibinfo {author}
  {\bibfnamefont {Y.}~\bibnamefont {Liu}}, \bibinfo {author} {\bibfnamefont
  {B.}~\bibnamefont {Niu}}, \bibinfo {author} {\bibfnamefont {D.~E.}\
  \bibnamefont {Leaird}}, \bibinfo {author} {\bibfnamefont {M.}~\bibnamefont
  {Qi}}, \ and\ \bibinfo {author} {\bibfnamefont {A.~M.}\ \bibnamefont
  {Weiner}},\ }\href {\doibase 10.1364/OE.24.000687} {\bibfield  {journal}
  {\bibinfo  {journal} {Opt. Express}\ }\textbf {\bibinfo {volume} {24}},\
  \bibinfo {pages} {687} (\bibinfo {year} {2016})}\BibitemShut {NoStop}%
\bibitem [{\citenamefont {Liu}\ \emph {et~al.}(2020)\citenamefont {Liu},
  \citenamefont {Tian}, \citenamefont {Lucas}, \citenamefont {Raja},
  \citenamefont {Lihachev}, \citenamefont {Wang}, \citenamefont {He},
  \citenamefont {Liu}, \citenamefont {Anderson}, \citenamefont {Weng},
  \citenamefont {Bhave},\ and\ \citenamefont {Kippenberg}}]{Liu2020}%
  \BibitemOpen
  \bibfield  {author} {\bibinfo {author} {\bibfnamefont {J.}~\bibnamefont
  {Liu}}, \bibinfo {author} {\bibfnamefont {H.}~\bibnamefont {Tian}}, \bibinfo
  {author} {\bibfnamefont {E.}~\bibnamefont {Lucas}}, \bibinfo {author}
  {\bibfnamefont {A.~S.}\ \bibnamefont {Raja}}, \bibinfo {author}
  {\bibfnamefont {G.}~\bibnamefont {Lihachev}}, \bibinfo {author}
  {\bibfnamefont {R.~N.}\ \bibnamefont {Wang}}, \bibinfo {author}
  {\bibfnamefont {J.}~\bibnamefont {He}}, \bibinfo {author} {\bibfnamefont
  {T.}~\bibnamefont {Liu}}, \bibinfo {author} {\bibfnamefont {M.~H.}\
  \bibnamefont {Anderson}}, \bibinfo {author} {\bibfnamefont {W.}~\bibnamefont
  {Weng}}, \bibinfo {author} {\bibfnamefont {S.~A.}\ \bibnamefont {Bhave}}, \
  and\ \bibinfo {author} {\bibfnamefont {T.~J.}\ \bibnamefont {Kippenberg}},\
  }\href {\doibase 10.1038/s41586-020-2465-8} {\bibfield  {journal} {\bibinfo
  {journal} {Nature}\ }\textbf {\bibinfo {volume} {583}},\ \bibinfo {pages}
  {385} (\bibinfo {year} {2020})}\BibitemShut {NoStop}%
\bibitem [{\citenamefont {Kondratiev}\ \emph {et~al.}(2021)\citenamefont
  {Kondratiev}, \citenamefont {Galiev},\ and\ \citenamefont
  {Lobanov}}]{KondratievSPIE21}%
  \BibitemOpen
  \bibfield  {author} {\bibinfo {author} {\bibfnamefont {N.~M.}\ \bibnamefont
  {Kondratiev}}, \bibinfo {author} {\bibfnamefont {R.~R.}\ \bibnamefont
  {Galiev}}, \ and\ \bibinfo {author} {\bibfnamefont {V.~E.}\ \bibnamefont
  {Lobanov}},\ }in\ \href {\doibase 10.1117/12.2588931} {\emph {\bibinfo
  {booktitle} {Nonlinear Optics and Applications XII}}},\ Vol.\ \bibinfo
  {volume} {11770},\ \bibinfo {editor} {edited by\ \bibinfo {editor}
  {\bibfnamefont {M.}~\bibnamefont {Bertolotti}}, \bibinfo {editor}
  {\bibfnamefont {A.~V.}\ \bibnamefont {Zayats}}, \ and\ \bibinfo {editor}
  {\bibfnamefont {A.~M.}\ \bibnamefont {Zheltikov}}},\ \bibinfo {organization}
  {International Society for Optics and Photonics}\ (\bibinfo  {publisher}
  {SPIE},\ \bibinfo {year} {2021})\ pp.\ \bibinfo {pages} {53 --
  58}\BibitemShut {NoStop}%
\bibitem [{\citenamefont {Kondratiev}\ \emph {et~al.}(2020)\citenamefont
  {Kondratiev}, \citenamefont {Lobanov}, \citenamefont {Lonshakov},
  \citenamefont {Dmitriev}, \citenamefont {Voloshin},\ and\ \citenamefont
  {Bilenko}}]{Kondratiev:20}%
  \BibitemOpen
  \bibfield  {author} {\bibinfo {author} {\bibfnamefont {N.~M.}\ \bibnamefont
  {Kondratiev}}, \bibinfo {author} {\bibfnamefont {V.~E.}\ \bibnamefont
  {Lobanov}}, \bibinfo {author} {\bibfnamefont {E.~A.}\ \bibnamefont
  {Lonshakov}}, \bibinfo {author} {\bibfnamefont {N.~Y.}\ \bibnamefont
  {Dmitriev}}, \bibinfo {author} {\bibfnamefont {A.~S.}\ \bibnamefont
  {Voloshin}}, \ and\ \bibinfo {author} {\bibfnamefont {I.~A.}\ \bibnamefont
  {Bilenko}},\ }\href {\doibase 10.1364/OE.411544} {\bibfield  {journal}
  {\bibinfo  {journal} {Opt. Express}\ }\textbf {\bibinfo {volume} {28}},\
  \bibinfo {pages} {38892} (\bibinfo {year} {2020})}\BibitemShut {NoStop}%
\bibitem [{\citenamefont {Guo}\ \emph {et~al.}(2016)\citenamefont {Guo},
  \citenamefont {Karpov}, \citenamefont {Lucas}, \citenamefont {Kordts},
  \citenamefont {Pfeiffer}, \citenamefont {Brasch}, \citenamefont {Lihachev},
  \citenamefont {Lobanov}, \citenamefont {Gorodetsky},\ and\ \citenamefont
  {Kippenberg}}]{Guo:16}%
  \BibitemOpen
  \bibfield  {author} {\bibinfo {author} {\bibfnamefont {H.}~\bibnamefont
  {Guo}}, \bibinfo {author} {\bibfnamefont {M.}~\bibnamefont {Karpov}},
  \bibinfo {author} {\bibfnamefont {E.}~\bibnamefont {Lucas}}, \bibinfo
  {author} {\bibfnamefont {A.}~\bibnamefont {Kordts}}, \bibinfo {author}
  {\bibfnamefont {M.~H.~P.}\ \bibnamefont {Pfeiffer}}, \bibinfo {author}
  {\bibfnamefont {V.}~\bibnamefont {Brasch}}, \bibinfo {author} {\bibfnamefont
  {G.}~\bibnamefont {Lihachev}}, \bibinfo {author} {\bibfnamefont {V.~E.}\
  \bibnamefont {Lobanov}}, \bibinfo {author} {\bibfnamefont {M.~L.}\
  \bibnamefont {Gorodetsky}}, \ and\ \bibinfo {author} {\bibfnamefont {T.~J.}\
  \bibnamefont {Kippenberg}},\ }\href {http://dx.doi.org/10.1038/nphys3893}
  {\bibfield  {journal} {\bibinfo  {journal} {Nature Physics}\ }\textbf
  {\bibinfo {volume} {13}},\ \bibinfo {pages} {94} (\bibinfo {year}
  {2016})}\BibitemShut {NoStop}%
\bibitem [{\citenamefont {Kondratiev}\ \emph {et~al.}(2017)\citenamefont
  {Kondratiev}, \citenamefont {Lobanov}, \citenamefont {Cherenkov},
  \citenamefont {Voloshin}, \citenamefont {Pavlov}, \citenamefont {Koptyaev},\
  and\ \citenamefont {Gorodetsky}}]{Kondratiev:17}%
  \BibitemOpen
  \bibfield  {author} {\bibinfo {author} {\bibfnamefont {N.~M.}\ \bibnamefont
  {Kondratiev}}, \bibinfo {author} {\bibfnamefont {V.~E.}\ \bibnamefont
  {Lobanov}}, \bibinfo {author} {\bibfnamefont {A.~V.}\ \bibnamefont
  {Cherenkov}}, \bibinfo {author} {\bibfnamefont {A.~S.}\ \bibnamefont
  {Voloshin}}, \bibinfo {author} {\bibfnamefont {N.~G.}\ \bibnamefont
  {Pavlov}}, \bibinfo {author} {\bibfnamefont {S.}~\bibnamefont {Koptyaev}}, \
  and\ \bibinfo {author} {\bibfnamefont {M.~L.}\ \bibnamefont {Gorodetsky}},\
  }\href {\doibase 10.1364/OE.25.028167} {\bibfield  {journal} {\bibinfo
  {journal} {Opt. Express}\ }\textbf {\bibinfo {volume} {25}},\ \bibinfo
  {pages} {28167} (\bibinfo {year} {2017})}\BibitemShut {NoStop}%
\bibitem [{\citenamefont {Voloshin}\ \emph {et~al.}(2021)\citenamefont
  {Voloshin}, \citenamefont {Kondratiev}, \citenamefont {Lihachev},
  \citenamefont {Liu}, \citenamefont {Lobanov}, \citenamefont {Dmitriev},
  \citenamefont {Weng}, \citenamefont {Kippenberg},\ and\ \citenamefont
  {Bilenko}}]{Voloshin2021}%
  \BibitemOpen
  \bibfield  {author} {\bibinfo {author} {\bibfnamefont {A.~S.}\ \bibnamefont
  {Voloshin}}, \bibinfo {author} {\bibfnamefont {N.~M.}\ \bibnamefont
  {Kondratiev}}, \bibinfo {author} {\bibfnamefont {G.~V.}\ \bibnamefont
  {Lihachev}}, \bibinfo {author} {\bibfnamefont {J.}~\bibnamefont {Liu}},
  \bibinfo {author} {\bibfnamefont {V.~E.}\ \bibnamefont {Lobanov}}, \bibinfo
  {author} {\bibfnamefont {N.~Y.}\ \bibnamefont {Dmitriev}}, \bibinfo {author}
  {\bibfnamefont {W.}~\bibnamefont {Weng}}, \bibinfo {author} {\bibfnamefont
  {T.~J.}\ \bibnamefont {Kippenberg}}, \ and\ \bibinfo {author} {\bibfnamefont
  {I.~A.}\ \bibnamefont {Bilenko}},\ }\href {\doibase
  10.1038/s41467-020-20196-y} {\bibfield  {journal} {\bibinfo  {journal}
  {Nature Communications}\ }\textbf {\bibinfo {volume} {12}},\ \bibinfo {pages}
  {235} (\bibinfo {year} {2021})}\BibitemShut {NoStop}%
\bibitem [{\citenamefont {Kondratiev}\ and\ \citenamefont
  {Lobanov}(2020)}]{KondratievBW2019}%
  \BibitemOpen
  \bibfield  {author} {\bibinfo {author} {\bibfnamefont {N.~M.}\ \bibnamefont
  {Kondratiev}}\ and\ \bibinfo {author} {\bibfnamefont {V.~E.}\ \bibnamefont
  {Lobanov}},\ }\href {\doibase 10.1103/PhysRevA.101.013816} {\bibfield
  {journal} {\bibinfo  {journal} {Phys. Rev. A}\ }\textbf {\bibinfo {volume}
  {101}},\ \bibinfo {pages} {013816} (\bibinfo {year} {2020})}\BibitemShut
  {NoStop}%
\bibitem [{\citenamefont {Godey}\ \emph {et~al.}(2014)\citenamefont {Godey},
  \citenamefont {Balakireva}, \citenamefont {Coillet},\ and\ \citenamefont
  {Chembo}}]{Chembo2014}%
  \BibitemOpen
  \bibfield  {author} {\bibinfo {author} {\bibfnamefont {C.}~\bibnamefont
  {Godey}}, \bibinfo {author} {\bibfnamefont {I.~V.}\ \bibnamefont
  {Balakireva}}, \bibinfo {author} {\bibfnamefont {A.}~\bibnamefont {Coillet}},
  \ and\ \bibinfo {author} {\bibfnamefont {Y.~K.}\ \bibnamefont {Chembo}},\
  }\href {\doibase 10.1103/PhysRevA.89.063814} {\bibfield  {journal} {\bibinfo
  {journal} {Phys. Rev. A}\ }\textbf {\bibinfo {volume} {89}},\ \bibinfo
  {pages} {063814} (\bibinfo {year} {2014})}\BibitemShut {NoStop}%
\bibitem [{\citenamefont {Herr}\ \emph {et~al.}(2014)\citenamefont {Herr},
  \citenamefont {Brasch}, \citenamefont {Jost}, \citenamefont {Wang},
  \citenamefont {Kondratiev}, \citenamefont {Gorodetsky},\ and\ \citenamefont
  {Kippenberg}}]{herr2014temporal}%
  \BibitemOpen
  \bibfield  {author} {\bibinfo {author} {\bibfnamefont {T.}~\bibnamefont
  {Herr}}, \bibinfo {author} {\bibfnamefont {V.}~\bibnamefont {Brasch}},
  \bibinfo {author} {\bibfnamefont {J.~D.}\ \bibnamefont {Jost}}, \bibinfo
  {author} {\bibfnamefont {C.~Y.}\ \bibnamefont {Wang}}, \bibinfo {author}
  {\bibfnamefont {N.~M.}\ \bibnamefont {Kondratiev}}, \bibinfo {author}
  {\bibfnamefont {M.~L.}\ \bibnamefont {Gorodetsky}}, \ and\ \bibinfo {author}
  {\bibfnamefont {T.~J.}\ \bibnamefont {Kippenberg}},\ }\href {\doibase
  10.1038/nphoton.2013.343} {\bibfield  {journal} {\bibinfo  {journal} {Nat.
  Photon.}\ }\textbf {\bibinfo {volume} {8}},\ \bibinfo {pages} {145} (\bibinfo
  {year} {2014})}\BibitemShut {NoStop}%
\bibitem [{\citenamefont {Bao}\ \emph {et~al.}(2014)\citenamefont {Bao},
  \citenamefont {Zhang}, \citenamefont {Matsko}, \citenamefont {Yan},
  \citenamefont {Zhao}, \citenamefont {Xie}, \citenamefont {Agarwal},
  \citenamefont {Kimerling}, \citenamefont {Michel}, \citenamefont {Maleki},\
  and\ \citenamefont {Willner}}]{Bao:14}%
  \BibitemOpen
  \bibfield  {author} {\bibinfo {author} {\bibfnamefont {C.}~\bibnamefont
  {Bao}}, \bibinfo {author} {\bibfnamefont {L.}~\bibnamefont {Zhang}}, \bibinfo
  {author} {\bibfnamefont {A.}~\bibnamefont {Matsko}}, \bibinfo {author}
  {\bibfnamefont {Y.}~\bibnamefont {Yan}}, \bibinfo {author} {\bibfnamefont
  {Z.}~\bibnamefont {Zhao}}, \bibinfo {author} {\bibfnamefont {G.}~\bibnamefont
  {Xie}}, \bibinfo {author} {\bibfnamefont {A.~M.}\ \bibnamefont {Agarwal}},
  \bibinfo {author} {\bibfnamefont {L.~C.}\ \bibnamefont {Kimerling}}, \bibinfo
  {author} {\bibfnamefont {J.}~\bibnamefont {Michel}}, \bibinfo {author}
  {\bibfnamefont {L.}~\bibnamefont {Maleki}}, \ and\ \bibinfo {author}
  {\bibfnamefont {A.~E.}\ \bibnamefont {Willner}},\ }\href {\doibase
  10.1364/OL.39.006126} {\bibfield  {journal} {\bibinfo  {journal} {Opt.
  Lett.}\ }\textbf {\bibinfo {volume} {39}},\ \bibinfo {pages} {6126} (\bibinfo
  {year} {2014})}\BibitemShut {NoStop}%
\bibitem [{\citenamefont {Jang}\ \emph {et~al.}(2021)\citenamefont {Jang},
  \citenamefont {Okawachi}, \citenamefont {Zhao}, \citenamefont {Ji},
  \citenamefont {Joshi}, \citenamefont {Lipson},\ and\ \citenamefont
  {Gaeta}}]{jang2021conversion}%
  \BibitemOpen
  \bibfield  {author} {\bibinfo {author} {\bibfnamefont {J.~K.}\ \bibnamefont
  {Jang}}, \bibinfo {author} {\bibfnamefont {Y.}~\bibnamefont {Okawachi}},
  \bibinfo {author} {\bibfnamefont {Y.}~\bibnamefont {Zhao}}, \bibinfo {author}
  {\bibfnamefont {X.}~\bibnamefont {Ji}}, \bibinfo {author} {\bibfnamefont
  {C.}~\bibnamefont {Joshi}}, \bibinfo {author} {\bibfnamefont
  {M.}~\bibnamefont {Lipson}}, \ and\ \bibinfo {author} {\bibfnamefont {A.~L.}\
  \bibnamefont {Gaeta}},\ }\href {\doibase 10.1364/OL.423654} {\bibfield
  {journal} {\bibinfo  {journal} {Opt. Lett.}\ }\textbf {\bibinfo {volume}
  {46}},\ \bibinfo {pages} {3657} (\bibinfo {year} {2021})}\BibitemShut
  {NoStop}%
\bibitem [{\citenamefont {Xue}\ \emph {et~al.}(2017)\citenamefont {Xue},
  \citenamefont {Wang}, \citenamefont {Xuan}, \citenamefont {Qi},\ and\
  \citenamefont {Weiner}}]{Xue:17}%
  \BibitemOpen
  \bibfield  {author} {\bibinfo {author} {\bibfnamefont {X.}~\bibnamefont
  {Xue}}, \bibinfo {author} {\bibfnamefont {P.-H.}\ \bibnamefont {Wang}},
  \bibinfo {author} {\bibfnamefont {Y.}~\bibnamefont {Xuan}}, \bibinfo {author}
  {\bibfnamefont {M.}~\bibnamefont {Qi}}, \ and\ \bibinfo {author}
  {\bibfnamefont {A.~M.}\ \bibnamefont {Weiner}},\ }\href {\doibase
  10.1002/lpor.201600276} {\bibfield  {journal} {\bibinfo  {journal} {Laser \&
  Photonics Reviews}\ }\textbf {\bibinfo {volume} {11}},\ \bibinfo {pages}
  {1600276} (\bibinfo {year} {2017})}\BibitemShut {NoStop}%
\bibitem [{\citenamefont {Karpov}\ \emph {et~al.}(2019)\citenamefont {Karpov},
  \citenamefont {Pfeiffer}, \citenamefont {Guo}, \citenamefont {Weng},
  \citenamefont {Liu},\ and\ \citenamefont {Kippenberg}}]{Karpov2019}%
  \BibitemOpen
  \bibfield  {author} {\bibinfo {author} {\bibfnamefont {M.}~\bibnamefont
  {Karpov}}, \bibinfo {author} {\bibfnamefont {M.~H.~P.}\ \bibnamefont
  {Pfeiffer}}, \bibinfo {author} {\bibfnamefont {H.}~\bibnamefont {Guo}},
  \bibinfo {author} {\bibfnamefont {W.}~\bibnamefont {Weng}}, \bibinfo {author}
  {\bibfnamefont {J.}~\bibnamefont {Liu}}, \ and\ \bibinfo {author}
  {\bibfnamefont {T.~J.}\ \bibnamefont {Kippenberg}},\ }\href {\doibase
  10.1038/s41567-019-0635-0} {\bibfield  {journal} {\bibinfo  {journal} {Nature
  Physics}\ }\textbf {\bibinfo {volume} {15}},\ \bibinfo {pages} {1071}
  (\bibinfo {year} {2019})}\BibitemShut {NoStop}%
\end{thebibliography}%

\end{document}


\title{Supplementary info -- Hybrid integrated dual-microcomb source}

\author{Nikita Yu. Dmitriev}
\thanks{These authors contributed equally to this work.}
\affiliation{Russian Quantum Center, Moscow, 143026, Russia}
\affiliation{Moscow Institute of Physics and Technology (MIPT), Dolgoprudny, Moscow Region, 141701, Russia}

\author{Sergey N. Koptyaev}
\thanks{These authors contributed equally to this work.}
\affiliation{Samsung R\&D Institute Russia, SAIT-Russia Laboratory, Moscow, 127018, Russia}

\author{Andrey S. Voloshin}
\thanks{These authors contributed equally to this work.}
\affiliation{Institute of Physics, Swiss Federal Institute of Technology Lausanne (EPFL), CH-1015 Lausanne, Switzerland}

\author{Nikita M. Kondratiev}
\thanks{These authors contributed equally to this work.}
\affiliation{Russian Quantum Center, Moscow, 143026, Russia}

\author{Valery E. Lobanov}
\affiliation{Russian Quantum Center, Moscow, 143026, Russia}

\author{Kirill N. Min'kov}
\affiliation{Russian Quantum Center, Moscow, 143026, Russia}

\author{Maxim V. Ryabko}
\affiliation{Samsung R\&D Institute Russia, SAIT-Russia Laboratory, Moscow, 127018, Russia}

\author{Stanislav V. Polonsky}
\affiliation{Samsung R\&D Institute Russia, SAIT-Russia Laboratory, Moscow, 127018, Russia}

\author{Igor A. Bilenko}
\affiliation{Russian Quantum Center, Moscow, 143026, Russia}
\affiliation{Faculty of Physics, M.V. Lomonosov Moscow State University, 119991 Moscow, Russia}

\maketitle

\section*{Supplementary Note 1: Microcomb parameters reproducibility}

\begin{figure}[ht!]
\includegraphics[width=\linewidth]{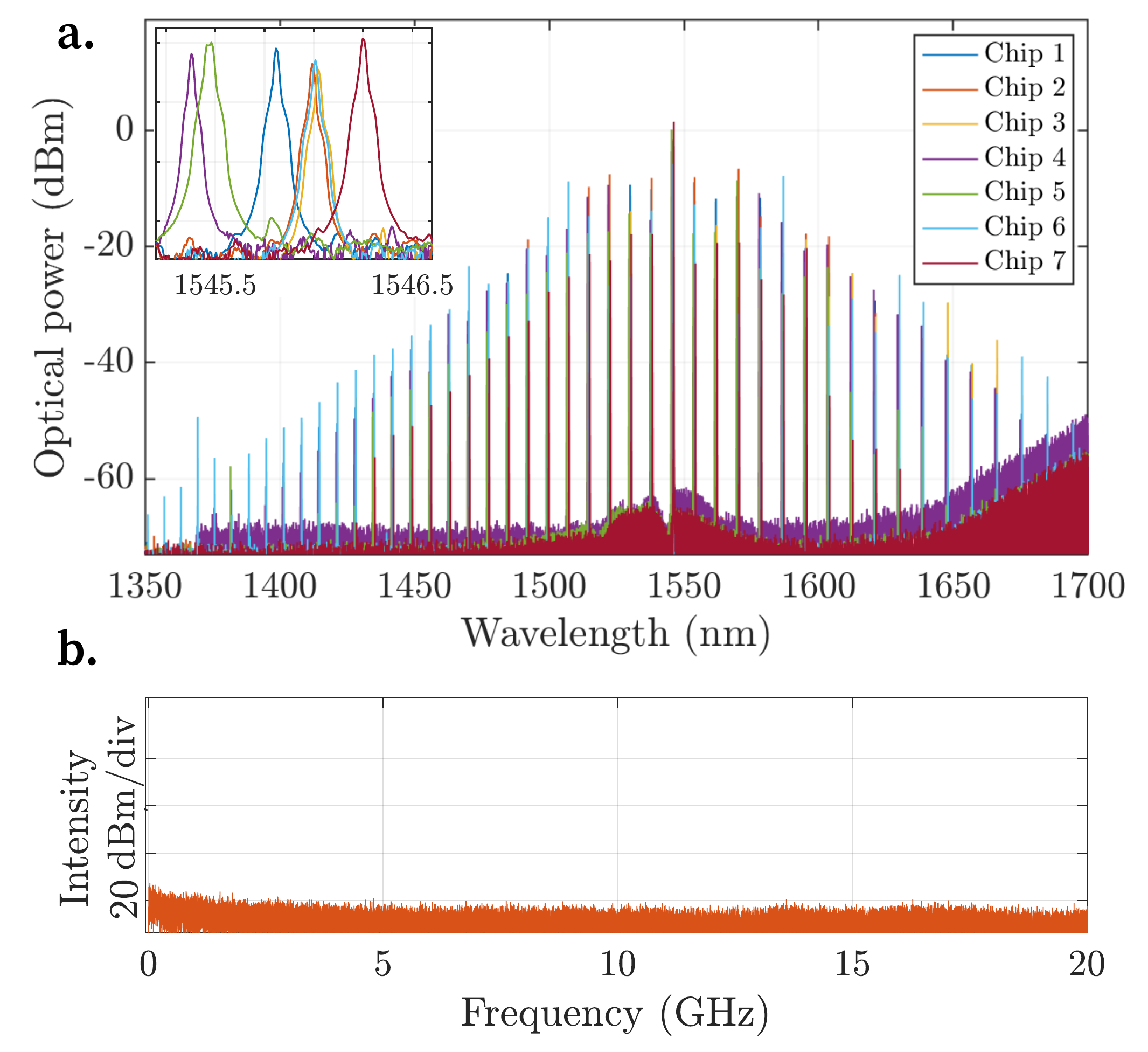}
\caption{ Soliton microcombs generated at different photonic chips \textbf{a.} Optical spectra of the generated soliton microcombs at 7 different chips with the same design and pumped with laser diode. Inset: zoom-in of the microcombs central line area.  \textbf{b.} RF spectrum of the generated microcombs demonstrating the absence of the low frequency noise}
\label{fig:fig5}
\end{figure}

In terms of commercially available end-product development for practical applications device characteristics reproducibility and yield ratio are extremely important. Concerning integrated microcomb or  dual-comb source considered here least reliable and most unpredictable element, responsible for device performance, is a photonic chip with high-Q SiN microresonator. 
The process used for photonic chip fabrication has some errors expressed in the deviation of geometric parameters and variation in the material refractive index. Despite such deviations are extremely small in absolute values, it can lead to significant variations in the microresonators' parameters such as 
coupling rate, microresonator eigenmodes' positions and dispersion characteristics. The microresonators' parameters affect generated microcombs or even significantly complicate its generation. In order to assess the fabrication deviations impact on device performance we examined 7 photonic chips with the same design from different wafer areas. As a result, we successfully generated microcombs using every chip we had and have not even notice significant difference in microcomb generation process from chip to chip. Spectrum of each microcomb was recorded using optical spectrum analyzer (OSA) (Fig.\ref{fig:fig5}(a)) and also to distinguish coherent (soliton) state low frequency noise of generated microcombs was observed using electrical spectrum analyzer (ESA) (Fig. \ref{fig:fig5}(b)). Comparison was made for microcombs generated in 1THz microresonators. 
All generated microcombs feature similar envelope, which indicates to similar dispersion landscape. The maximum observed deviation between microcomb lines position corresponding to same serial number is about 0.9 nm corresponding to $\sim$112 GHz (Fig. \ref{fig:fig5}(a), inset). Such deviations are usually compensated using powerful micro-heaters or piezo-elements above microresonator, which allows to change its FSR and shift eigenfrequency up to hundreds of GHz \cite{Xue:16,Liu2020}. However, micro-heaters used in this experiment enable to shift eigenfrequencies only up to tens of GHz. With regard to this fact, only several photonic chips from the set enable microcombs matching to observe dual-comb signal in RF range. 

Thus, we have obtained microcomb generation with photonic chips from different wafer areas and thereby have demonstrated that fabrication process deviations do not lead to inability of microcombs generation in case of using self-injection locked laser diode for pumping. Also, using 1 THz FSR microresonator we have estimated deviations of eigenfrequencies position across the wafer and selected photonic chips which provide microcombs matched to observe  RF dual-comb signal.

\section*{Supplementary Note 2: Laser diodes comparison}
As a part of conducted experiment, we have compared two different types of laser diodes (LDs), Fabry-Perot (FP) and distributed feedback (DFB), in terms of practical use. 
Single mode multi-frequency FP diodes manufactured by Seminex Corp., USA, used in the experiment features $\sim$ 35 GHz longitudinal mode spacing, 1535 nm central wavelength and $\sim$ 200 mW peak output optical power at 500 mA of the injection current. The single frequency DFB diodes have 1545 nm wavelength, side-mode suppression ratio (SMSR) of $\sim$ -50 dB and peak optical power of $\sim$ 100 mW at 400 mA. Each type of the LDs was used in turn for microcomb generation using the same microresonators. The final results obtained for both types of the LDs looks similar and there was no dramatic differences in generated microcombs' properties. However, during the process of alignment and adjustments of the experimental setup different types of LDs demonstrate extremely different behaviour. Here we describe the features of application for both types of the LDs, as well as present comparative analysis in terms of practical application, describing its advantages and drawbacks.

DFB features single frequency emission, that  makes the process of the alignment and microresonator's mode excitation is much more predictable and well managed. 
Using FP one faces with modes competition and mode-hopping effects, since a number of LD longitudinal mode can be locked to several different microresonator eigenfrequencies. Thus, microcomb generation process with FP LD is much more sensitive to alignment. Any minor translation, which does not affect microcomb generation in case of DFB LD pumping, can cause significant changes due to mode-hopping or even destroy microcomb generation in case of FP LD use. 

However, often it is easier to find and excite the mode using exactly FP LD when it comes to microresonator with large FSR, for example 1 THz. Also, as a rule, FP LD in comparison with DFB LD features more output power  and higher power efficiency, that sometimes plays a crucial role. Indeed, power is a critical parameter for observing nonlinear effects, such as generation of an optical comb, especially when it comes to complex optical schemes with optical splitters. Moreover, powerful DFB LDs in comparison with FP are much more expensive and less available. In fact, the accessibility, price and interchangeability of elements are very important in the context of the commercial product development. 

Thus, it is more expedient to use more affordable and powerful FP diodes for commercial product development and DFB diodes for research purposes on experimental stands in laboratory. Both of these diodes allows "turn-key" comb generation and applicable
for packaged device.

\section*{Supplementary Note 3: Self-injection locking and thermal effects}
Most of the thermal effects are suppressed as the laser frequency becomes locked to the microresonator and laser-microresonator detuning becomes fixed. If the microresonator frequency is disturbed by the thermal effects, the generation frequency also changes, maintaining the comb generation regime \cite{KondratievSPIE21}. We repeat this important derivation in more details. The frequency comb generation is described by the nonlinear equation of the form
\begin{align}
    \label{resonator_eq}
    \dot a_\mu=-(1-i\zeta-id_{\Sigma_\mu})a_\mu+iS_\mu+f\delta_\mu,
\end{align}
where $a$ is the microresonator mode field,
$d_{\Sigma_\mu}= d_2\mu^2+d_3\mu^3...$ is the integrated dispersion, $\zeta=2(\omega_{\rm gen}-\omega_0)/\kappa$ is the normalized pump frequency detuning ($\omega_{\rm gen}$ is the generation frequency, $\omega_0$ and $\kappa$ are the pumped WGM frequency and linewidth),
$S_\mu$ is the nonlinear sum and $f$ is the normalized pump \cite{Kondratiev:20}. If we turn to the thermal-influenced equations they are modified as following \cite{Guo:16}:
\begin{align}
    \label{thermal_res_eq}
    \dot a_\mu=&-(1-i\zeta-i\theta-id_{\Sigma_\mu})a_\mu+iS_\mu+f\delta_\mu,\\
    \dot \theta=&\tilde\kappa_\theta\left(r_\theta P_a -\theta\right),
\end{align}
where $\theta$ is normalized thermal frequency shift, $\tilde\kappa_\theta$ and $r_\theta$ are the thermal to optical decay rate and nonlinearity ratios and $P_a=\sum|a_\mu|^2$ is the total intracavity intensity. Comparing equations \eqref{resonator_eq} and \eqref{thermal_res_eq} we can see that the soliton generation is governed by the effective detuning $\zeta_{\rm eff}=\zeta+\theta$. Note that negative thermorefraction coefficient corresponds to negative $r_\theta$ and, consequently negative $\theta$. In stationary regime $\theta=r_\theta P_a$, so we can see that the nonlinear resonance "tilts" to the lower frequencies, while the soliton steps do not shift significantly (the bigger shift, the more the soliton number) and can become inaccessible \cite{Guo:16}.

In the SIL regime the system should be completed with two more equation systems - for the backward wave amplitudes $b_\mu$ (note that the thermal equation should also be modified to have the full microresonator power as a source) and for the laser amplitude $a_l$. To perform the numerical modeling the laser medium gain model also should be specified (we use the normalized carrier concentration $N_g$ rate equations). The final system can be written as following \cite{KondratievSPIE21}
\begin{align}
\frac{dN_g}{d\tau} =& \frac{g}{ g_0} \frac{f^2}{\tilde\kappa_{\rm WGR}^2}(\tilde\kappa_l-N_g  |a_l|^2)+\tilde\kappa_N(\tilde\kappa_l - N_g),\\
\label{Model_las_n}
\frac{da_l}{d\tau}=&\left(-i\xi_0-iv_\xi \tau+\alpha^c_g N_g-\tilde \kappa_l\right)a_l-\frac{\tilde K_0}{f}b_0 e^{-i\psi_s}\\
\label{Model_a}
\frac{da_\mu}{d\tau}=&-(1-id_{\Sigma_\mu}-i\theta) a_\mu+i\beta b_{\mu}+i S_\mu^a +f \delta_{\mu 0},\\  
\label{Model_b}
\frac{db_\mu}{d\tau}=&-(1-id_{\Sigma_\mu}-i\theta) b_\mu+i\beta^* a_\mu+i S_\mu^b\\
\label{Model_T}
\frac{d\theta}{d\tau}=&\frac{\kappa_\theta}{\kappa}\left(r_\theta(P_a+P_b)-\theta\right),  
\end{align}
where the $\tau=\kappa t/2$ is normalized time,
$g/ g_l$ is the laser gain to Kerr nonlinearity rate ratio,
$\tilde\kappa_{\rm WGR}$ is the coupling of the microresonator normalized to the microresonator mode linewidth,
$\tilde\kappa_{l}$ and $\tilde\kappa_{N}$ are the laser field and carrier relaxation rates normalized to the microresonator mode linewidth,
$\alpha_g^c=(1+i\alpha_g)$ and $\alpha_g$ is the Henry factor,
$\psi_s$ is the locking phase,
$\xi_0$ is the initial laser detuning and 
$v_\xi=v_{f}[{\rm Hz/s}]\frac{8\pi}{\kappa^2}$ is the scan speed in half-linewidth per reversed half-linewidth,
$\tilde K_0=\frac{K_0}{2\beta\sqrt{1+\alpha_g^2}}$ is the resonator to laser coupling coefficient, defined by the stabilization coefficient,
$K_0=8\eta\beta\tilde\kappa_{do}$ is the zero-detuning stabilization coefficient,
$\tilde\kappa_{do}$ is the ratio of the laser output mirror coupling rate to the WGM linewidth,
$\beta$ is the normalized scattering rate in microresonator. 
It can be noted that $K_0$ also approximately equal to triple locking width in units of WGM linewidth in linear regime. Solving these equations numerically one may see that the generated comb regime does not depend on the value of the thermal nonlinearity up to significant values.

Some more insight can be obtained using the tuning curve approach \cite{Kondratiev:17}. We can see \cite{KondratievSPIE21} that the resulting tuning curve for $\zeta_{\rm eff}$ coincides with the solution without thermal influence. Using nonlinear shifts $\bar \zeta=\zeta_{\rm eff}+\delta\zeta_{\rm nl}$, ${\bar\beta}^2=\delta\beta_{\rm nl}^2+\beta^2$ and $\bar \xi=\xi+\delta\zeta_{\rm nl}$ for the tuning curve over the laser cavity detuning $\xi$ we can obtain an implicit expression\cite{Voloshin2021}:
\begin{align}
\delta\beta_{\rm nl}&=\frac{2\alpha_x-1}{2}|f|^2\frac{1+(\bar\zeta+\delta\beta_{\rm nl})^2-\beta^2}{(1+{\bar\beta}^2-{\bar \zeta}^2)^2+4{\bar \zeta}^2},\\
\label{tune_intermediate}
\xi&=\zeta_{\rm eff}+\frac{K_0}{2}\frac{2\bar\zeta\cos\bar\psi+(1+\bar\beta^2-\bar\zeta^2)\sin\bar\psi}{(1+\bar\beta^2-\bar\zeta^2)^2+4\bar\zeta^2},\\
\label{zetaeff}
\zeta_{\rm eff}&=\bar\zeta-\frac{2\alpha_x+1}{2}|f|^2\frac{1+(\bar\zeta+\delta\beta_{\rm nl})^2+\beta^2}{(1+{\bar\beta}^2-{\bar \zeta}^2)^2+4{\bar \zeta}^2},
\end{align}
where 
$\bar\psi=\psi_0(\omega_m\tau_s)+\frac{\kappa\tau_s}{2}\zeta$ is the locking phase
($\tau_s$ is the roundtrip time from the laser to the microresonator).
From the equations \eqref{tune_intermediate}-\eqref{zetaeff} we can see that exactly the $\zeta_{\rm eff}$ is the detuning that is stabilized, while the generation detuning $\zeta$ will follow the temperature change. Furthermore, in the locked state this detuning becomes fixed to the detuning near
\begin{align}
    \label{zeta0}
    \zeta_{\rm eff}^0=
    \begin{cases}
        -\frac{3}{2}f^2, &f\ll1\\
        -3\left(\frac{f^2}{2}\right)^{1/3}+\left(\frac{f^2}{2}\right)^{-1/3}, &f>1\\
    \end{cases}
\end{align}
This expression is obtained for $\alpha_x=1$, low $\beta\ll1$ and $\bar\psi=0$. We choose $\bar\zeta=0$, which guarantees it being in the locked state. This is a good estimation if the pump is not very high ($f<2$), where the tuning and resonance curves are more-or-less symmetric, without self-intersections and good stabilization can be achieved. The sample solutions of \eqref{tune_intermediate}-\eqref{zetaeff} for different $f$ are shown in Supplementary Fig. \ref{sfig:Nltunecurves}. It can be shown that $\zeta_{\rm eff}^0$ always lies inside the soliton generation region $\zeta_{\rm eff} \in[-3(f/2)^{2/3}+(f/2)^{-2/3}/4;-\pi^2f^2/8]$ \cite{KondratievBW2019, Chembo2014}. Unfortunately, the lower boundary of the soliton existence region approximation is not very good near $f=1$ as they are based on the resonance curve bistability, which appears only after $f=1.24$.

\begin{figure}[t!]
\includegraphics[width=0.9\linewidth]{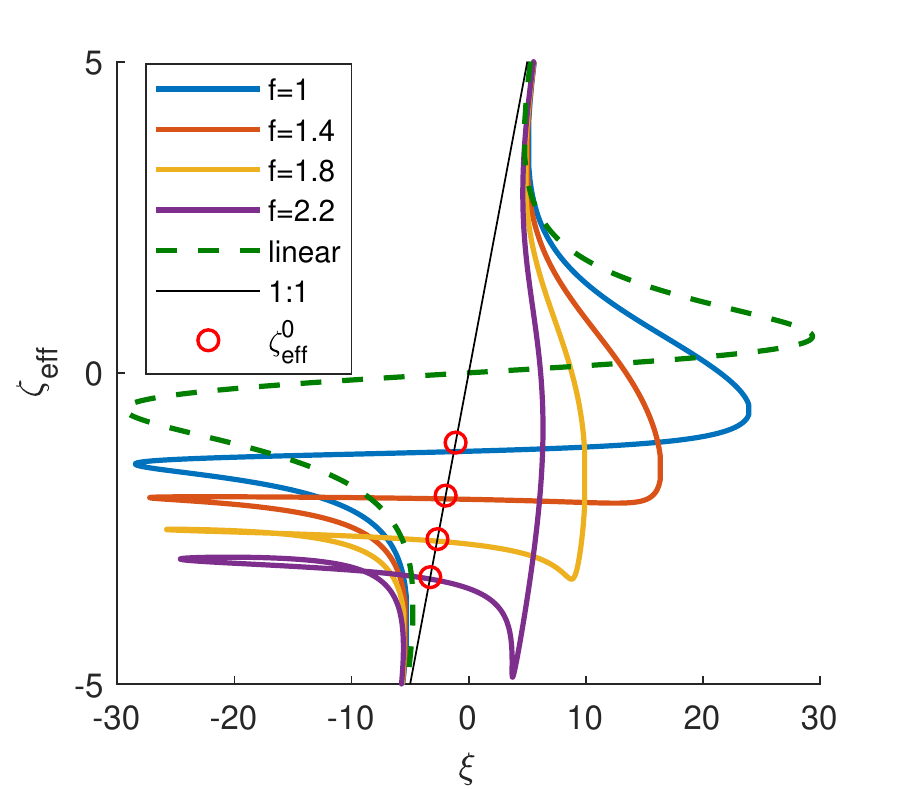}
\caption{Sample solutions of \eqref{tune_intermediate}-\eqref{zetaeff} for different pump amplitude $f$ for parameters $\alpha_x=1$, $\bar\psi=0$, $\beta=0.11$, $K_0=88$ together with locked effective detunings \eqref{zeta0} (red circles).}
\label{sfig:Nltunecurves}
\end{figure}

\section*{Supplementary Note 4: Comb envelope approximation and power efficiency}

The generated comb for single soliton can be calculated as following \cite{herr2014temporal}:
\begin{align}
\label{comblines}
    a_\mu&=\sqrt{\frac{d_2}{2}}\,{\rm sech}\left(\frac{\pi\mu}{2}\sqrt{\frac{d_2}{-\zeta_{\rm eff}}}\right)e^{i\psi^{\rm sol}}\\
    \psi^{\rm sol}&=\arctan\frac{\sqrt{-2\pi^2f^2\zeta_{\rm eff}-16\zeta_{\rm eff}^2}}{-4\zeta_{\rm eff}}.
\end{align}
Note that the total comb (or soliton) power $P_{\rm soliton}=2\sqrt{-\zeta_{\rm eff}d_2}/\pi$ does not depend on the pump amplitude
\begin{align}
f=\sqrt{\frac{8g}{\kappa^3}}\sqrt{\frac{\eta\kappa}{\tau_{\rm WGM}}}A_{l}|_{\rm coupler}=\sqrt{\frac{8cn_2Q^2\eta P_{\rm coupler}}{\omega n_g^2V_{\rm eff}}},
\end{align}
meaning that the higher the pump, the lower the comb generation efficiency. In the formula
$g$ is nonlinearity rate, $\kappa$ is the WGM linewidth, 
$\eta=\kappa_c/\kappa$ is the normalized coupling rate, 
$\tau_{\rm WGM}$ is the WGM round-trip time, 
$A_{l}|_{\rm coupler}$ and $P_{\rm coupler}$ are the pump field amplitude and power inside the coupler,
$n_2$ is the nonlinear index,
$Q$ is the loaded quality factor,
$n_g$ is the WGM group index,
$V_{\rm eff}$ is the WGM mode volume.
It is convenient to introduce the parametric instability threshold power $P_{\rm th}$, so that $f=\sqrt{P_{\rm coupler}/P_{\rm th}}$. Upon reaching this power the nonlinear generation process starts.
To turn to the output comb amplitude we have to recall the nonlinearity normalization
\begin{align}
A_\mu^{\rm out}=-\sqrt{\eta\kappa\tau_{\rm WGM}} a_\mu\sqrt{\frac{\kappa}{2g}}=\frac{2\eta}{f}a_\mu A_{l}|_{\rm coupler}.
\end{align}
Using the threshold power we can write out a simple expression for the output comb line power in the form $P_\mu^{\rm out}=4\eta^2|a_\mu|^2P_{\rm th}$ and the total power
\begin{align}
\label{Pouttotal}
P_{\rm soliton}^{\rm out}=\frac{8\eta^2}{\pi}\sqrt{-d_2\zeta_{\rm eff}}P_{\rm th}.
\end{align}
This brings out an nontrivial fact that lowering the threshold power also reduces the output comb power.

The central line in the output light also interferes with the background of nonlinear resonance curve
\begin{align}
A_{\rm CW}^{\rm out}=\left(1-\frac{2\eta}{1-i\zeta_{\rm eff}-ia_{\rm CW}^2}\right)A_{l}|_{\rm coupler},
\end{align}
and $A_{\rm CW}^2$ is the squared absolute value of the stationary nonlinear resonance. In our case we take the lower branch of bistability region
\begin{align}
a_{\rm CW}^2=&-\frac{2}{3}\zeta_{\rm eff}-\frac{2}{3}\sqrt{\zeta_{\rm eff}^2-3}\times\nonumber\\
&{\rm cosh} \frac{1}{3}{\rm arccosh}\frac{-2\zeta_{\rm eff}^3-18\zeta_{\rm eff}-27f^2}{2(\zeta_{\rm eff}^2-3)^{3/2}}
\end{align}

The power efficiency can be calculated in two ways. First is the "pump to total comb" or "pump to solitons" ($\eta_{\rm p2s}$) efficiency \cite{Bao:14,jang2021conversion}, where the total comb power \eqref{Pouttotal} is divided by the pump power in the coupler $P_{\rm coupler}$. It is usually used in theoretical calculations giving out more tidy formulas. The second way can be referred to as "pump to comb sidebands" or "pump to comb" ($\eta_{\rm p2c}$), when the central (pumped) line is excluded \cite{Xue:17}:
\begin{align}
\label{eff_p2c}
\eta_{\rm p2c}=\frac{4\eta^2}{f^2}\left(\sum |a_\mu|^2-|a_0|^2\right)=\frac{4\eta^2}{f^2}\left(\frac{2}{\pi}\sqrt{-d_2\zeta_{\rm eff}}-\frac{d_2}{2}\right),
\end{align}
where the first summand is effectively "pump to solitons" $\eta_{\rm p2s}$ efficiency and the subtracted term is the central line power. This is convenient in experiment  not to bother with the interference with the pump. It can also be more informative for those more interested in sideband power than in the total comb power. In the main article we use this variant.
At maximum detuning $\zeta_{\rm max}=-\pi^2f^2/8$ the first summand of \eqref{eff_p2c} is
\begin{align}
\label{eff_p2s}
\eta^{\rm max}_{\rm p2s}\approx\frac{4\eta^2}{f}\sqrt{\frac{d_2}{2}}=\eta^{3/2}\sqrt{\frac{d_2\kappa  n_g^2 V_{\rm eff}}{cn_2Q P_{\rm coupler}}},
\end{align}
which coincides with expression for pump-to-soliton efficiency in \cite{Bao:14,jang2021conversion}. 
We can see that even at maximum detuning, that grows with the pump amplitude, the power efficiency scales down. In the SIL regime the efficiency is lower as the detuning is locked to lower value \eqref{zeta0}. Nevertheless the high efficiency can be maintained due to the possibility of lower pump power usage. 

The self-injection locking usually leads to the multi-soliton state, which also increases the pump-to-comb efficiency. To model the $N$-soliton comb we multiply the \eqref{comblines} with
\begin{align}
\Phi_\mu=\sum_{k=1}^{N} \exp{i\mu\phi_k},
\end{align}
where $\phi_k$ is the $k$-soliton angle position on the circumference. If the solitons are equidistant (so-called perfect solitonic crystal \cite{Karpov2019}) the comb will have $N$-FSR spacing. However, if the solitons have random angle distribution we get 1-FSR spaced comb and its sech$^2$ envelope experience non-smooth modulation. The total comb power (and the pump-to-comb efficiency) grows linearly with the number of solitons until the mutual distance between them is less than 5 soliton widths ($N_{\rm sat}\approx 2\pi/(10\sqrt{-d_2/\zeta_{\rm eff}}{\,\rm arccosh}\sqrt{2})$) \cite{herr2014temporal}. Up to this moment the pump-to-sidebands efficiency saturates (see Supplementary Fig. \ref{sfig:Supp_fig_eff}). We should also note that this number is usually inaccessible as the detuning is usually scanned from red to blue side and the soliton formation occurs at low $\zeta$ and their number does not grow. So the practical estimation for maximum soliton number would be $N_{\rm max}\approx \sqrt{1/d_2}$ \cite{Karpov2019}.

\begin{figure}[t!]
\includegraphics[width=0.9\linewidth]{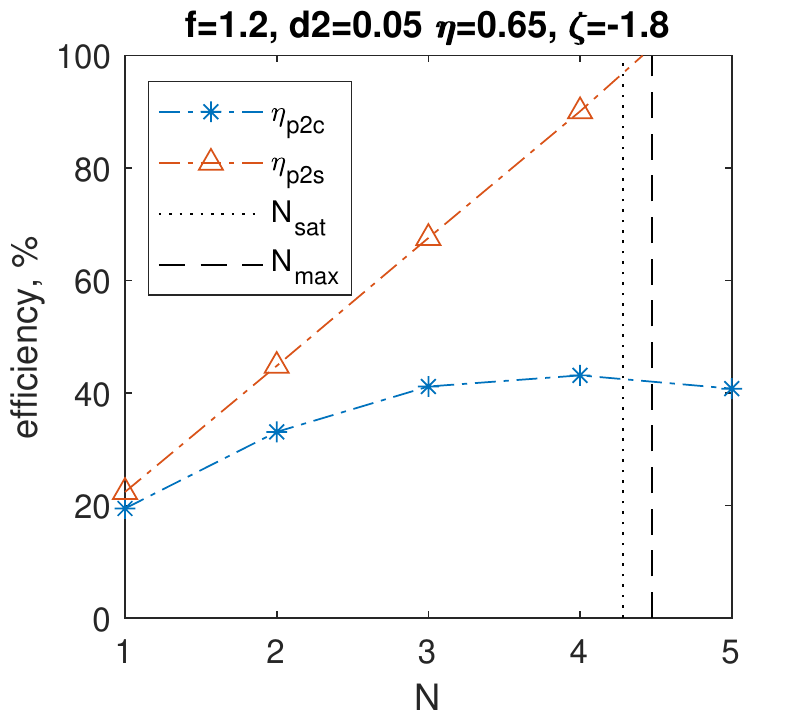}
\caption{Pump to soliton \eqref{eff_p2s} and pump to comb sidebands \eqref{eff_p2c} efficiency dependence on the soliton number for parameters, close to 1 THz chip. The former is shown with red triangles, the latter -- with blue asterisk. Vertical dashed and dotted lines show the 5-width and maximum soliton number respectively.}
\label{sfig:Supp_fig_eff}
\end{figure}
To sum up this section, the comb power does not depend on the pump power directly. Though both the maximal and the locked detunings do depend on the pump power, these dependencies are weaker than the linear one and the pump-to-comb efficiency decreases with it. This creates the illusion that using the low power laser and reducing the threshold we can make a very energy-efficient device.
However, the second point is that the comb power does depend on the nonlinearity threshold power. So, using the above strategy we end up with negligible output signal. This brings us to rather counter-intuitive conclusion, that for best performance the threshold power should be increased (meaning lower Q-factors or less nonlinearity, for example) and matched with the used laser pump power. This immediately brings out a trade-off problem as the lower Q-factor means less stabilization and wider beatnote.
Another way to increase the pump-to-comb efficiency is the second order dispersion coefficient increase. Unfortunately, this brings us to a trade-off problem for the comb width.

\bibliography{bibliography}